\documentclass[referee,fleqn,usenatbib]{mnras}

\usepackage[T1]{fontenc}
\usepackage{ae,aecompl}

\usepackage{graphicx}	
\usepackage{amsmath}	
\usepackage{amssymb}	
\usepackage{float}
\usepackage{booktabs}
\usepackage{rotating}
\usepackage{natbib}
\usepackage{fmtcount}
\usepackage{threeparttablex}
\usepackage{multirow}
\usepackage{morefloats}
\usepackage{dblfloatfix}
\usepackage{dcolumn}
\usepackage{amsmath}
\usepackage{latexsym}
\usepackage{longtable}
\usepackage{lipsum} 
\usepackage{textcomp}
\usepackage{gensymb}
\usepackage{mathtools}
\usepackage{afterpage}

\title[SN 2015ba in IC~1029]{SN 2015ba: a Type IIP supernova with a long plateau.}

\author[Raya Dastidar et al.]{Raya Dastidar$^{1,2}$\thanks{E-mail: rayadastidar@aries.res.in, rdastidr@gmail.com},
 Kuntal Misra$^{1}$, G. Hosseinzadeh$^{3,4}$, A. Pastorello$^{5}$,\newauthor
M. L. Pumo$^{5,6,7}$, S. Valenti$^{8}$, C. McCully$^{3,4}$, L. Tomasella$^{5}$, I. Arcavi$^{3,4,9}$, \newauthor N. Elias-Rosa$^{5}$, Mridweeka Singh$^{1,10}$, Anjasha Gangopadhyay$^{1,10}$, D. A. Howell$^{3,4}$, \newauthor Antonia Morales-Garoffolo$^{11}$, L. Zampieri$^{5}$, Brijesh Kumar$^{1}$, M. Turatto$^{5}$, \newauthor S.  Benetti$^{5}$, L. Tartaglia$^{12}$, P. Ochner$^{5}$, D. K. Sahu$^{13}$, G.C. Anupama$^{13}$ \newauthor and S. B. Pandey$^{1}$
\\
$^{1}$Aryabhatta Research Institute of observational sciencES, Manora Peak, Nainital 263 001 India\\
$^{2}$Department of Physics \& Astrophysics, University of Delhi, Delhi-110 007\\
$^{3}$Las Cumbres Observatory, 6740 Cortona Dr., Suite 102, Goleta, CA 93117-5575, USA\\
$^{4}$Department of Physics, University of California, Santa Barbara, CA 93106-9530, USA\\
$^{5}$INAF Osservatorio Astronomico di Padova, Vicolo dell'Osservatorio 5, 35122 Padova, Italy\\
$^{6}$Universit$\grave{a}$ degli Studi di Catania, DIEEI and DFA, Via Santa Sofia 64, I-95123 Catania, Italy\\
$^{7}$INFN-Laboratori Nazionali del Sud, Via Santa Sofia 62, I-95123 Catania, Italy\\
$^{8}$Department of Physics, University of California, 1 Shields Ave, Davis, CA 95616-5270, USA\\
$^{9}$Einstein Fellow\\
$^{10}$Pt.Ravi Shankar Shukla University, Raipur 492 010,  India\\
$^{11}$Department of Applied Physics, University of C$\acute{a}$diz, Campus of Puerto Real, E-11510 C$\acute{a}$diz, Spain\\
$^{12}$The Oskar Klein Centre, Department of Astronomy, AlbaNova, SE 106 91, Stockholm, Sweden\\
$^{13}$Indian Institute of Astrophysics, Koramangala, Bangalore 560 034, India
}
\date{Accepted XXX. Received YYY; in original form ZZZ}

\pubyear{2018}

\begin{document}
\label{firstpage}
\pagerange{\pageref{firstpage}--\pageref{lastpage}}
\maketitle

\begin{abstract}
We present optical photometry and spectroscopy from about a week after explosion to $\sim$272 d of an atypical Type IIP supernova, SN 2015ba, which exploded in the edge-on galaxy IC~1029. SN 2015ba is a luminous event with an absolute {\it V}-band magnitude of $-$17.1 $\pm$ 0.2 mag at 50~d since explosion and has a long plateau lasting for $\sim$123 d. The distance to the SN is estimated to be 34.8 $\pm$ 0.7 Mpc using the expanding photosphere and standard candle methods. High-velocity H Balmer components constant with time are observed in the late-plateau phase spectra of SN 2015ba, which suggests a possible role of circumstellar interaction at these phases. Both hydrodynamical and analytical modelling suggest a massive progenitor of SN 2015ba with a pre-explosion mass of 24-26 M$_\odot$. However, the nebular spectra of SN 2015ba exhibit insignificant levels of oxygen, which is otherwise expected from a massive progenitor. This might be suggestive of the non-monotonical link between O-core masses and the zero-age main-sequence mass of pre-supernova stars and/or uncertainties in the mixing scenario in the ejecta of supernovae.

\end{abstract}




\section{Introduction}

Type IIP Supernovae (SNe IIP, hereafter) originate from precursor stars that had retained a substantial amount of their hydrogen layers \citep[greater than $\sim$ 3-5 M$_{\odot}$, e.g.][]{2009ARA&A..47...63S,2013MNRAS.434.3445P} before exploding as core-collapse SNe (CCSNe), and are characterized by a plateau in the light curve typically lasting $\sim$100 d \citep[e.g. SNe 1999em, 2004et, 2014cx][]{2003MNRAS.346...97N,2012ApJ...756L..30A,2016ApJ...832..139H}. During this phase, the {\it VRI}-band luminosity remains constant to within $\leq$ 0.5 mag \citep{2011MNRAS.412.1441L, 2014MNRAS.445..554F} and P-Cygni features in the spectra become conspicuous. The plateau phase is ascribed to the recombination wave that moves inward in mass coordinates, while the photosphere stays at roughly the same radius and temperature, resulting in the release of internal energy at a roughly constant rate.  This phase ends when all of the hydrogen envelope has recombined, and is followed by a sudden 2-5 mag drop in brightness in \textless{} 30 days \citep{2010ApJ...715..833O}. A recent study of SNe II by \cite{2016MNRAS.459.3939V} suggests a 1.0-2.6~mag drop in magnitude in luminous SNe IIP (e.g. 1.6 mag in SN 2004et) from the plateau to the radioactive tail, while their less luminous, lower velocity cousins exhibit a larger drop ranging between 3 to 5~mag (e.g. 3.83~mag in SN 2005cs). The SN then enters the nebular phase, with the light curve being powered by the decay of $^{56}$Co to $^{56}$Fe, which is characterized by an exponential decay. 

Early theoretical works of \cite{1971Ap&SS..10...28G}, \cite{1976ApJ...207..872C}, \cite{1977ApJS...33..515F} and \cite{1980ApJ...237..541A} suggest that SNe IIP stem from red supergiant (RSG) stars. At present, there are about a dozen confirmed cases of RSGs as SN progenitors detected from direct imaging. The first discovered was the 8-9 M$_{\odot}$ RSG precursor of SN 2003gd \citep{2003PASP..115.1289V,2004Sci...303..499S,2005MNRAS.359..906H}, while the latest was the 12.5 $\pm$ 1.2 M$_{\odot}$ progenitor to SN 2012aw \citep{2016MNRAS.456L..16F}. According to modern stellar evolution models \citep{2003ApJ...591..288H,2004MNRAS.353...87E}, all stars with initial masses in the range 8 $\lesssim$ M $\lesssim$ 25-30 M$_{\odot}$ are expected to end up as RSGs, subsequently exploding as core-collapse SNe. However, the direct detection of progenitors in the archival pre-explosion images constrained the SN IIP precursor masses close to the theoretical low-mass limit of the core-collapse process viz. 8.5 (e.g. SN 2003gd) to 16.5 M$_{\odot}$ (e.g. SNe 2009kr, 2012ec) \citep[][and references therein]{2009ARA&A..47...63S, 2015PASA...32...16S}. This has also been inferred from nucleosynthetic yields from analysis of nebular spectra  \citep{2010MNRAS.405.2113D,2014MNRAS.439.3694J}, while hydrodynamical models yield larger upper limits \citep{2008A&A...491..507U,2009A&A...506..829U,2011ApJ...741...41P,  2011ApJ...729...61B, 2017MNRAS.464.3013P}. Nevertheless, for the recent event SN 2015bs, which exploded in a low-metallicity environment,  the Zero Age Main Sequence (ZAMS) mass is estimated to be 17-25 M$_{\odot}$\citep{2018NatAs.tmp...53A} through comparison to nebular phase spectral models and is much higher than that deduced for normal Type II SN using similar methods.

Large scale synoptic surveys are discovering large number of SNe events every year, out of which SNe IIP constitute nearly 50\% of all CCSNe \citep{2011MNRAS.412.1441L,2017ApJ...837..121G}. A number of studies have been undertaken to constrain the physical properties of these explosions such as the ejected mass, explosion energy and pre-supernova radius by modelling the observed SN properties, such as their light curves and spectra. A detailed study of SNe IIP has also been done by \cite{2003ApJ...582..905H}, suggesting that these events exhibit a wide range of plateau luminosities ($-$15 \textless{} M$_V$ \textless{} $-$18), ejecta masses (14 - 56 M$_\odot$), kinetic energies (0.6 - 5.5 x 10$^{51}$ erg), $^{56}$Ni masses (0.0016 - 0.26 M$_\odot$) and ejecta velocities (2000 to 8000 km s$^{-1}$). A continuity in the observable properties of SNe is seen from faint, low-velocity, nickel-poor events such as SN 1997D \citep{1998ApJ...498L.129T,  2004MNRAS.347...74P, 2003MNRAS.338..711Z} to bright, high-velocity, nickel-rich objects like SN 1992am \citep{2003ApJ...582..905H}. This was further confirmed with the analysis of samples of low and standard luminosity SNe IIP by several authors \citep{2004MNRAS.347...74P,2014MNRAS.439.2873S,2014MNRAS.441..671A,2014MNRAS.442..844F,2017MNRAS.464.3013P,2018MNRAS.475.1937T}. 

The early time spectra of SNe IIP in most cases show a blue continuum, with few broad P-Cygni features. A blue notch in the absorption component of H Balmer lines has been observed in few SNe II, which is attributed to Si II/ Ba II $\lambda$6497 when observed at early phases, and to high velocity H I when observed at late phases, possibly arising from weak circumstellar interaction \citep{2000ApJ...545..444B,2002PASP..114...35L,2007ApJ...662.1136C,2014MNRAS.438L.101V,2017ApJ...850...89G}. The spectra obtained at late times are also an useful diagnostic to probe the core of the progenitor star. \cite{2012MNRAS.420.3451M} and \cite{2012A&A...546A..28J,2014MNRAS.439.3694J} compared their spectral synthesis models with the nebular spectra of SNe II, and estimated progenitor masses which are in good agreement with the mass estimates from direct detection of progenitor. More recently, \cite{2017MNRAS.467..369S} carried out the analysis of 91 nebular spectra of 38 SNe IIP suggesting that SNe IIP with more massive progenitors have an extended oxygen layer that is well-mixed with the hydrogen layer. 

Although, the observed properties of SNe IIP seem to follow a continuous trend, a handful of peculiar events display rare characteristics in their light curves or spectra. 
SN 2016X, showcased a long rise time to the {\it R}-band peak \citep[12.6 $\pm$ 0.5 d,][]{2018MNRAS.475.3959H} as compared to normal SNe IIP ($\sim$ 7 d). SN 2009ib, an intermediate luminosity SN IIP, had a longer plateau phase, more massive ejecta and more $^{56}$Ni-rich material compared to intermediate luminosity SNe \citep{2015MNRAS.450.3137T}. SN 2009bw, a SN IIP with interaction signatures, recorded the fastest drop from photospheric to the nebular phase \citep[$\sim$2.2 mag in 13 d,][]{ 2012MNRAS.422.1122I}. Despite the bright peak (M$_V$ = $-$18 mag)  and plateau magnitudes, the radioactive tail of SN 2007od is fainter than most SNe IIP \citep{2011MNRAS.417..261I}. The explosion energy, ejecta mass and $^{56}$Ni mass of SN 2009js is found to be similar to sub-luminous SNe, while exhibiting a significantly higher luminosity, comparable to those of intermediate luminosity events \citep{2013ApJ...767..166G}. The early spectra of SN 2013am featured relatively narrow P-Cygni profiles and prominent Ca II and Ba II lines, similar to the low luminosity and low velocity SN 2005cs, while with a shorter plateau phase and brighter light curve tail, the photometric properties of this event deviated from those of SN 2005cs \citep{2014ApJ...797....5Z,2018MNRAS.475.1937T}. These controversial properties of SNe IIP motivated the study of each of them in greater detail.

In this paper, we present the detailed analysis of SN 2015ba, a relatively bright SN IIP, which exploded in the nearly edge-on galaxy IC~1029. The paper is structured as follows: Section \ref{sec2} presents the data and a short overview of the reduction procedure; in Section \ref{sec3}, the distance, explosion epoch and the reddening towards SN 2015ba is extensively discussed. The light and colour curves are illustrated in Section \ref{sec4}. The distance to SN 2015ba is estimated using the Expanding Photosphere Method (EPM) and Standard Candle Method (SCM) in Section \ref{sec5}. The temperature and radius evolution is discussed in Section \ref{sec6}. The spectral evolution, SYN++ modelling and line velocities are discussed in Section \ref{sec7}. The bolometric light curve, $^{56}$Ni mass yield and the progenitor properties derived using analytical and hydrodynamical modelling are given in Section \ref{sec8}. Finally, a short summary of the work is presented in Section \ref{sec9}.

\section{SN 2015ba: Data and Reduction}
\label{sec2}
SN 2015ba was discovered by Koichi Itagaki (Teppo-cho, Yamagata, Japan) on 2015 November 28.8071 UT (JD = 2457355.31) in the galaxy IC 1029 at an unfiltered magnitude of 16.7 mag (CBET 4209\footnote{www.cbat.eps.harvard.edu/iau/cbet/004200/CBET004209.txt}). The precise location of the SN was given by T. Yusa and Osaki to be around 19$^{\prime\prime}$E, 43$^{\prime\prime}$S of the centre of IC 1029. \cite{2015ATel.8353....1B} obtained an optical spectrum on 2015 December 2.13 UT showing weak and shallow P-Cygni profiles of the Balmer lines superimposed on a blue continuum. This is consistent with a SN II spectrum at early phase. The expansion velocities reported from the measurement of the position of  the minima of the H$\alpha$ and H$\beta$ Balmer lines were around 8900 and 8400 km s$^{-1}$, respectively. The details of SN 2015ba and its host galaxy IC 1029 are given in Table \ref{tab:sn15ba_ic1029_detail}.

Our observing campaign of SN 2015ba was triggered 3 days after the discovery, using instruments equipped with broadband {\it BVRI} and {\it ugriz} filters listed in Table \ref{tab:details_instrument_detectors}. High cadence photometric data were obtained up to 266 d from discovery, beyond which the SN was below the detection limit of 1-2 m class telescopes. The images were pre-processed and reduced as discussed in Appendix \ref{phot} and night to night zero-points calculated using the local standards (see Figure \ref{fig:id_chart} and Table \ref{tab:local}) were applied to obtain the final SN magnitudes as listed in Table \ref{photometry}. The spectroscopic monitoring was conducted at 27 epochs from a number of facilities as listed in Table \ref{tab:details_instrument_detectors} and the log of spectroscopic observations is presented in Table \ref{tab:spectra_log}.

\section{Parameters of SN 2015ba}
\label{sec3}
\subsection{Distance}
Using the recessional velocity $v_{Vir}$ = 2664 $\pm$ 3 km s$^{-1}$ given in HyperLeda \citep{2014AA...570A..13M}\footnote{http://leda.univ-lyon1.fr/} and Hubble constant H$_0$ = 73.24 $\pm$ 1.74 km s$^{-1}$ Mpc$^{-1}$ \citep{2016ApJ...826...56R}, we obtain a Virgo infall distance of about 36.4 $\pm$ 0.9 Mpc. Further, we implement the Expanding Photosphere Method (EPM) and the Standard Candle Method (SCM) to estimate the distance from early photometric and spectroscopic observations. This will be discussed in detail in Section \ref{sec5}. The EPM and SCM provides a distance of 36.6 $\pm$ 1.9 Mpc and 30.1 $\pm$ 1.4 Mpc, respectively. We take the weighted mean of distances estimated using the three methods (listed in Table \ref{ave_dist}), and we hereafter adopt a distance of 34.8 $\pm$ 0.7 Mpc, corresponding to a distance modulus $\mu$ = 32.73 $\pm$ 0.04 mag.

\subsection{Explosion epoch}
\label{exp_ep}
We use the SNID code \citep{2007ApJ...666.1024B} to constrain the explosion epoch of SN 2015ba, which has been tested in the works of \cite{2014ApJ...786...67A} and \cite{2017ApJ...850...89G}. This code matches the SN spectrum to a library of spectral templates, and uses cross-correlation technique to constrain the explosion epoch. Since most spectral lines lie at the blue wavelengths, we perform our fits in the wavelength range 3500 to 6000 \AA. The quality of the fit is represented by the \textquoteleft rlap\textquoteright{} parameter, with higher value implying a better correlation. From the best three matches, we derive a mean value of 12.5 $\pm$ 7.3 d from explosion for the spectrum obtained on 2015 December 4 (JD 2457360.67), which provides 2015 November 21.5 UT (JD 2457348.2 $\pm$ 7.3) as the explosion date. The best three matches with the rlap parameter, are listed in Table \ref{expl_epoch}.

The EPM also gives an approximate explosion epoch (see Section \ref{sec5} for details), as 2015 November 23.2 UT (JD 2457349.7 $\pm$ 1.0), which is consistent with that estimated using SNID.  Therefore, we adopt throughout the paper 2015 November 23 (JD  2457349.7 $\pm$ 1.0) as the most reliable explosion epoch ($t_0$) of SN 2015ba.

\subsection{Extinction}
\label{ext} 
In order to gain insights on the true nature of an event and to derive physical parameters from photometry, it is important to estimate the Galactic and the host extinction in the SN direction. The Galactic extinction in the direction of IC 1029 corresponds to a colour excess of E(B$-$V) = 0.0153 $\pm$ 0.0003 mag \citep{2011ApJ...737..103S}. There are various methods for a crude estimation of host galaxy extinction. In our case, we inspect three methods: i) fitting a blackbody function to the early spectral energy distribution (SED) of the SN, ii) estimating from the equivalent width (EW) of Na I D line, and iii) using the \textquoteleft colour method\textquoteright{} suggested by \cite{2010ApJ...715..833O}.

\begin{figure}
	\begin{center}
		\includegraphics[scale=0.70]{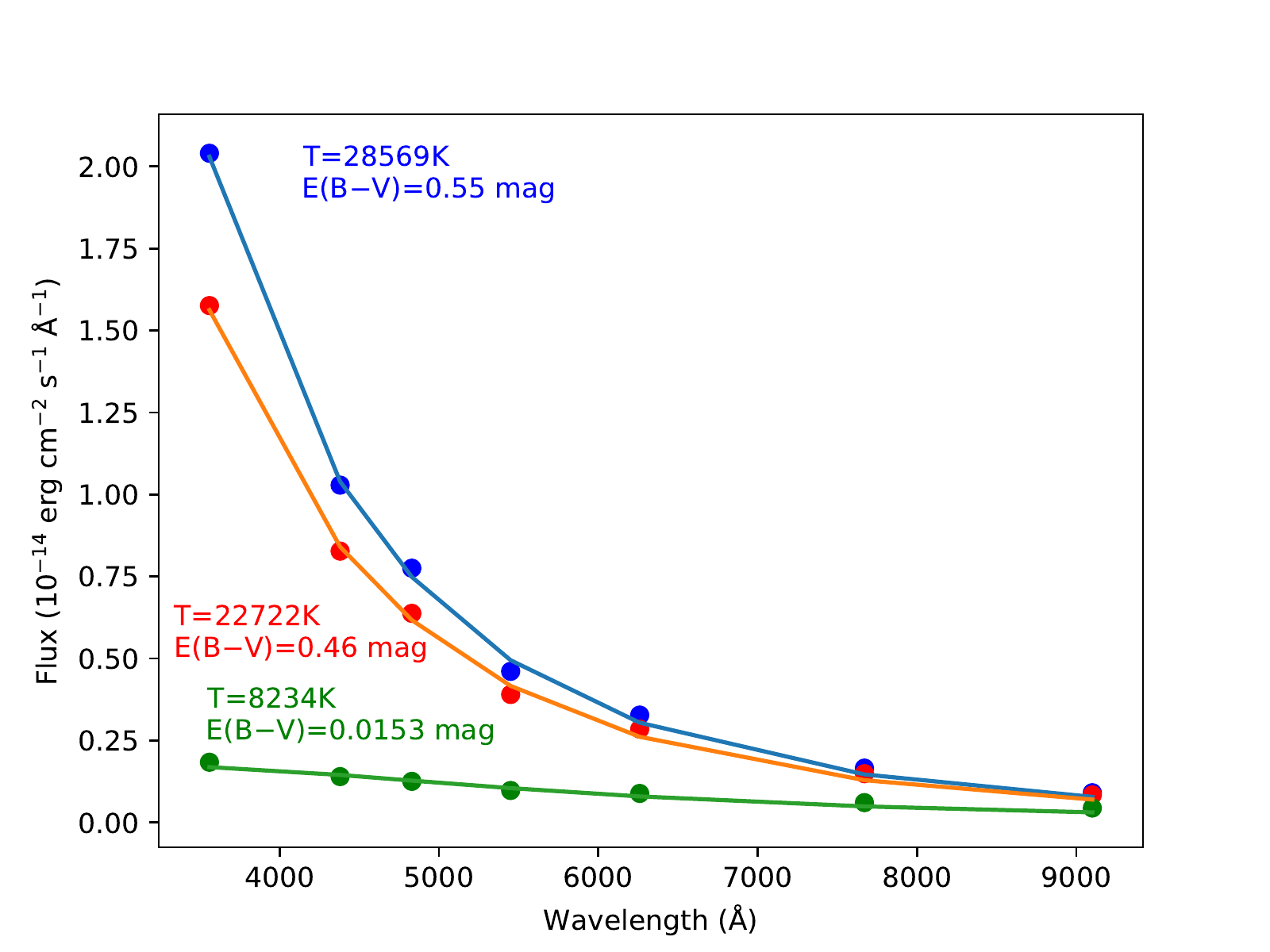}
	\end{center}
	\caption{The SED of SN 2015ba at 10 d with varying reddening is fitted by a blackbody function to derive the best fit temperature. We used E(B$-$V)=0.0153 mag, corresponding to Galactic extinction, E(B$-$V)=0.46 mag corresponding to Galactic plus host galaxy extinction (as discussed in text) and E(B$-$V)=0.55 mag for which the best fit temperature becomes unphysical. The filled circles denote the fluxes in {\it uBgVriz} corrected for the corresponding reddening and the continuous lines are the best-fit blackbody models.}
	\label{fig:EBV}
\end{figure}

The comparison of the early spectral shape of a classical SN to a blackbody function can provide an upper bound of the colour excess, as discussed by \cite{1996ApJ...466..911E}. With increasing extinction values, the blackbody spectra will start to deviate from the continuum, establishing an upper limit to the reddening. More recently, theoretical modelling of \cite{2006A&A...447..691D} and \cite{2011ApJ...729...61B} suggest that the maximum likely temperature in the early phases of a SN II is 25-30 kK. We fit a Planck function to the SED constructed from the magnitudes of SN 2015ba obtained on 2015 December 3.21 UT (+10 d since explosion) by varying the amount of reddening. We used the reddening law of \cite{1989ApJ...345..245C} with a total-to-selective extinction ratio (R$_V$) of 3.1 to deredden the observed fluxes. The best fit temperature corresponding to each value of the colour excess is shown in Figure \ref{fig:EBV}. This exercise sets an upper limit of E(B$-$V) to 0.55 mag above which the temperature becomes unphysical.

\begin{figure}
	\begin{center}
	\includegraphics[scale=0.54, trim={4cm 3.8cm 20cm 4.5cm}]{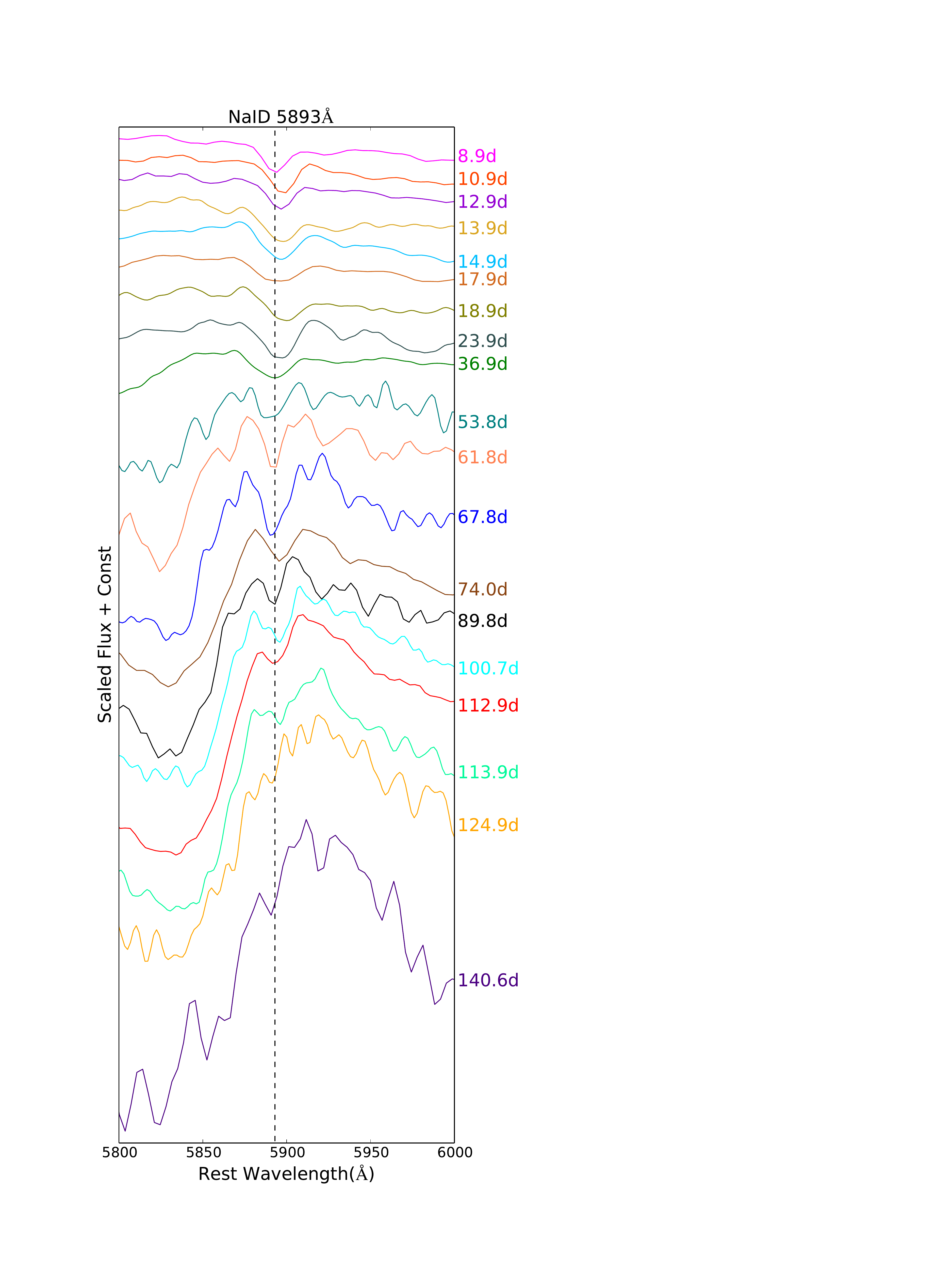}
	\end{center}
	\caption{The narrow Na I D absorption dip due to host galaxy, IC 1029. The spectra on 8.9, 12.9, 14.9, 36.9 and 74.0 day with SNR \textgreater{} 50 at 0.6 $\micron$ are used to estimate the host galaxy extinction.}
	\label{fig:EBV_Na}
\end{figure}

The presence of a narrow Na I D line in the spectra of SN 2015ba at the redshift of the host galaxy (Figure \ref{fig:EBV_Na}) is a possible indication of extinction within the host. Several empirical relations exist to correlate the EW of Na I D to the colour excess E(B$-$V). The EW of the Na I D lines in five spectra with good signal-to-noise ratio are tabulated in Table \ref{15ba_EW}. However, most of these relations show a large scatter and are unreliable with low-resolution spectra \citep{2011MNRAS.415L..81P}. \cite{1997A&A...318..269M} showed that the lines saturate for EW of Na I D $\gtrsim$ 0.6 \AA, hence leading to an underestimate of the colour excess. Nevertheless, we have listed all the existing relations in Table \ref{ebv} and calculated the corresponding E(B$-$V) values.

The third method implemented in the calculations of the colour excess E(B$-$V) is the \textquoteleft colour method\textquoteright{}. This predicts that all SNe IIP should reach the same intrinsic colour towards the end of the plateau phase. It assumes that the opacity of a SN IIP is dominated by electron-scattering, and will attain the recombination temperature of hydrogen at the end of plateau phase. Hence, the difference in the observed and the intrinsic colours may be attributed to reddening due to dust along the line of sight. Using a library of SN II spectra, \cite{2010ApJ...715..833O} computed a conversion factor between E(V$-$I) and $A_V$ and suggested the following prescription:
\begin{gather}
A_V(V-I) = 2.518[(V-I)-0.656]\\
\sigma(A_V) = 2.518 \sqrt{\sigma_{(V-I)}^2 + 0.0053^2 + 0.0059^2}
\end{gather}
We calculate the weighted mean of $(V-I)$ colours (corrected for Galactic extinction) on 112.3 d, 115.3 d and 119.3 d from the explosion corresponding to the end of the plateau phase, which results in $(V-I)$ = 1.20 $\pm$ 0.01 mag. The estimated $A_{V(host)}$ is 1.37 $\pm$ 0.20 and E(B$-$V)$_{host}$ = 0.44 $\pm$ 0.06 mag.  

The reddening estimate using the equivalent width and the colour method are consistent with each other within the errors, and are listed in Table \ref{ebv}. We, therefore, compute the weighted mean of the host galaxy reddening as E(B$-$V)$_{host}$ = 0.44 $\pm$ 0.05 mag. The total E(B$-$V) due to host and Milky Way is 0.46 $\pm$ 0.05 mag, which is adopted throughout the paper.

\section{Temporal evolution of SN 2015ba Light Curve}
\label{sec4}
A careful study of the light curve properties is essential for the characterization of the event and constrain the properties of the explosion.  By comparing the light curve parameters such as the mid-plateau absolute magnitude in{\it V}-band with those of other archetypal SNe IIP, one can sub-classify the events as sub-luminous, normal or over-luminous. Moreover, the steepness of the plateau indicates the extent of the thermalization of the ejecta, with a flatter plateau indicating efficient thermalization and higher $^{56}$Ni mass yield. Hydrodynamical modelling of \cite{2011ApJ...729...61B} showed that extensive mixing of $^{56}$Ni is required to produce flatter plateaus. For the comparison, we have selected a sample of SNe IIP with variable luminosities, plateau lengths and plateau slopes. SN 2005cs \citep{2009MNRAS.394.2266P} is a low-luminosity ($M_V$ = $-$14.83 mag) Type IIP event while DLT16am \citep{2018ApJ...853...62T} is one of the brightest ($M_V$ = $-$17.76 mag) SNe IIP; the plateau length of SN 2007od is $\sim$ 25 days in {\it R}-band \citep{2011MNRAS.417..261I} while the plateau length of SN 2009ib is 118 days in {\it V}-band \citep{2015MNRAS.450.3137T}; the $^{56}$Ni mass yield of SN 2005cs is  \textless{} 0.003 M$_\odot$ \citep{2009MNRAS.394.2266P} while for SN 2004et M($^{56}$Ni) = 0.06 M$_\odot$ \citep{2006MNRAS.372.1315S,2007MNRAS.381..280M,2010MNRAS.404..981M}. Further, we include the bright SNe 2007od and 2009bw, the normal luminosity SNe 1999em, 1999gi, 2004dj, 2012aw, 2013ab, 2014cx, ASASSN-14gm and 2016X, and the intermediate luminosity SNe  2008in, 2009N in the sample. Our sample includes a total of 17 events. The parameters of the comparison sample are listed in Table \ref{parameter_SNIIP_sample}. 

\subsection{Main Light Curve features}
\label{light_curve_features}
\begin{figure}
	\begin{center}
		\includegraphics[scale=1.00, width=1.0\textwidth,clip, trim={0.8cm 0cm 0cm 0cm}]{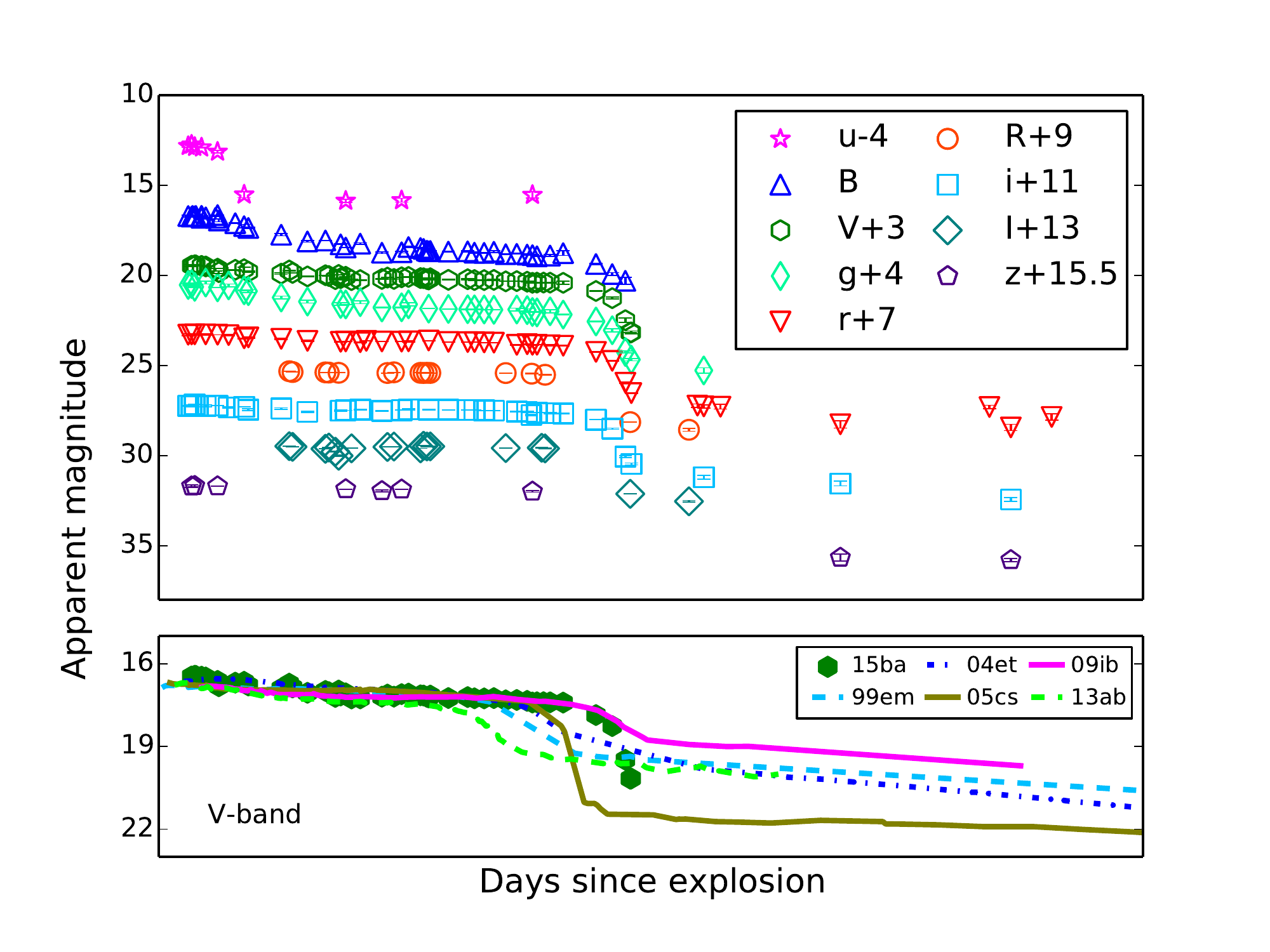}
	\end{center}
	\caption{{\it Top panel:} Broadband {\it uBVgrRiIz} light curves of SN 2015ba shifted arbitrarily for clarity. {\it Bottom panel:} The {\it V}-band light curve of SN 2015ba compared with other SNe IIP. }
	\label{fig:light_curve}
\end{figure}

SN 2015ba was discovered $\sim$5.6 d after explosion (t$_0$ taken from Section \ref{exp_ep}), caught dropping off from the peak and settling onto the plateau. The {\it BVRI} and {\it ugriz} light curves ranging from 8 to 272 d since explosion are shown in the top panel of Figure \ref{fig:light_curve}. We have limited number of measurements at late epochs, and observations were obtained only with the {\it griz} filters. We converted the {\it g}-band data in the tail to {\it V}- and {\it B}-band using the relations (1) and (6) given in \cite{2006A&A...460..339J}. In the bottom panel of Figure \ref{fig:light_curve}, we show the {\it V}-band light curve of SN 2015ba along with other SNe IIP. While the plateau length of SN 2015ba ($\sim$123 d in {\it V}-band) is similar to that of SN 2009ib  \citep[$\sim$118 d in {\it V}-band,][]{2015MNRAS.450.3137T}, the magnitude drop from the plateau to the nebular phase is larger for SN 2015ba ($\sim$3 mag). The decline rate in the {\it r}-band of SN 2015ba in the first 50 days is $\sim$0.5 mag, consistent with the criterion suggested by \cite{2011MNRAS.412.1441L}, that the drop in {\it R}-band magnitude in the first 50 days for SNe IIP is $\leq$ 0.5 mag. The {\it B}-band decline rate of SN 2015ba is 2.9 mag 100 d$^{-1}$ which is also typical of SNe IIP \citep{1994A&A...282..731P}. The mid-plateau absolute magnitude of SN 2015ba in {\it V}-band (M$_V^p$) is $-$17.1 $\pm$ 0.2~mag, well within the magnitude limits of normal Type IIP SNe \citep[$-$18 < M$_V^p$ < $-$15,][] {1994A&A...282..731P}. While, all these numbers suggest that SN 2015ba belong to the plateau sub-group of Type II events, the decline rate in the {\it V}-band for SN 2015ba in the first 50 days is 0.69 $\pm$ 0.05 mag and is higher than that used by \cite{2014MNRAS.442..844F} to filter out Type IIP events from their sample.

Further, we carried out a comparison of the plateau lengths of SNe IIP sample by fitting the analytic function  provided in \cite{2016MNRAS.459.3939V}, as given below:
\begin{equation}
y(t) = \frac{-a_0}{1+e^{(t-t_{PT})/w_0}} + (p_0 \times t) + m_0
\end{equation}
where {\it t} is the time from explosion in days, {\it t$_{PT}$} is the time in days from explosion to the transition point between the end of the plateau phase and start of the radioactive tail, {\it a$_0$} is the depth of the drop from plateau to radioactive tail, {\it w$_0$} is the slope of the drop and {\it p$_0$} constrains the slope before and after the drop.
We fitted this function to the {\it V}-band light curve of SN 2015ba and the best fit value of {\it a$_0$} is 3.01 $\pm$ 0.09 mag in $\sim$20 d, which corresponds to the upper limit of the range of 2-3 mag drop for typical SNe IIP \citep{2010ApJ...715..833O}, while this exceeds the range (1-2.6 mag) suggested by \cite{2016MNRAS.459.3939V}. The same exercise is done for the comparison sample, and the best fit values are listed in Table \ref{lc_par_comp}. The $t_{PT}$ of our sample range from 95 d in SN 2016X to $\sim$ 141 d in SN 2015ba. The drop in magnitude from plateau to nebular phase is maximum for SN 2005cs (4.07 $\pm$ 0.05 mag) and minimum for SN 2016X (1.33 $\pm$ 0.04 mag).
  
\begin{figure} 
\begin{center}
\includegraphics[scale=1.0,width=0.73\textwidth,clip, trim={0.3cm 0cm 0cm 0cm}]{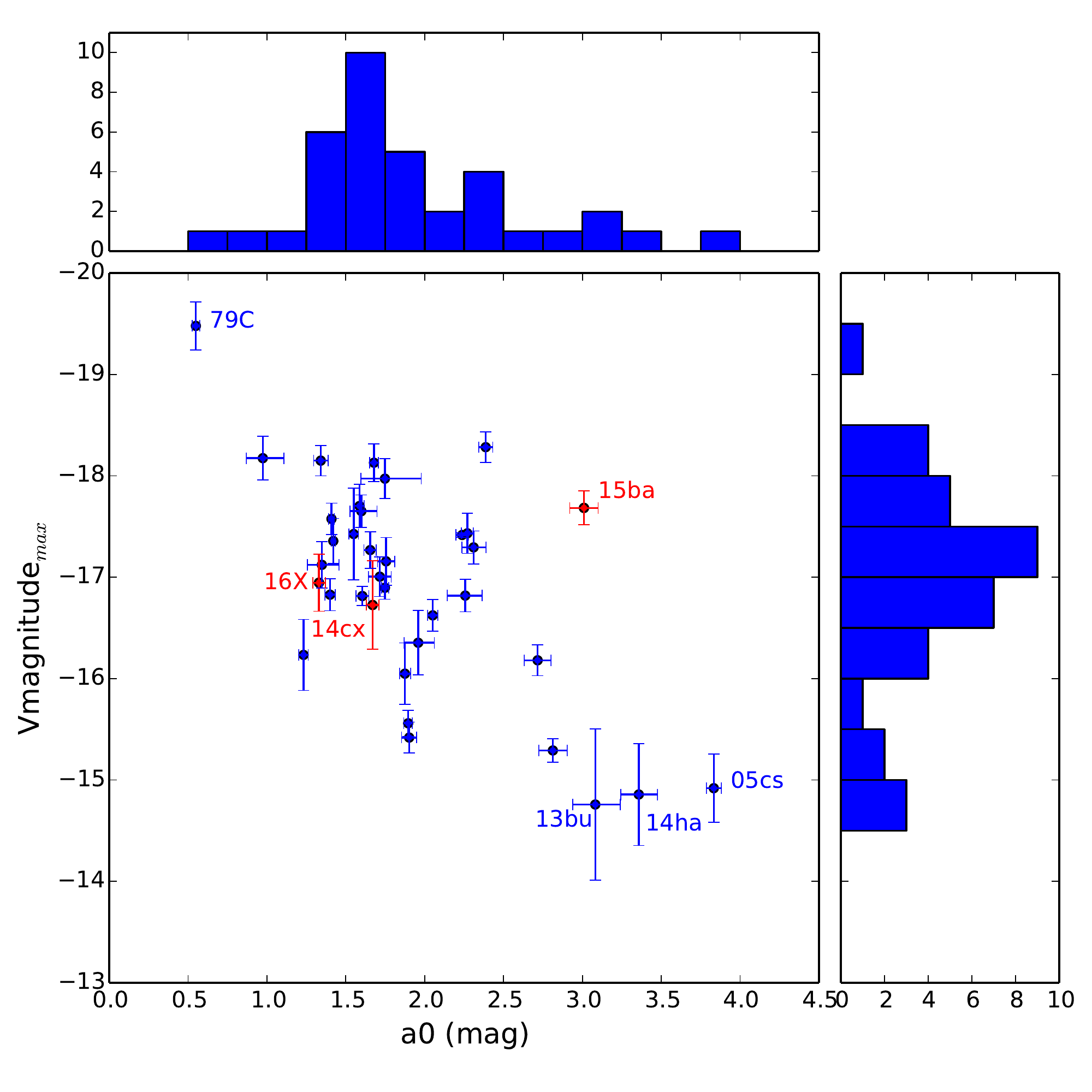}
\end{center}
\caption{Plot of absolute peak magnitude in {\it V}-band and the depth of drop from plateau to nebular phase (a$_0$). The sample of SNe from \citet{2016MNRAS.459.3939V} are marked in blue, while the newly added SNe are marked in red.}
\label{fig:valenti_fig06}
\end{figure}

We reproduced the plot between absolute peak {\it V} magnitude and a$_0$ from \cite{2016MNRAS.459.3939V} along with three additional SNe IIP (SNe 2014cx, 2016X and 2015ba; marked in red) from our sample (Figure \ref{fig:valenti_fig06}). The three objects in \cite{2016MNRAS.459.3939V} sample which shows high a$_0$ are SNe 2005cs, 2013bu and ASASSN-14ha, which are also low-luminosity SNe IIP. As noted by \cite{2016MNRAS.459.3939V}, the two parameters (a$_0$ and Vmagnitude$_{max}$) seem to correlate, however, the correlation is largely affected by the significant drop of the three faint SNe and the small drop of SN 1979C. 

\subsection{Colour curves}
\begin{figure}
\begin{center}
\includegraphics[scale=0.75]{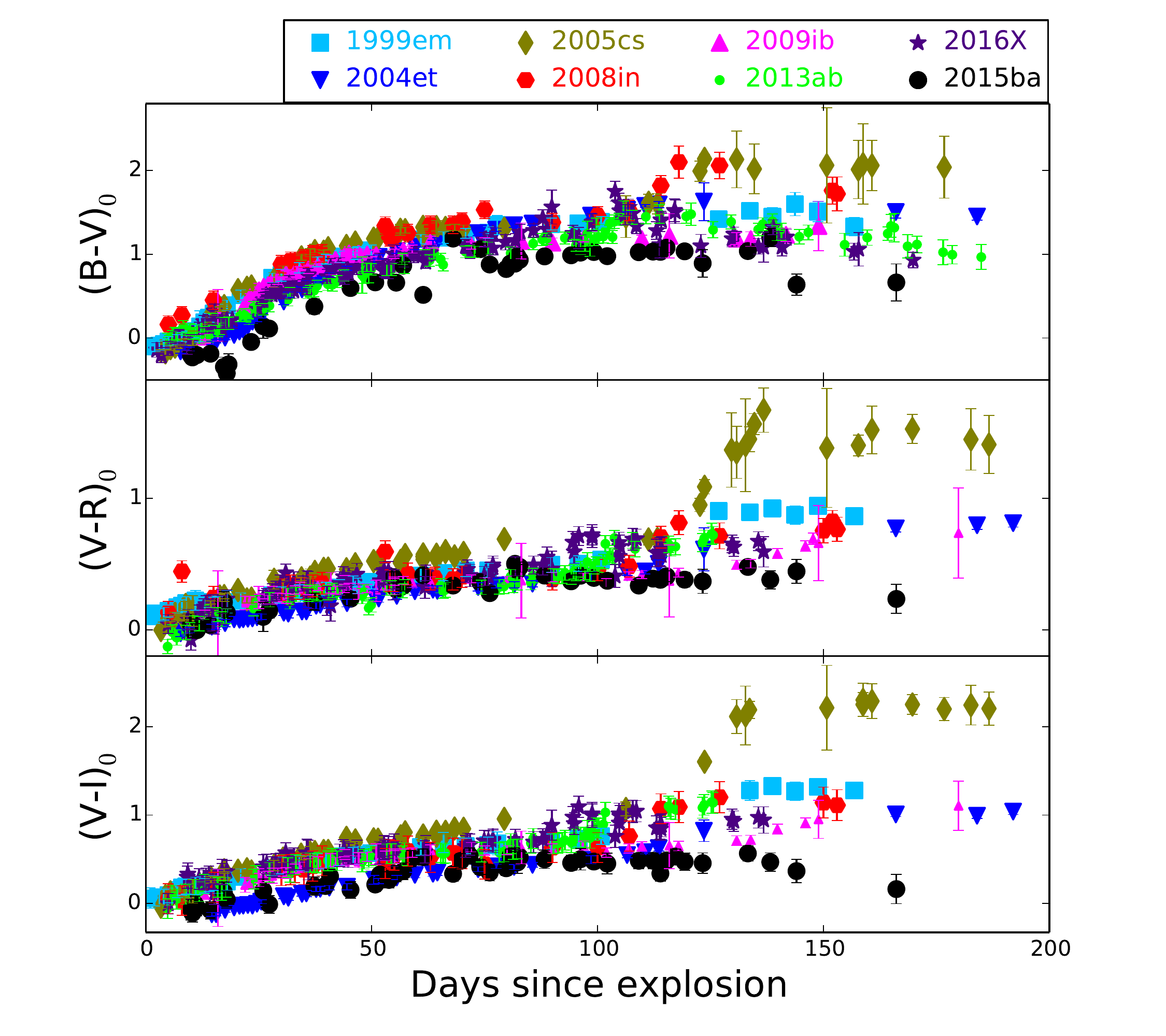}
\end{center}
\caption{Comparisons of the $(B-V)_0$, $(V-R)_0$ and $(V-I)_0$ colour curves of SN 2015ba with other SNe IIP.}
\label{fig:color_curve}
\end{figure}

The temporal evolution of reddening-corrected broadband colours provide important clues to the dynamics of SN ejecta. The reddening-corrected colour evolution [$(B-V)_0$, $(V-R)_0$ and $(V-I)_0$] of SN 2015ba is shown in Figure \ref{fig:color_curve}, with those of other well-studied Type IIP SNe. The $(B-V)_0$ colour gradually becomes redder by about $\sim$1 mag in the first $\sim$50 d, as a consequence of the expansion and cooling of the ejecta. The $(V-R)_0$ and $(V-I)_0$ colours evolve more slowly, with \textless{} 0.5 mag in 50 d. In the plateau phase, the colours show negligible evolution with time, signifying a nearly constant temperature during this phase. After $\sim$144 d, when the SN enters the nebular phase, the colours become bluer. The overall evolution is similar to typical SNe IIP with the colours being marginally bluer than those of the comparison SNe IIP.

\section{Distance}
\label{sec5}

There are several redshift independent distance estimates listed in NED for the host galaxy IC~1029. However, the values are inconsistent and vary over a large range (31.1 - 105~Mpc). For this reason, we carry out an independent estimate of the distance to SN 2015ba using the EPM and SCM prescriptions. The EPM is a variant of the Baade-Wesselink method to estimate SN distances \citep{1974ApJ...193...27K}. It compares the linear and angular radius of homologously expanding optically thick SN ejecta to compute the SN distance. The SCM \citep{2002ApJ...566L..63H} is based on the correlation between the SN brightness and the expansion velocity in the middle of the plateau. It needs less input data, but requires calibration via SNe with well-established distances. We discuss these methods in details below.

\subsection{Expanding Photosphere Method (EPM)}
The early phase of a SN, when the ejecta is fully ionized and electron scattering dominates the total opacity at the photosphere, can be approximated to be radiating as a diluted blackbody. The angular radius of the expanding ejecta at any time {\it t} can then be approximated as:
\begin{equation}
 \theta = \frac{R}{D} = \sqrt\frac{f_\lambda 10^{0.4A_\lambda}}{\zeta_\lambda^2(T_c)\pi B_\lambda(T_c)} 
 \end{equation}
where $B_\lambda$ is the Planck function at colour temperature $T_c$, $f_\lambda$ is the flux density received at Earth, $A_\lambda$ is the extinction, and $\zeta_\lambda(T_c)$ is the colour temperature dependent \textquotedblleft dilution factor\textquotedblright{} and $R = v(t-t_0$), where $(t-t_0)$ is the time since explosion.

Recasting this equation in terms of broadband photometry by integrating over the filter response function, we get
\begin{equation}
m_\lambda = -5 log(\zeta_\lambda\theta)+A_\lambda+b_\lambda
\end{equation}
where $b_\lambda$ is the convolution of the filter response function with the blackbody model $\pi B_\lambda$.

\cite{2001ApJ...558..615H} computed $b_\lambda(T)$, by fitting a polynomial in the temperature range of 4000-25000K, as
\begin{equation}
b_{\lambda}(T) = \sum_{i=1}^5 C_{i}(\lambda)\bigg( \frac {10^{4} K}{T}\bigg)^i
\end{equation}
The coefficients $C_i$'s are listed in Table 13 of \cite{2001ApJ...558..615H}.

Minimizing the quantity, 
\begin{equation}
\epsilon = \sum_{\lambda\in S} [  m_\lambda +5 log(\zeta_\lambda\theta)-A_\lambda-b_\lambda]^2 
\end{equation}
we get $\zeta_\lambda\theta$ and $T_c$ simultaneously. The errors in $\zeta_\lambda\theta$ and $T_c$ are estimated by randomly mixing the magnitudes with uncertainties drawn from their normal distribution, thereby generating a sample of 1000 data points. The standard deviation of the result gives the error for these quantities. 
 
\begin{figure}
\begin{center}
\includegraphics[scale=0.5]{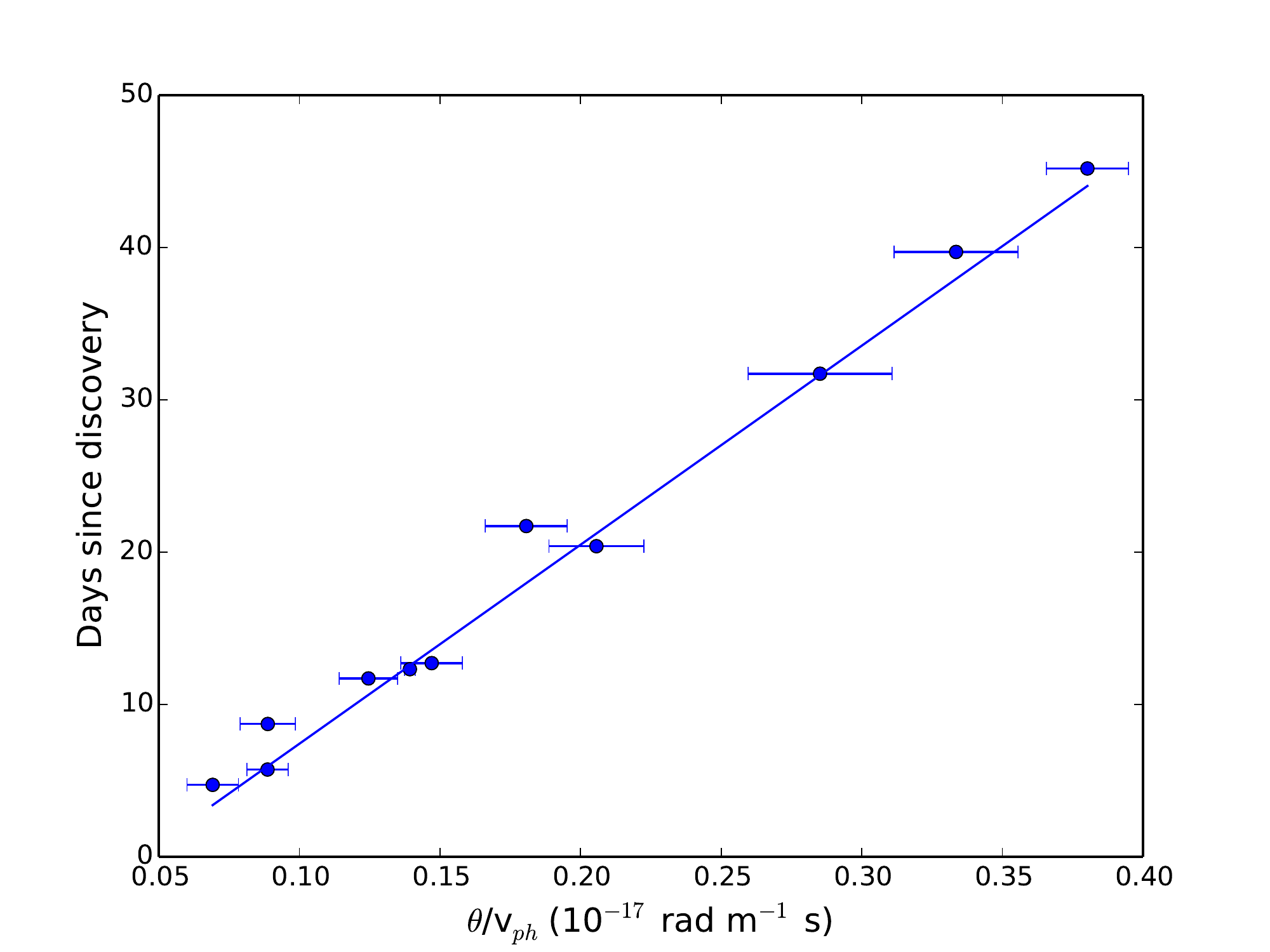}
\end{center}
\caption{Distance determination with the EPM for SN 2015ba, using \citet{2005A&A...439..671D} prescription for the dilution factors and the {\it BVI} observations. The distance and the explosion epoch are obtained from the best fit.} 
\label{fig:EPM}
\end{figure}

 The dilution factors $\zeta$ can be expressed as a function of $T_c$ from atmospheric model calculations. \cite{2005A&A...439..671D} used second order polynomial fits to the distribution of dilution factors versus blackbody colour temperatures, 
\begin{equation}
 \zeta_{S}(T_{S}) = \sum_{i=0}^2 a_{S,i} \bigg(\frac {10^{4} K}{T_{S}}\bigg)^i 
\end{equation}
We estimated $\theta$ adopting the coefficients from \cite{2005A&A...439..671D} and the values obtained are listed in Table \ref{epm}. The photospheric velocity $v_{ph}$ is obtained from the spectral modelling of He I 5876$\lambda$ line up to 13 d after discovery and Fe II 5169$\lambda$ line up to 40 d after discovery. We generate  a sample of 1000 data points from the normal distribution of the uncertainty in the photospheric velocity. To each of this data set, cubic spline interpolations is performed to account for the missing velocity values for the corresponding epoch of photometric data. From the standard deviation of these values,we obtain the errors. 
We perform a linear fit to $t$ versus $\theta/v_{ph}$ to estimate the distance from the slope and the time of outburst $t_0$ from the y-intercept, following the expression: 
\begin{equation}
t =D(\theta/v_{ph})+t_0
\end{equation}
The linear fit to the data is shown in Figure \ref{fig:EPM} and the distance and explosion epoch estimates from the fit are 36.6 $\pm$ 1.9 Mpc and 5.6 $\pm$ 1.0 days before the discovery date (2015 November 28.8 UT, JD 2457355.3), respectively.

\subsection{Standard Candle Method}
The SCM has several versions \citep{2006ApJ...645..841N,2009ApJ...694.1067P,2010ApJ...715..833O} with different SNe samples, but, yielding consistent results. We will discuss two versions and use their average value as the SCM distance estimate. The value of Hubble constant adopted below is H$_0$ = 73.24 $\pm$ 1.74 km s$^{-1}$ Mpc$^{-1}$ \citep{2016ApJ...826...56R}.

i) The version of \citet{2009ApJ...694.1067P} used the measured brightness, expansion velocity of a sample of 34 SNe with known distances to calibrate the equation
\begin{eqnarray}
\label{poz_eq}
\mathcal{M_I} - \alpha \cdot \log \left( {v_{Fe}(50\mathrm{d}) \over 5000} \right) +R_I((V-I) - (V-I)_0) - m_I \nonumber\\
= -5 \cdot \log(H_0D)
\end{eqnarray}
where $\mathcal{M_I}$ = $-$1.615 $\pm$ 0.08 mag, $\alpha$ = 4.4 $\pm$ 0.6, $R_I$ = 0.8 $\pm$ 0.3 and $(V-I)_0$ = 0.53~mag. Note, that the $R_I$ value used here differs from the \cite{1989ApJ...345..245C} value ($R_I$ = 1.36). We measure the $I$-band magnitude of SN~2015ba on day +50 as $m_I$ = 16.07$\pm$ 0.04~mag, $(V-I)$ = 0.954$\pm$0.01 mag and $V_
{FeII}$ =  3332$\pm$ 55 km~s$^{-1}$ and obtain a distance D = 28.8$\pm$2.6 Mpc.

ii) \cite{2010ApJ...715..833O} used the magnitudes and velocities measured 30 days before the middle of the transition phase (t$_{PT}$) in the relation of \cite{2002ApJ...566L..63H}
\begin{eqnarray}
\label{oliv_eq}
m + \alpha \cdot \log \left( {v_{Fe} \over 5000} \right) -\beta(V-I) =  5 \cdot \log(H_0D) + zp
\end{eqnarray}
and calibrated the formula using a sample of 37 nearby SNe IIP. The coefficients $\alpha$, $\beta$ and $zp$ are different for {\it V} and {\it I} bands. We determine $t_{PT}-t_0$ = 140.7 d after explosion (see Section \ref{light_curve_features}). 30d before this date, we measure the values v(110.7d) = 2252$\pm$54 km s$^{-1}$, m$_V$ = 17.31$\pm$0.02 mag and m$_I$ = 16.16$\pm$0.02 mag. We estimate the distances to be 29.9$\pm$2.5 Mpc and 31.2$\pm$2.3 Mpc using equation \ref{oliv_eq} for {\it V} and {\it I} bands respectively. However, note that K-correction has not been applied in this case.

The SCM distance estimates using the above two methods are consistent with each other. The weighted SCM distance is 30.1 $\pm$~1.4 Mpc. We list the distance estimates obtained via recessional velocity, EPM and SCM in Table \ref{ave_dist}. The weighted mean of these distances is 34.8 $\pm$ 0.7~Mpc and has been adopted throughout the paper.

\section{Temperature and Radius}
\label{sec6}
\begin{figure}
	\begin{center}
		\includegraphics[scale=0.6]{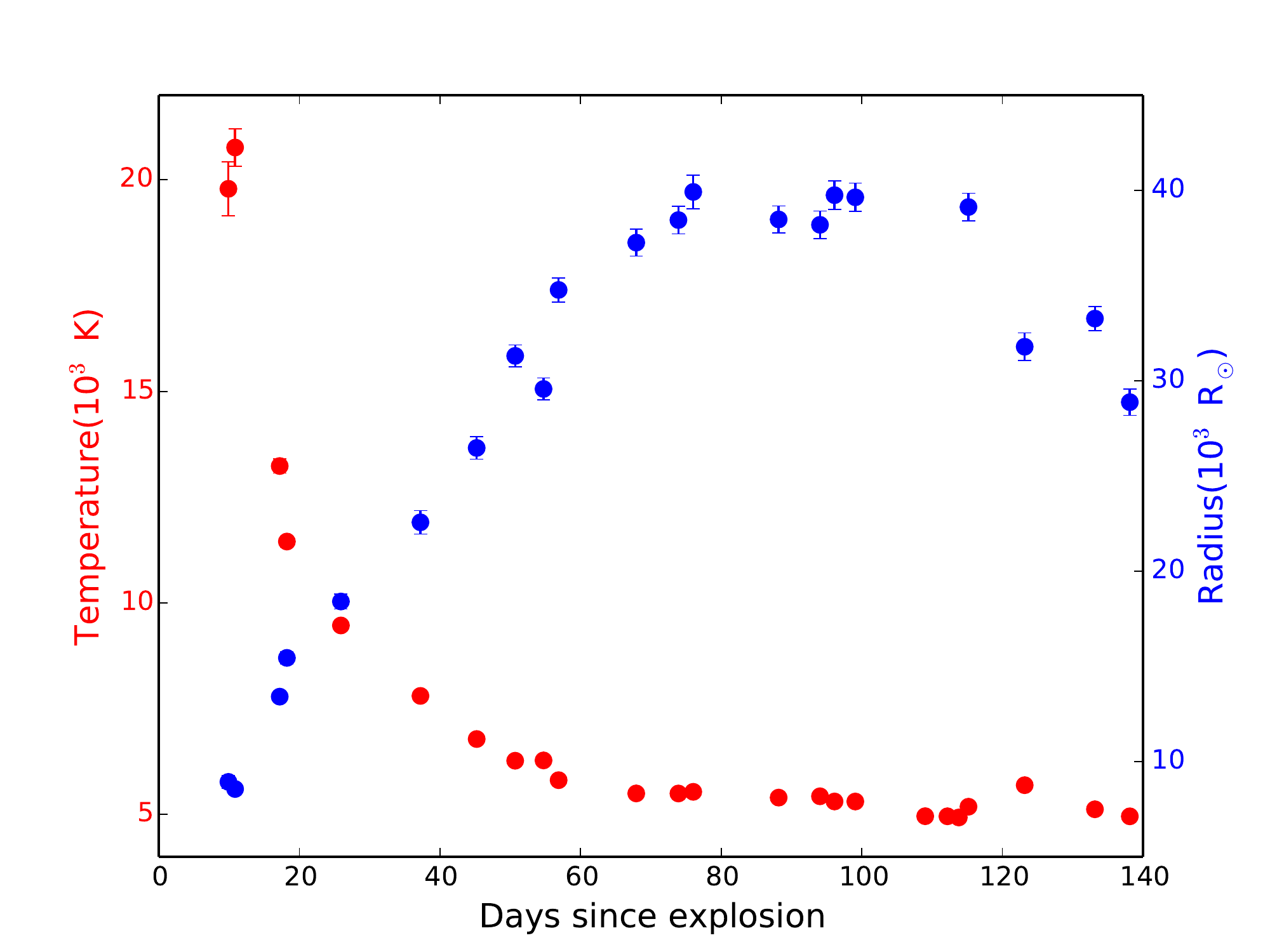}
	\end{center}
	\caption{Evolution of the temperature and the radius for SN 2015ba.}
	\label{fig:temp_rad_plot}
\end{figure}

We also determine the photospheric temperature and the radius evolution by constructing SEDs with the {\it BVI} photometric fluxes at each epoch and fitting it with a Planck function until $\sim$ 140~d. We have also applied the dilution factor correction \citep{2005A&A...439..671D} corresponding to the {\it BVI} filter set to the blackbody radii to estimate the photospheric radii. As shown in Figure \ref{fig:temp_rad_plot}, the temperature drops from 20000 K at 10 d to 6300 K at 50 d since explosion due to rapid adiabatic cooling, and thereafter it declines very slowly to 5300 K at 100 d since explosion. The slow decline of the late phase is due to the low opacity of the ejecta which favours cooling through photon energy diffusion. The photospheric radius shows a constant growth up to $\sim$ 70 d and then  remains nearly constant until 115 d. During the plateau phase, due to the declining electron density, the photosphere remains contiguous to the receding recombination front on top of the expanding envelope, leaving the photospheric radius nearly constant. Afterwards, the radius starts to decrease up to 140 d, beyond which the SED cannot be compared to a blackbody any longer.

\section{Spectral Analysis}
\label{sec7}
Figure \ref{fig:spectra_plot} presents the spectral evolution of SN 2015ba from $\sim$ 9 d to $\sim$ 141 d after explosion, with a preliminary identification of absorption features. This is necessary to get an idea of the overall structure of the envelope with the revelation of deeper layers at later times. Prominent features of H are visible throughout the evolution. We attempt a  detailed line identification at two phases, 8.9 d and 53.8 d  spectrum using the SYN++ \citep{2011PASP..123..237T} spectral code.

SYN++ is an evolution of SYNOW, which uses the Sobolev approximation to produce synthetic spectra of SNe during the photospheric phase. SYN++ assumes that spectral lines are formed via resonance scattering above a sharp photosphere and the ejecta are homologously expanding. The location of the photosphere is expressed in velocity coordinates as v$_{ph}$ (in km s$^{-1}$).
With the optical depths for each species, line strengths are computed assuming Boltzmann excitation (i.e., local thermodynamic equilibrium) using a specified excitation temperature T$_{exc}$ (in K). Non-local thermodynamic equilibrium effects are partially accounted for by allowing different T$_{exc}$ values for each species, that can be different from the photospheric temperature T$_{phot}$. The latter is used only for computing the blackbody radiation emitted by the photosphere.

\begin{figure} 
	\begin{center}
		\includegraphics[scale=0.95,clip, trim={1cm 0cm 1.5cm 0cm}]{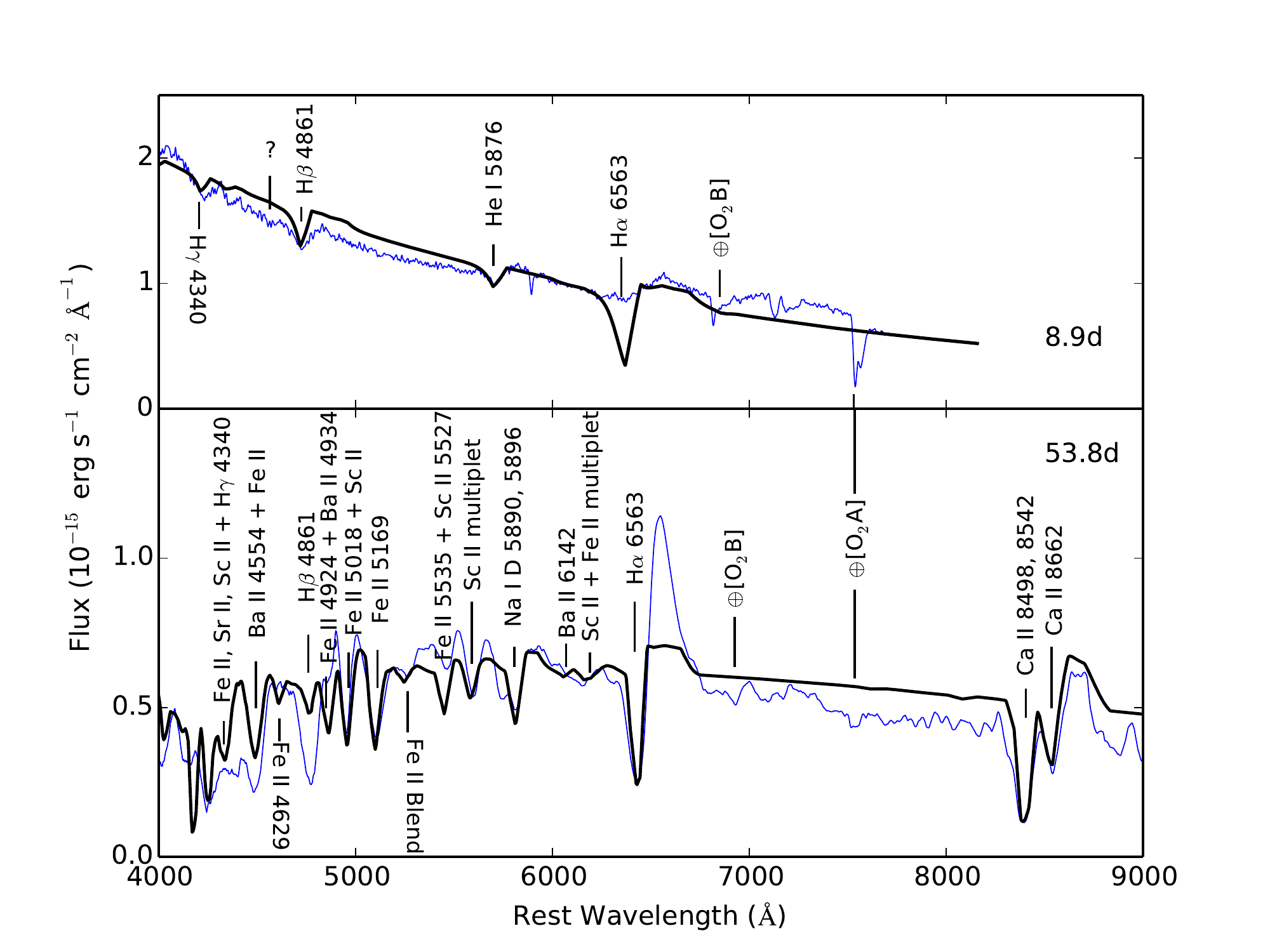}
	\end{center}
	\caption{SYN++ modelling of the 8.9 and 53.9~d spectra of SN 2015ba. Model spectra are shown with thick solid (black) lines, while the observed ones are in thin (blue) solid line.}
	\label{fig:spectra_2015ba}
\end{figure}

The early spectrum mostly exhibits broad H and He I lines superposed on a blue continuum, which are fairly well-reproduced in the synthetic spectrum. There is a weak but conspicuous absorption feature at $\sim$ 4560 \AA{}, marked by \textquoteleft ?\textquoteright{} and identified from the modelling as He II 4686. With time, multiple metal lines gain prominence, resulting in line blendings. The SYN++ modelling aids in identifying most of the blended features in the plateau phase spectrum at 53.8~d (Figure \ref{fig:spectra_2015ba}). The P-Cygni profiles to the blue side of H$\alpha$ profile are identified as Sc II and Ba~II $\lambda$6142. However, there is a discrepancy between the observed and the modelled spectra at shorter wavelengths, which could perhaps improve with the addition of other species such as Ti II lines. 
Although, the SYN++ fit reproduces most of the features, it fails to reproduce the emission component of H$\alpha$ line. The net emission from H$\alpha$ line is a characteristic feature of SN II spectra and is not reproducible in such simplified models without assuming an ad hoc net line emission \citep[as reported in][]{1990sjws.conf..149J}.

\subsection{Early phase features}
\begin{figure}
	\begin{center}
		\includegraphics[scale=1.0, width=.83\textwidth,clip, trim={0cm 0cm 0cm 0cm}]{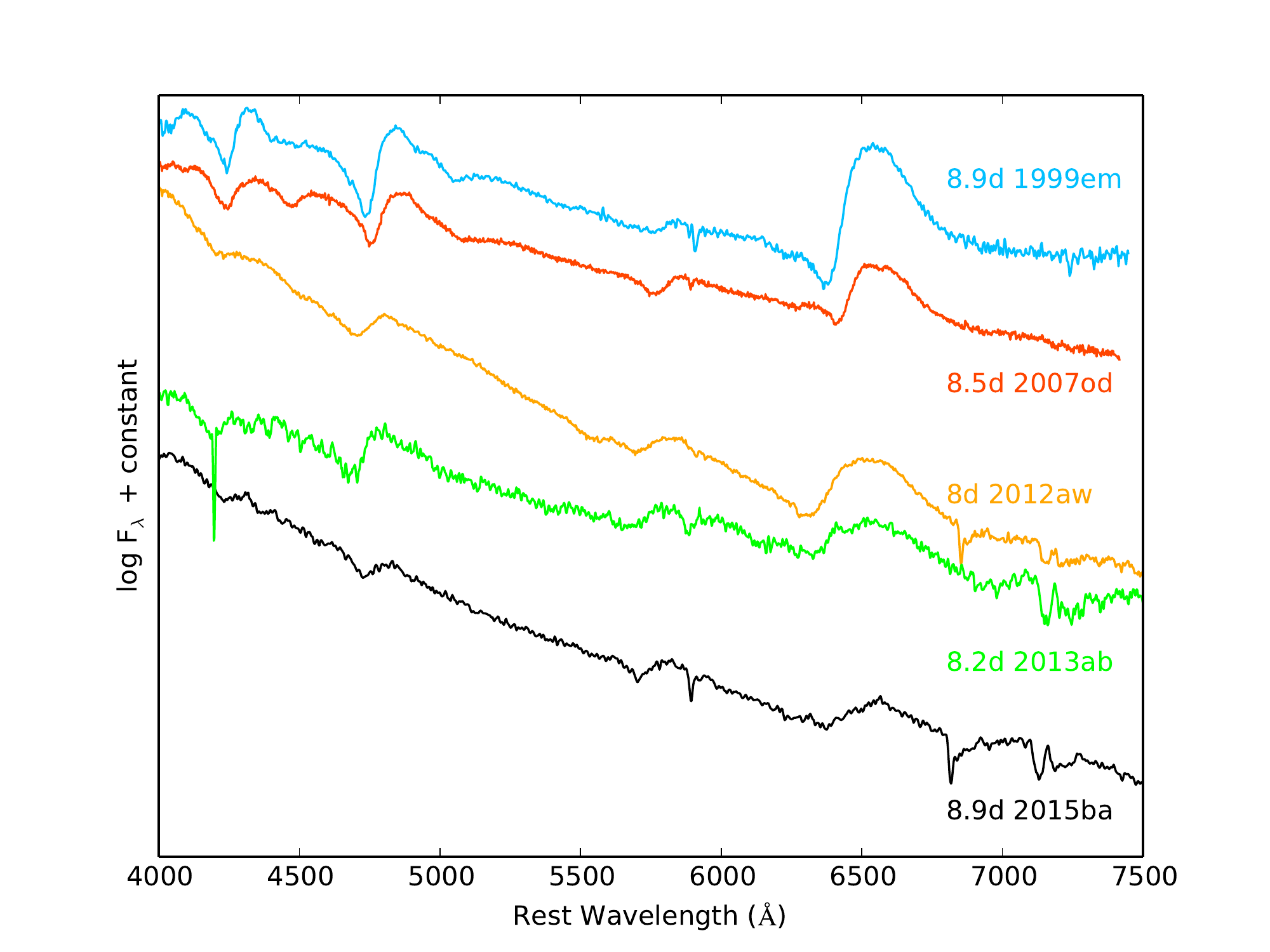}
	\end{center}
	\caption{Comparison of an early spectrum (+8d) of SN 2015ba with other archetypal Type IIP SNe at similar phases.}
	\label{fig:spectra_comp_early}
\end{figure}

The early spectra (8.9-12.9 d) are characterized by a few broad P-Cygni profiles superimposed on a blue continuum. The most prominent are the H and He I $\lambda$5876 lines. H$\alpha$ has a weaker absorption component as compared to other H I lines. The 8.9 d spectrum modelled using SYN++ suggests the absorption feature at $\sim$ 4560 \AA{} arises from the excitation of He. We compared the 8.9 d spectra of SN 2015ba with coeval epoch spectra of SNe 1999em, 2007od, 2012aw and 2013ab in Figure \ref{fig:spectra_comp_early}. The P-Cygni profiles are quite well-developed in the spectra of SNe 1999em, 2007od and 2012aw, while those of SNe 2013ab and  2015ba have a more featureless continuum. This is possibly due to temperature difference in the initial phases among the SNe. The spectrum from 12.9 d to 18.9 d marks the transition from the early to the plateau phase. The He I feature  completely disappears after 18.9 d, and later on metal lines start to develop.

\subsection{Plateau phase}
\begin{figure}
	\begin{center}
		\includegraphics[scale=1.0, width=.83\textwidth,clip, trim={7.5cm 4.5cm 29cm 7.6cm}]{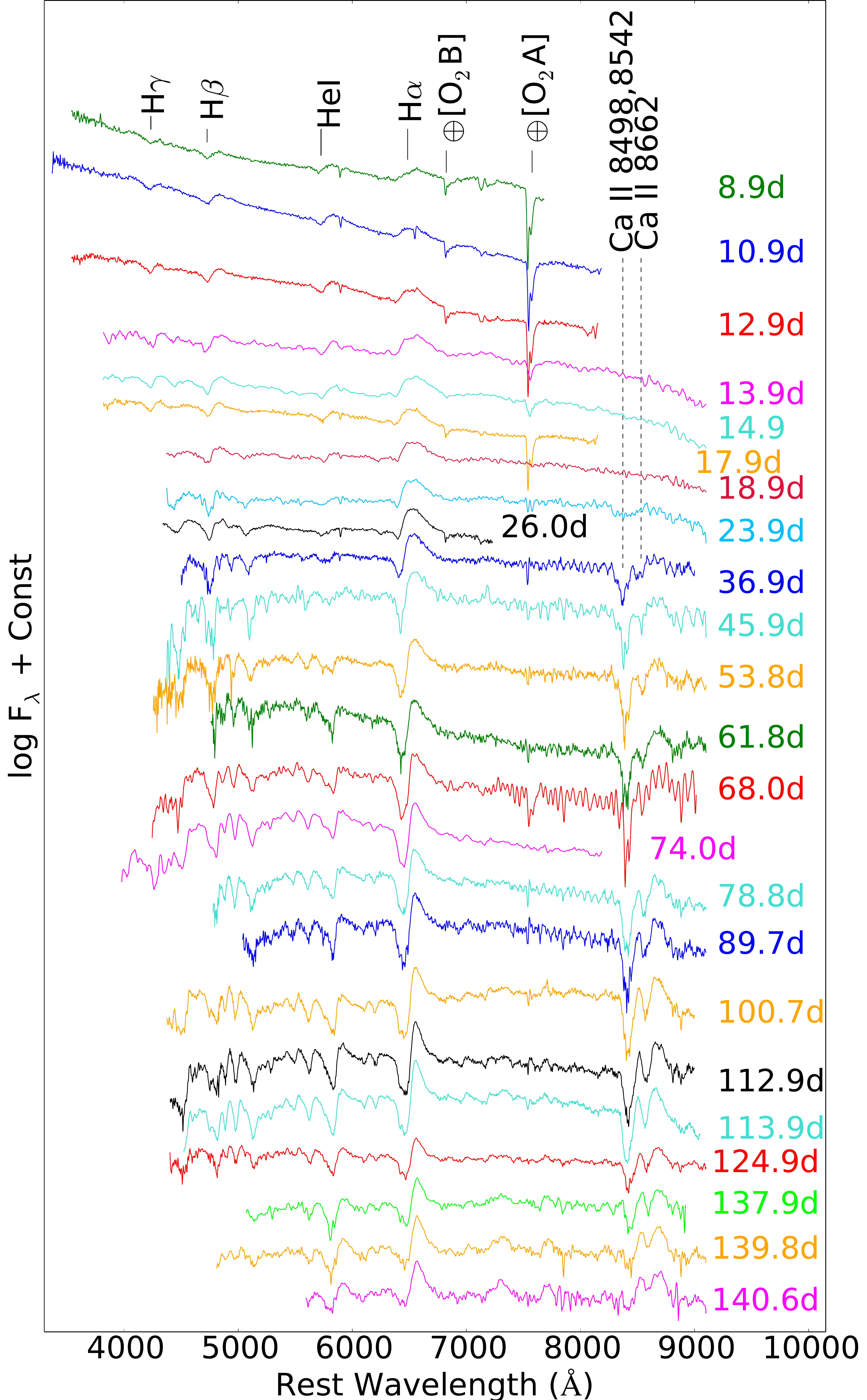}
	\end{center}
	\caption{Doppler-corrected spectra of SN 2015ba from 8.9 to 140.6 d since explosion. The symbol $\oplus$ indicates the position of the most prominent telluric absorption bands.}
	\label{fig:spectra_plot}
\end{figure}

\begin{figure}
	\begin{center}
		\includegraphics[scale=1.0, width=.83\textwidth,clip, trim={0cm 0cm 0cm 0cm}]{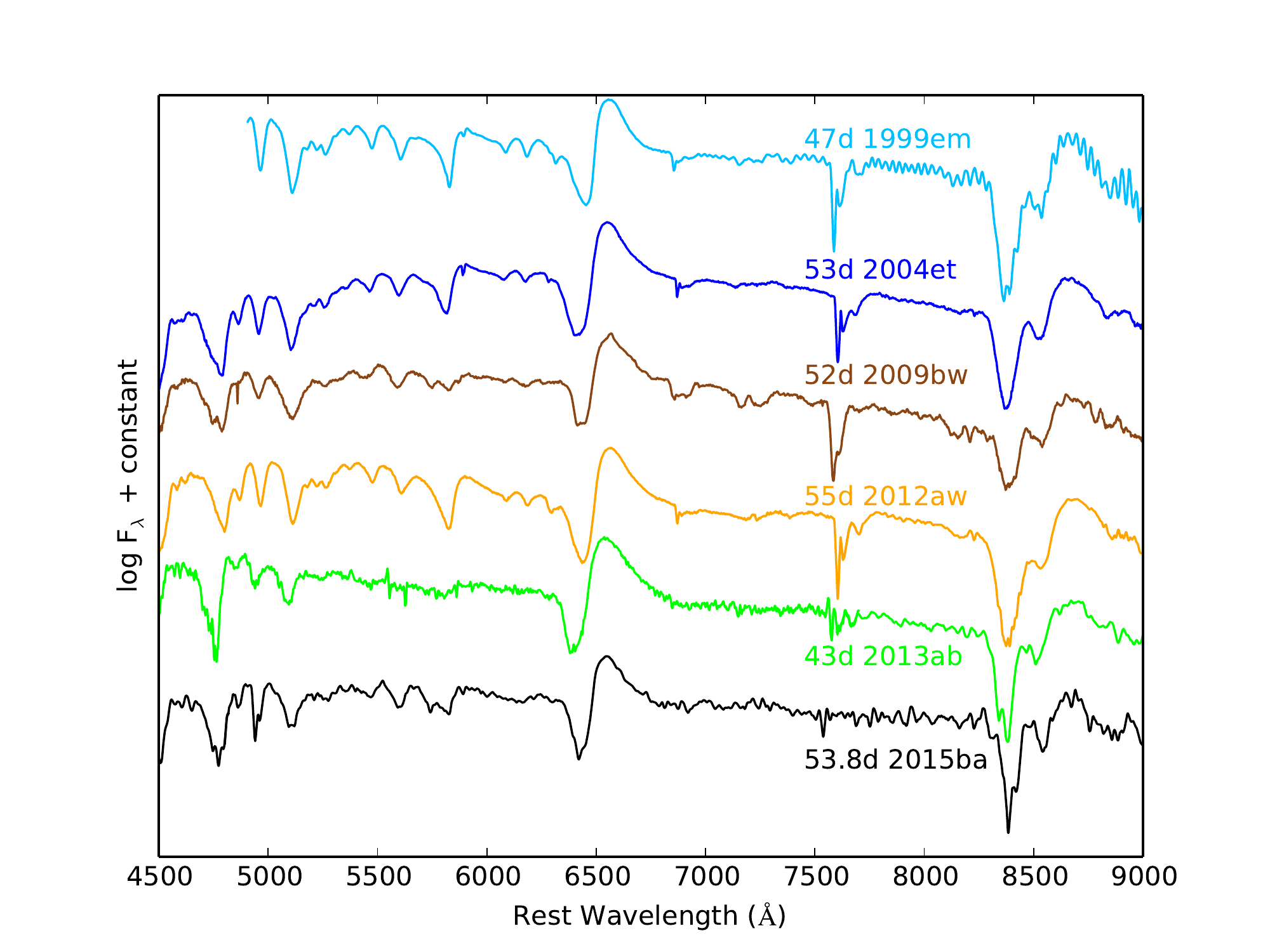}
	\end{center}
	\caption{Comparison of a mid-plateau spectrum (53.8~d) of SN 2015ba with other archetypal SNe IIP at similar phases.}
	\label{fig:spectra_comp_plateau}
\end{figure}

As the SN expands and cools, the opacity drops and the continuum becomes redder. Metal lines develop, the emission component of H$\alpha$ becomes stronger, while the other H I Balmer lines attenuate. In place of He I 5876, Na I D line develops and become stronger with time. The spectra evolves very slowly during the plateau (36.9 to 124.9 d, Figure \ref{fig:spectra_plot}). Fe II $\lambda$5169 is feebly visible in the 18.9 d spectrum and becomes prominent after day 23.9. Other lines of Fe II (Fe II $\lambda\lambda$ 5018, 4924) appear in the 23.9 d spectrum, but are weak. These lines become prominent from the 36.9 d spectrum. Ba II lines appear with the 26 d spectrum. The Ca~II IR triplet ($\lambda\lambda$ 8498, 8542, 8662) becomes visible from day 23.9. Thereafter, a number of P-Cygni profiles develop in the spectra. We compare the mid-plateau phase spectrum of SN 2015ba with those of other SNe IIP in Figure \ref{fig:spectra_comp_plateau}. The spectra of SN 2015ba shows remarkable similarity in features with other SNe IIP, only the strength of the Ba II, Sc II and Fe II lines are weaker in SNe 2015ba, 2013ab and 2009bw.

\subsection{Nebular Phase}
\begin{figure}
	\begin{center}
		\includegraphics[scale=1.0, width=.83\textwidth]{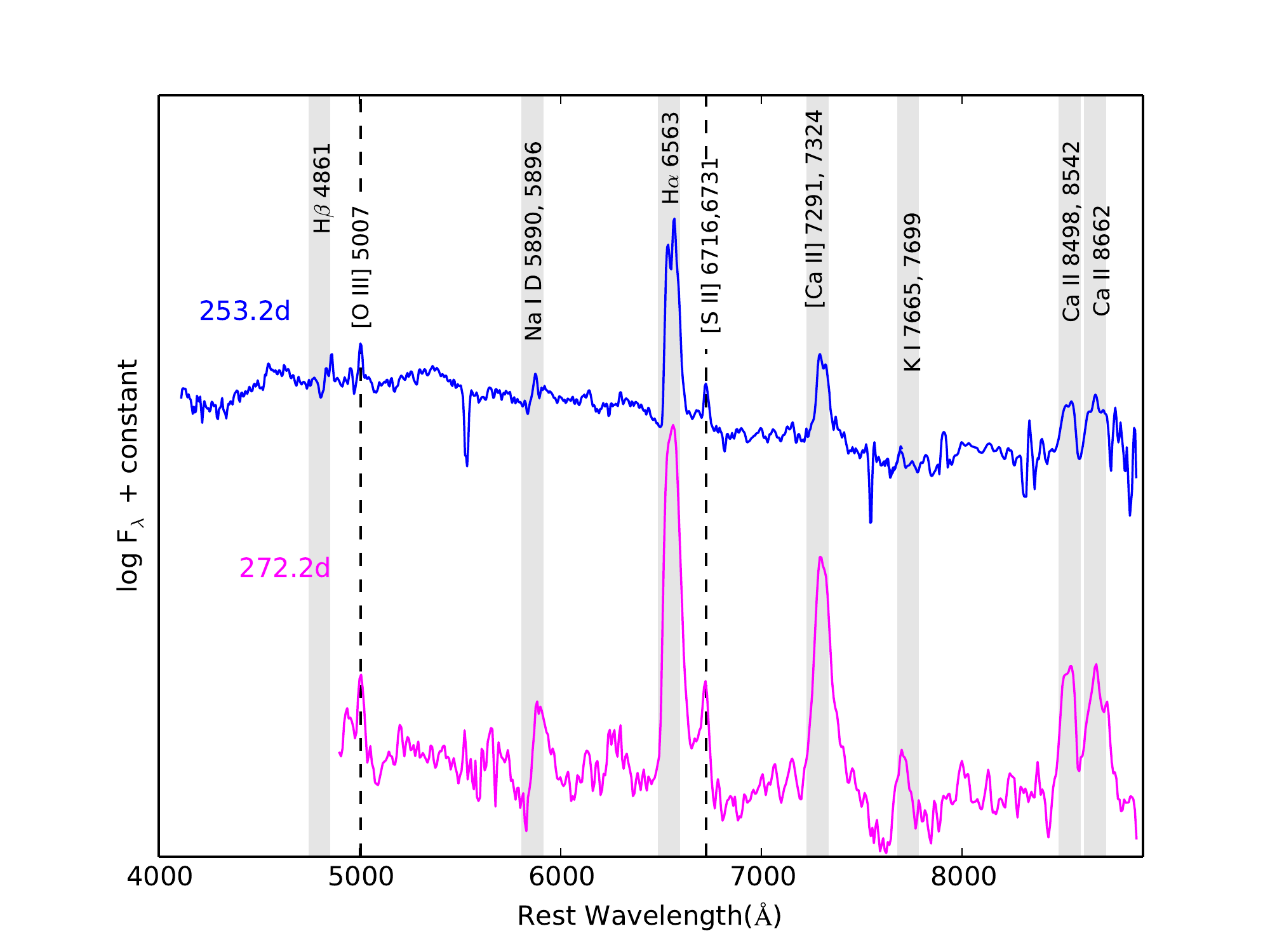}
	\end{center}
	\caption{Doppler-corrected nebular phase spectra of SN 2015ba. The emission lines from the galaxy are shown with dashed lines.}
	\label{fig:spectra_second_plot}
\end{figure}

\begin{figure}
	\begin{center}
		\includegraphics[scale=1.0, width=.83\textwidth,clip, trim={0cm 0cm 0cm 0cm}]{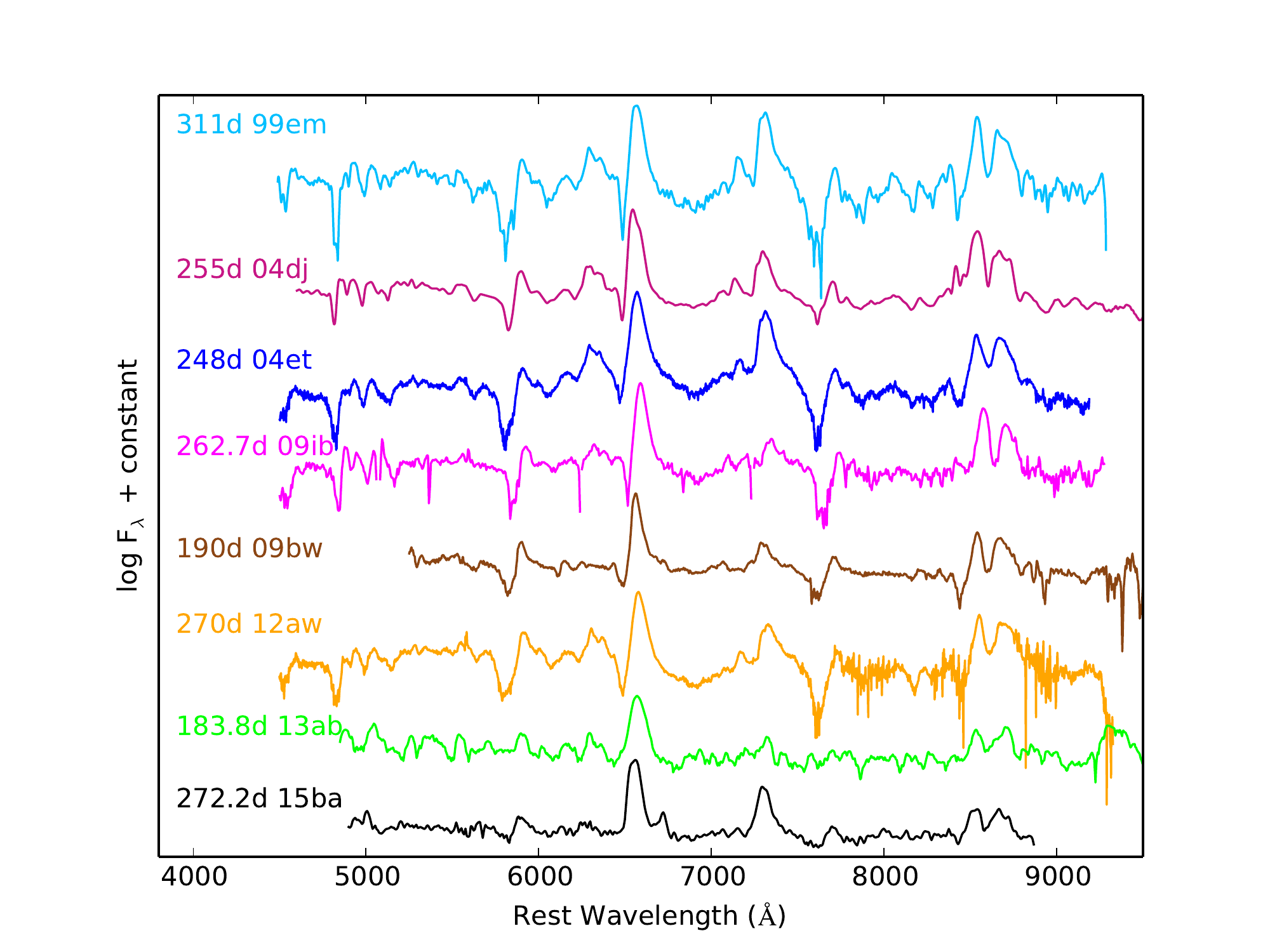}
	\end{center}
	\caption{Comparison of the nebular spectrum (272.2 d) of SN 2015ba with those of other archetypal Type IIP SNe.}
	\label{fig:spectra_comp_nebular}
\end{figure}

In the nebular phase, owing to the optically thin ejecta, the absorption component of the P-Cygni profiles vanishes. The 253 and 272~d spectra obtained with the 10.4~m GTC, shown in Figure \ref{fig:spectra_second_plot}, mainly exhibit H, Na I D and Ca II infrared triplet features along with some nebular lines. The H$\alpha$ line is the most prominent, while the other H Balmer lines, though detectable, are much weaker. The Balmer lines exhibit double-peaked profiles in the 253~d spectrum, with the red peak at the rest wavelength. The double peaked feature in H$\alpha$ may possibly be attributed to noise in the spectra, but a similar feature is detected for H$\beta$. This supports that it is an intrinsic feature. \cite{2015MNRAS.448.2482J} suggested that double peaked profiles may arise from dust in the circumstellar medium at very late epochs ($\gtrsim$ 400 d). However, at the epochs ranging from 200-300 d, \cite{2015MNRAS.448.2482J} state that dust will have negligible effect on the optical spectra of SNe IIP. We, therefore, disfavour an effect of dust, and suggest this to be an artifact due to background line contamination. The appearance of [O III] 5007 and [S II] 6716, 6731 emission lines from the host galaxy are clear signatures that the background is contaminating the SN spectra at late phases. Forbidden lines of [Ca~II] 7291, 7324 and the K I $\lambda\lambda$7665, 7699 also becomes prominent in the nebular spectra. The plethora of permitted and forbidden emission lines of iron from various multiplets generates several blends. The comparison of the nebular phase spectrum of SN 2015ba at 272 d with late-time spectra of other SNe IIP is shown in Figure \ref{fig:spectra_comp_nebular}. Most of the features in the nebular spectra in the comparison SNe IIP are similar, however, the forbidden [O I] 6300,6364 doublet is conspicuous in the comparison spectra and is absent in SNe 2015ba and 2009bw. Also, the absorption components in SN 2015ba is smaller with respect to the comparison sample. 
 
\subsection{Evolution of the line profiles}
\begin{figure}
	\begin{center}
		\includegraphics[scale=1.0, width=.95\textwidth,clip, trim={1.7cm 0cm 0cm 0cm}]{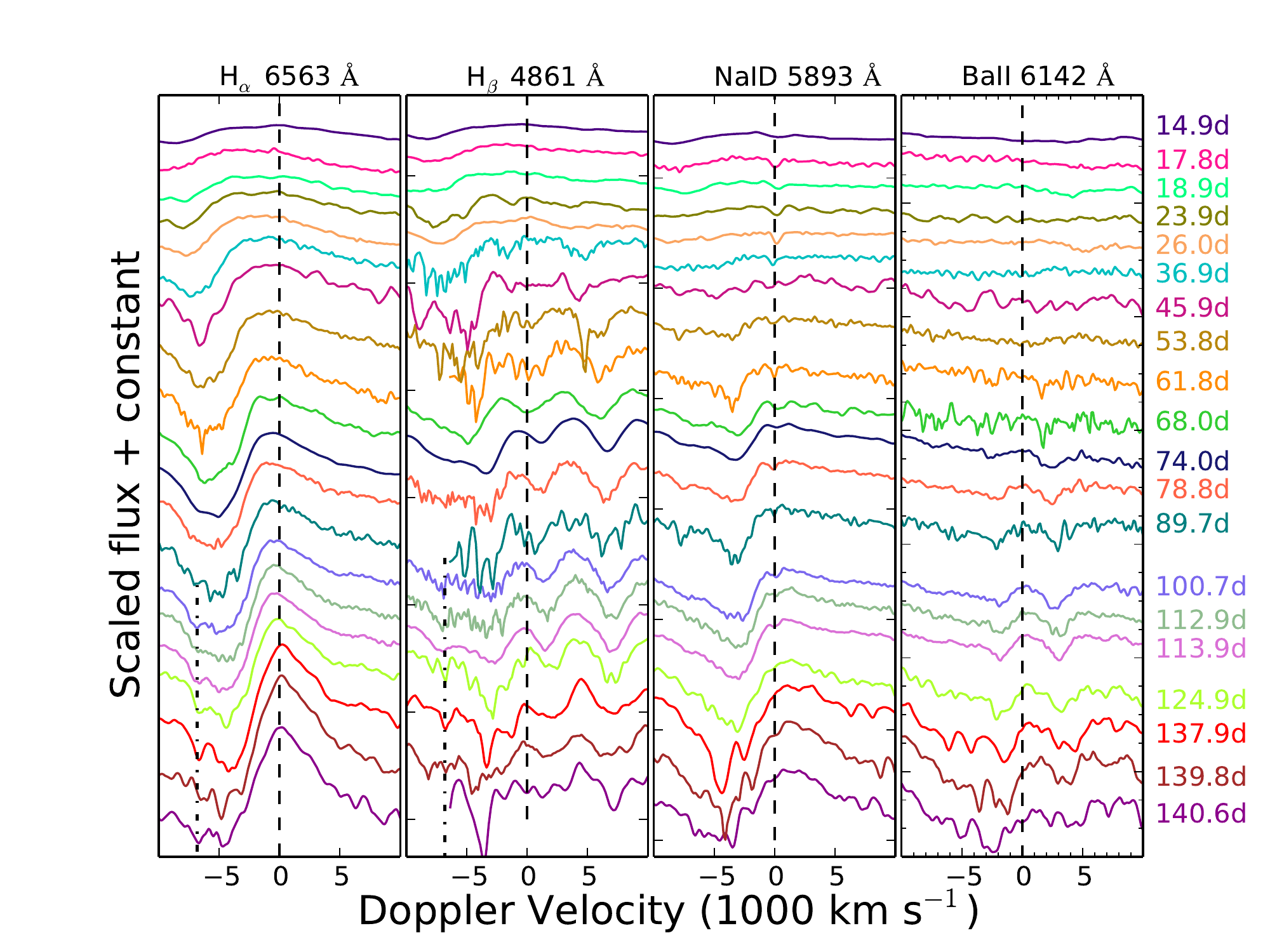}
	\end{center}
	\caption{Evolution of H$\alpha$,  H$\beta$, Na I D and Ba II 6142 \AA{} emission line profiles during the photospheric phase. The high velocity (HV) feature conspicuous from the 89.7 d spectrum up to the 140.6 d spectrum is shown with dot-dashed lines.}
	\label{fig:spectra_line_evolution}
\end{figure}

The evolution of H$\alpha$, H$\beta$, Na I D and Ba II $\lambda$6142 lines from 14.9 to 140.6~d is shown in Figure \ref{fig:spectra_line_evolution} in the velocity domain and centered at the rest wavelengths of these lines. The H$\alpha$ feature is broad and weak at early phases and becomes stronger and narrower at late phases. The profiles of H$\beta$, Na I D and Ba II $\lambda$6142 are hardly discernible at early phases and only gain prominence at later phases. The He I 5876 line is visible at early phases at the position of Na I D line, and gradually attenuates while the Na I D line emerges. A blueshifted emission profile is seen for H$\alpha$ as well as H$\beta$ line, which gradually shifts to redder wavelengths, reaching zero velocity at the end of plateau phase. As pointed out by \cite{1988SvAL...14..334C} and later by \cite{1990sjws.conf..149J}, the blueshifted emission peak results from the diffuse reflection by the photosphere of the resonance photons emitted in the SN atmosphere that are not absorbed because of the dominant electron scattering opacity at early phases. However, \cite{2014MNRAS.441..671A} carried out an analysis of a sample of SNe IIP and linked the blueshift to the steep density structure of the ejecta, which gives rise to strong occultation effects in the receding part of the ejecta, shifting the emission peak towards the blue. Such blueshifted emission lines have been successfully modelled by \cite{2005A&A...437..667D} using the non-LTE model CMFGEN, taking into account the steep density profile of ejecta. Since H$\alpha$ is the strongest emission line in SN II spectra, the velocity offset is more conspicuous in this line as compared to other emission lines, although the peaks of other emission lines are also blueshifted by a smaller amount.

\begin{figure}
	\begin{center}
		\includegraphics[scale=1.0, width=.75\textwidth,clip, trim={2.0cm 0.3cm 2.0cm 0cm}]{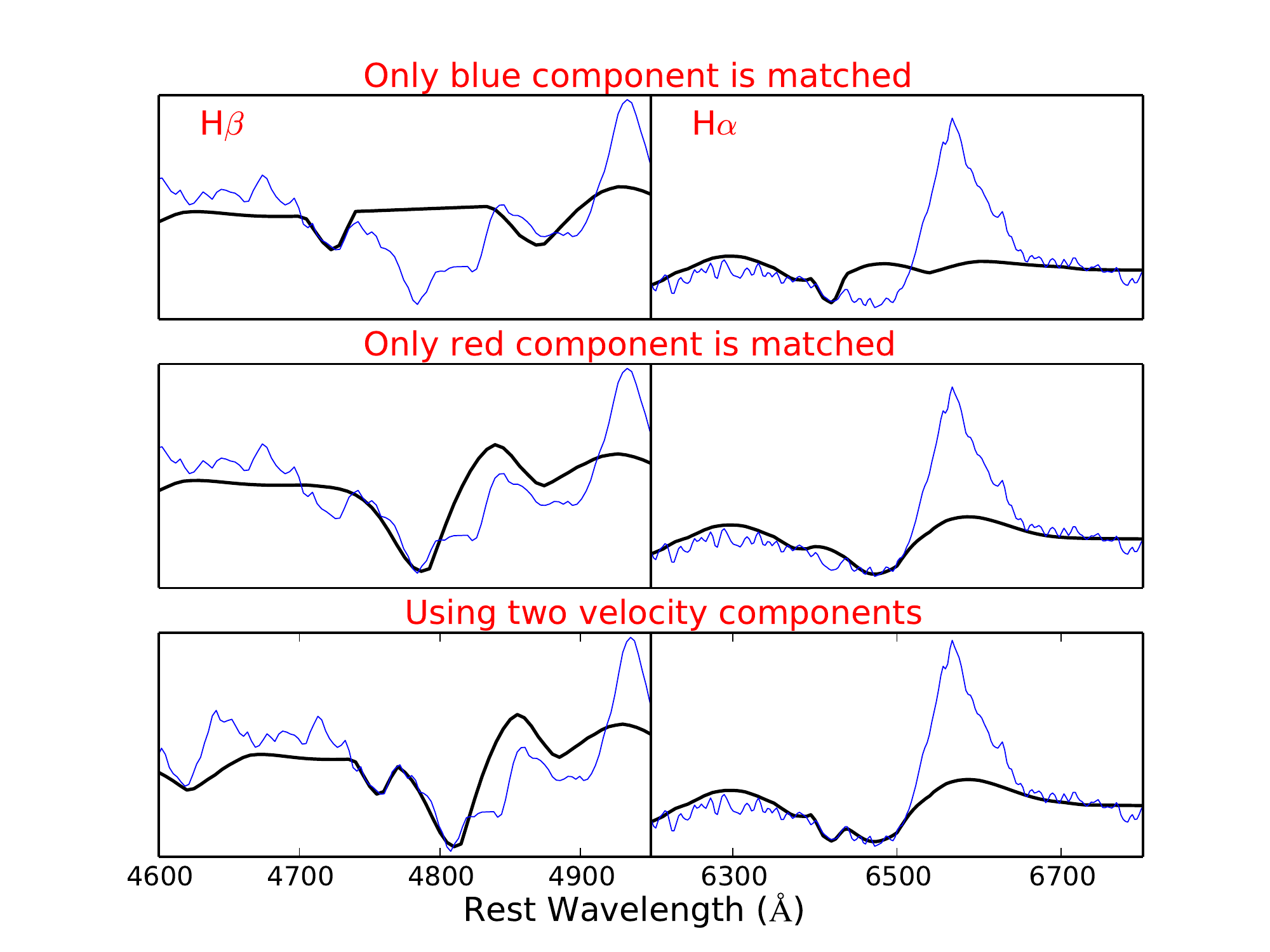}
	\end{center}
	\caption{SYN++ modelling of the 137.9 d spectrum showing fits to the high velocity (HV) notch in H$\alpha$ (right panel) and H$\beta$ (left panel). A single high velocity component (6800 km s$^{-1}$) is used to match the blue dip (top panel), a single low velocity component (3100 km s$^{-1}$ for H$\beta$ and 4500 km s$^{-1}$ for H$\alpha$) is used to match the red dip (middle panel) and two velocity components are used to fit both the absorption dips (bottom panel).}
	\label{fig:syn++_HV}
\end{figure}

We also observe a non-evolving, narrow absorption component  to the blue side of H I lines from the 89.7 to 140.6 d spectra of SN 2015ba (marked with dash-dotted lines in Figure \ref{fig:spectra_line_evolution}). This component is most conspicuous in H$\alpha$ trough at $\sim$ 6420 \AA{} and is detectable in the blue wings of H$\beta$ at $\sim$ 4750 \AA{} in the 124.9 and 137.9 d spectra. This component has been observed at the mid-plateau epochs in several SNe II \citep{2000ApJ...545..444B,2002PASP..114...35L,2007ApJ...662.1136C,2012MNRAS.422.1122I,2014MNRAS.438L.101V,2015ApJ...806..160B,2017ApJ...850...89G} and has been attributed to either Si II 6355 \AA{} or high velocity (HV) hydrogen lines as a result of weak interaction between the SN ejecta and the RSG wind. However, this feature does not seem to evolve with time in SN 2015ba like other species in the shell, hence it is not arising from Si II. \cite{2007ApJ...662.1136C} suggested that the blue shoulder in the absorption profile of H I lines originate from the excitation of the cool, dense shell formed at the interface of the SN ejecta and CSM. Considering this feature to be HV components of H$\alpha$ and H$\beta$, both of these are observed at about 6800 km s$^{-1}$ from the rest wavelength of these lines, which supports the identification of these components as HV Balmer lines. We are also able to model this feature in the 137.9~d spectrum with a high velocity component of H I in the SYN++ modelling as shown in Figure \ref{fig:syn++_HV}, which further strengthens our assumption.

\subsection{Velocity Evolution}
The photospheric velocity of the ejecta traces the velocity of propagation of the recombination wave which in turn depends on the physical conditions of the ejecta such as the temperature, density and extent of mixing of elements in the ejecta. \cite{2005A&A...439..671D} described the photosphere as the layer where the continuum optical depth is $\sim$ 2/3. However, there is no single prescribed line which can accurately represent the true photospheric layer and its velocity. We used Fe II 5169 line as a proxy for the photospheric velocity in the plateau phase \citep{2005A&A...439..671D} while at early phases, He I 5876 line is used as suggested in \cite{2012MNRAS.419.2783T}. The velocity is determined from the shift in the absorption minima of these lines or through spectral modelling. We also compute the radial velocity from the absorption minima of H$\alpha$, H$\beta$ and Sc II lines. These lines form in the outer layers of the envelope, exhibiting higher expansion velocities than the Fe II lines. However, we also note the drop in velocity of H$\beta$ at later phases, indicative of the presence of hydrogen in the deeper ejecta layers. 

\begin{figure} 
	\begin{center}
		\includegraphics[scale=0.7,clip, trim={1.2cm 0.2cm 2.0cm 1.5cm}]{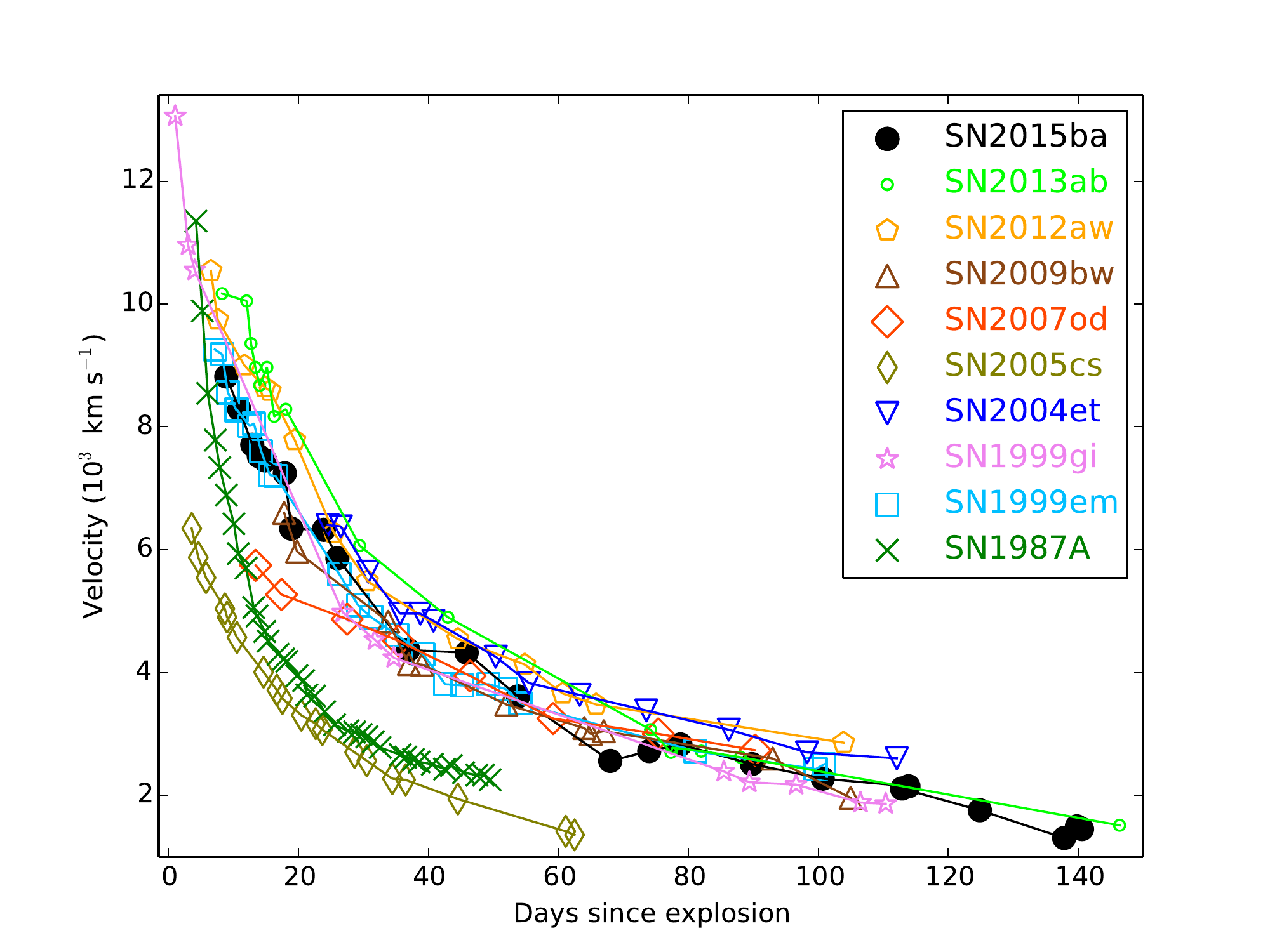}
	\end{center}
	\caption{The evolution of photospheric velocity ($v_{ph}$) of SN 2015ba is compared with those of other well-studied SNe. The $v_{ph}$ plotted here are absorption trough velocities (from He I at early phases and Fe II $\lambda$5169 at late phases).}
	\label{fig:vel_comp}
\end{figure}

The photospheric velocity evolution of SN 2015ba is compared to other SNe IIP in Figure \ref{fig:vel_comp}. The velocities derived from the absorption minima of He I/ Fe II line have been used for all SNe to maintain consistency. While the velocity profile of SN 2015ba is similar to other SNe IIP, the velocities are significantly higher than SNe 1987A and 2005cs. The velocities of SN 2015ba are, however, more similar to SNe 1999em and 1999gi and are about $\sim$ 500~km~s$^{-1}$ lower than SNe 2004et, 2012aw and 2013ab. 

\section{Explosion Properties}
\label{sec8}
\subsection{Bolometric light curve}
\label{sec8.1}
The bolometric light curve, with contributions ranging from the ultraviolet (UV) to the infrared (IR) domains, is a useful tool to derive the physical parameters of the explosion such as the synthesized $^{56}$Ni mass, and the mass and kinetic energy of the ejecta. However, a well-sampled data set over all wavelengths is usually rare. Therefore, we construct a pseudo-bolometric light curve by integrating the fluxes in {\it BVri} bands only. 

We convert the dereddened {\it BVri} magnitudes of SN 2015ba to fluxes using the zero points adopted from \cite{1998A&A...333..231B} and \cite{1996AJ....111.1748F}. We integrate over the filter bandpass using these monochromatic fluxes implementing a trapezoidal integration rule, and for each epoch we obtain the pseudo-bolometric flux. Adopting a distance of 34.8 $\pm$ 0.7 Mpc, we compute the pseudo-bolometric luminosity (L$_{BVri}$). 

\begin{figure}
  \begin{center}
		\includegraphics[scale=0.9]{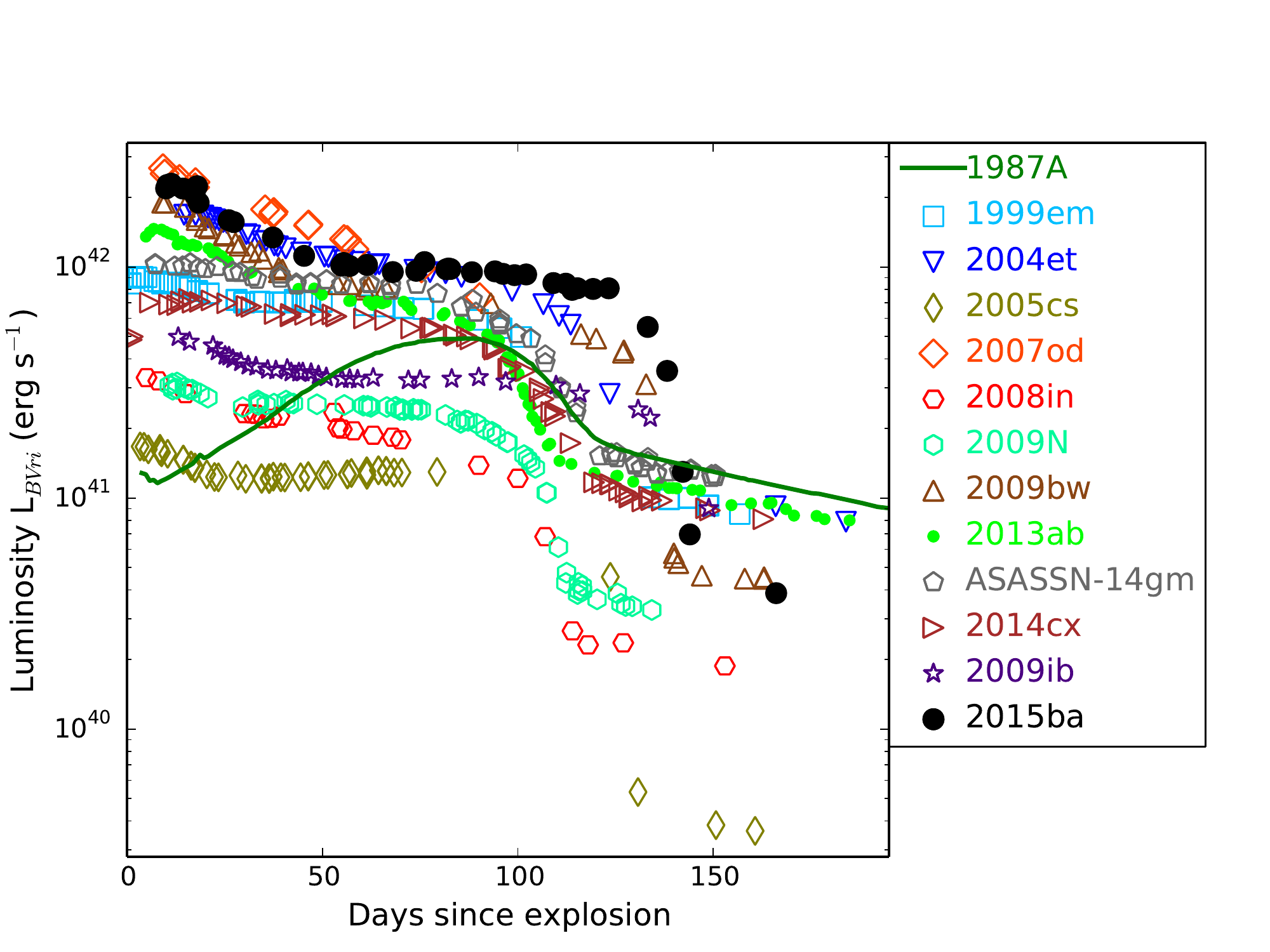}
	\end{center}
	\caption{{\it BVri} pseudo-bolometric light curve of SN 2015ba shown along with those of a sample of Type IIP SNe. {\it r}/{\it R} and {\it i}/{\it I} bands were used for the comparison sample depending on the availability of observations in these filters.}.
	\label{fig:final_bolometric_plot}
\end{figure}

The pseudo-bolometric light curve of SN 2015ba is compared with those of the SNe sample in Figure \ref{fig:final_bolometric_plot}. We generate the pseudo-bolometric light curves of all comparison SNe using the method described above for consistency. The luminosity of SN 2015ba in the plateau phase is comparable with that of the luminous SN 2004et and ASASSN-14gm. However, the constant luminosity phase is extended for SN 2015ba, possibly indicating a massive hydrogen envelope. Due to limited multiband data in the radioactive tail phase, we are unable to estimate the pseudo-bolometric luminosities for SN 2015ba at late phases.

We construct the true bolometric light curve, using the direct integration module in SuperBoL \citep{2017PASP..129d4202L} which is based on the method given in  \cite{2009ApJ...701..200B}. With this method, the IR contribution is obtained by fitting a blackbody spectrum to the observed fluxes redward of 5000\AA{} \citep[as suggested in][]{2014MNRAS.437.3848L} and then integrating the best fit blackbody curve from the longest observed wavelength to $\lambda$ = $\infty$.  The missing UV flux is obtained by integrating the blackbody  function from the shortest observed flux to f$_\lambda$ = 0 at $\lambda$ = 2000\AA. 
We cannot adopt this method to estimate the UV contribution to the bolometric light curve in the nebular phase since this method is only valid during the optically thick phases (up to $\sim$ 115 days in SN 2015ba) when the SED of the SN can be approximated to a blackbody. So, we estimate the UV contribution at late phases with a linear function from the effective wavelength of the shortest observed filter to f$_\lambda$ = 0 at $\lambda$ = 2000\AA. The UV (2000 \AA - 3500 \AA) and IR contribution (8200 - 21900 \AA) in SN 2015ba, thus obtained, is $\sim$ 36\% and $\sim$ 10\%, respectively on day 11. The UV contribution falls to $\sim$ 10\% and the IR contribution rises to 40\% by day 120.

The contributions in the UV and IR are added to the pseudo-bolometric luminosities in order to obtain the true bolometric luminosities. The true bolometric luminosities are used in the analytical and hydrodynamical modelling discussed later in Sections \ref{analy} and \ref{hydro} respectively.
  
\subsection{$^{56}$Ni mass}
\label{ni_mass}
The nebular phase light curve is thought to be powered by the radioactive decay of $^{56}$Co to $^{56}$Fe resulting in the emission of gamma rays and positrons. As the ejecta in the nebular phase are still opaque to gamma rays and the $^{56}$Co is the radioactive decay product of $^{56}$Ni, the luminosity is proportional to the $^{56}$Ni nucleosynthesized at the time of shock breakout. We determine the mass of $^{56}$Ni using the bolometric luminosity by adopting two empirical methods discussed below. 

\subsubsection{$^{56}$Ni mass from tail luminosity}
Assuming that the deposition of $\gamma$-photons and positrons emitted from radioactive decay of $^{56}$Co results in the thermalization of energy in the ejecta, $^{56}$Ni mass can be independently estimated from the tail luminosity, as described by  \cite{2003ApJ...582..905H}:
\begin{equation}
 M_{Ni} = 7.866 \times 10^{-44} \times L_t exp\Bigg[\frac{(t-t_0)/(1+z) - 6.1}{111.26}\Bigg] M_\odot 
\end{equation}
where $t_0$ is the explosion time, 6.1 days is the half-life time of $^{56}$Ni and 111.26 days is the e-folding time of the $^{56}$Co decay. The tail luminosity $L_t$ is computed at the only epoch available at the tail ($\sim$166 d) from the {\it V}-band magnitude corrected for distance, extinction, and bolometric correction factor of 0.26 $\pm$ 0.06 mag during the nebular phase \citep{2003ApJ...582..905H} using the following equation:
\begin{equation}
\label{equ14}
 log (L_t) = -0.4 [ V_t - A_V(tot) + BC] + 2 log D -3.256
 \end{equation}
 where D is the distance in cm and BC is the bolometric correction in the nebular phase. The value of $L_t$ obtained with equation \ref{equ14} is 9.9$\pm$1.8 $\times$ 10$^{40}$ erg s$^{-1}$, which corresponds to $^{56}$Ni mass of 0.032 $\pm$ 0.006 M$_\odot$.

\subsubsection{$^{56}$Ni mass from nebular spectra}
We also deduce the $^{56}$Ni mass from FWHM of the H$_\alpha$ emission line in the late time spectra of SN 2015ba using the relation from \cite{2012MNRAS.420.3451M}:
\begin{equation}
\label{equ15}
M(^{56}Ni) = A \times 10^{B.FWHM_{corr}} M_\odot
\end{equation}
where A = 1.81$^{+1.05}_{-0.68} \times 10^{-3}$, B = 0.0233 $\pm$ 0.0041 and FWHM$_{corr}$ is the FWHM of line profiles corrected for the spectral resolution of the instruments used to obtain the spectrum given by
\begin{equation}
\label{equ16}
FWHM_{corr} = \sqrt{FWHM_{obs}^2 - FWHM_{instr}^2}
\end{equation}
 We obtain the FWHM by fitting a gaussian function to the emission lines in the spectra at 253 and 272 d. For the instrumental FWHM (FWHM$_{instr}$), we measured the FWHM of [O I] 6300\AA {} sky emission line, which was found to be 18.17 $\pm$ 1.71 \AA{} and 27.94 $\pm$ 1.48 \AA{} on 253 and 272 d respectively. The FWHM of H$_\alpha$ emission line is 60.68 $\pm$ 1.83 \AA{} and 65.68 $\pm$ 2.16 \AA{} and using equation \ref{equ16}, we obtain the corrected FWHM to be 57.89 $\pm$ 2.50 \AA{} and 59.44 $\pm$ 2.62 \AA{} on 253 and 272 d respectively. Substituting the corrected FWHM in equation \ref{equ15}, we obtain the average $^{56}$Ni mass from the two late time spectra to be  0.042$^{+0.026}_{-0.020}$ M$_\odot$. 
 
We find that the $^{56}$Ni mass obtained using the above two methods yields consistent results within the errors. The weighted mean of ejected $^{56}$Ni is 0.032 $\pm$ 0.006 M$_\odot$.

\subsection{Analytical modelling}\label{analy}
The physical parameters of SN 2015ba such as the explosion energy, ejected mass and initial radius can be estimated using the two-component semi-analytical modelling of \cite{2014A&A...571A..77N} and \cite{2016A&A...589A..53N} incorporating a dense inner core and an extended low mass envelope. This technique uses the radiative diffusion model to include the effect of the recombination front in the ejecta which was originally developed by \cite{1989ApJ...340..396A}. The main assumptions in the extended model of \cite{2014A&A...571A..77N} is a homologously expanding spherical ejecta and constant Thomson scattering opacity in a given ionized layer. The photon diffusion equation in the ejecta is then solved considering the recombination energy, the radioactive energy from the decay of  $^{56}$Ni and  $^{56}$Co and the effect of gamma ray leakage. We fit this model to the true bolometric light curve (see Section \ref{sec8.1}) and derive the best fit parameters. One should note, however, that independent values of opacity, ejected mass (M$_{ej}$) and kinetic energy (E$_{kin}$) cannot be estimated using this technique; only the degenerate combinations of M$_{ej}$$\kappa$ or E$_{kin}$$\kappa$ can be constrained.

We used a constant density profile for the core and the shell. The best fit ejecta mass, progenitor radius and explosion energy are 22 M$_\odot$, 4.8 $\times$ 10$^{13}$ cm ($\sim$ 690 R$_\odot$), and 2.3 foe respectively and the total mass of the progenitor, assuming 1.3-2.0 M$_\odot$ remnant mass, is 24~M$_\odot$. The mass of $^{56}$Ni estimated from the fit is 0.032 M$_\odot$, which is consistent with the value derived from tail luminosity.

\begin{figure}
\includegraphics[scale=0.8,trim= 2cm 8.5cm 1cm 0.3cm ]{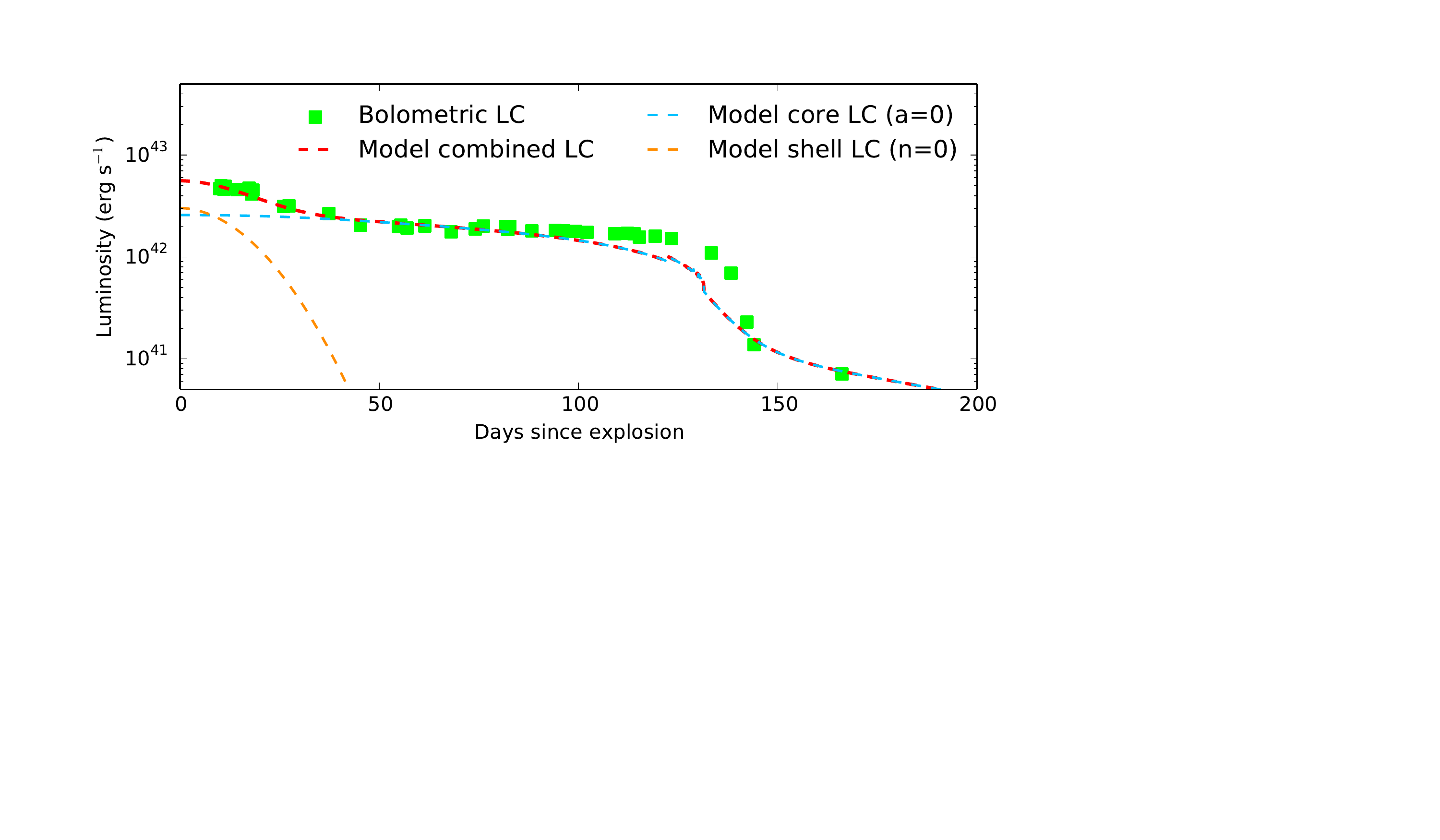}
\caption{The best fit analytical model LC of SN 2015ba using \citet{2016A&A...589A..53N}. The derived model parameters from the fitting are $E$= 2.3 foe, $R$= 4.8 $\times$ 10$^{13}$ cm, and $M_{ej}$= 22 M$_{\odot}$.} 
\label{fig:model_15ba_ana}
\end{figure}

The parameters of the outer shell can also be estimated using this model, like the radius of the H-envelope (R = 2.6 $\times$ 10$^{13}$ cm). Although the poor sampling of the early light curve makes this value not reliable enough for consideration, yet we consider this value as a lower limit for this explosion. The data along with the best fit model is shown in Figure \ref{fig:model_15ba_ana}. The parameters of the shell and core are listed in Table \ref{Nagy}. 

\subsection{Hydrodynamical modelling}\label{hydro}
We derive the SN progenitor's physical properties at the time of the explosion (namely, the ejected mass $M_{ej}$, the progenitor radius $R$ and the total --- kinetic plus thermal --- energy $E$) by means of a well-tested radiation-hydrodynamical modelling procedure already applied to many other observed SNe \citep[e.g.~SNe 2007od, 2009bw, 2009E, 2012A, 2012aw, 2012ec, 2013ab, 2013am, 2014cx, and OGLE-2014-SN-073; see][respectively]{2011MNRAS.417..261I,2012MNRAS.422.1122I,2012A&A...537A.141P,2013MNRAS.434.1636T,2014ApJ...787..139D,2015MNRAS.448.2312B,2015MNRAS.450.2373B,2018MNRAS.475.1937T,2016ApJ...832..139H,2017NatAs...1..713T}
 
A complete description of this procedure can be found in \cite{2017MNRAS.464.3013P}, here we recall that it includes the hydrodynamical modelling of all the main SN observables (i.e. bolometric light curve, evolution of line velocities and the temperature at the photosphere) so as to constrain the SN progenitor's physical properties through a simultaneous $\chi^2$ fit of these observables against model calculations. Two codes are employed for computing the models. The first one is the semi-analytic code described in \cite{2003MNRAS.338..711Z}, which solves the energy balance equation for an ejecta of constant density in homologous expansion. The second one is the general-relativistic, radiation-hydrodynamics Lagrangian code presented in \cite{2010MSAIS..14..123P} and \cite{2011ApJ...741...41P}, which was specifically tailored to simulate the evolution of the physical properties of SN ejecta and the behavior of the main SN observables up to the nebular stage, taking into account both the gravitational effects of the compact remnant and the heating effects due to the decay of the radioactive isotopes synthesized during the explosion. The semi-analytic code is used for a preparatory analysis aimed at individuating the parameter space describing the SN progenitor at the time of the explosion and, consequently, to guide the more realistic, but time consuming simulations performed with the general-relativistic, radiation-hydrodynamics Lagrangian code.

\begin{figure}
\includegraphics[scale=0.6,trim= 1.2cm 1cm 5cm 0.3cm]{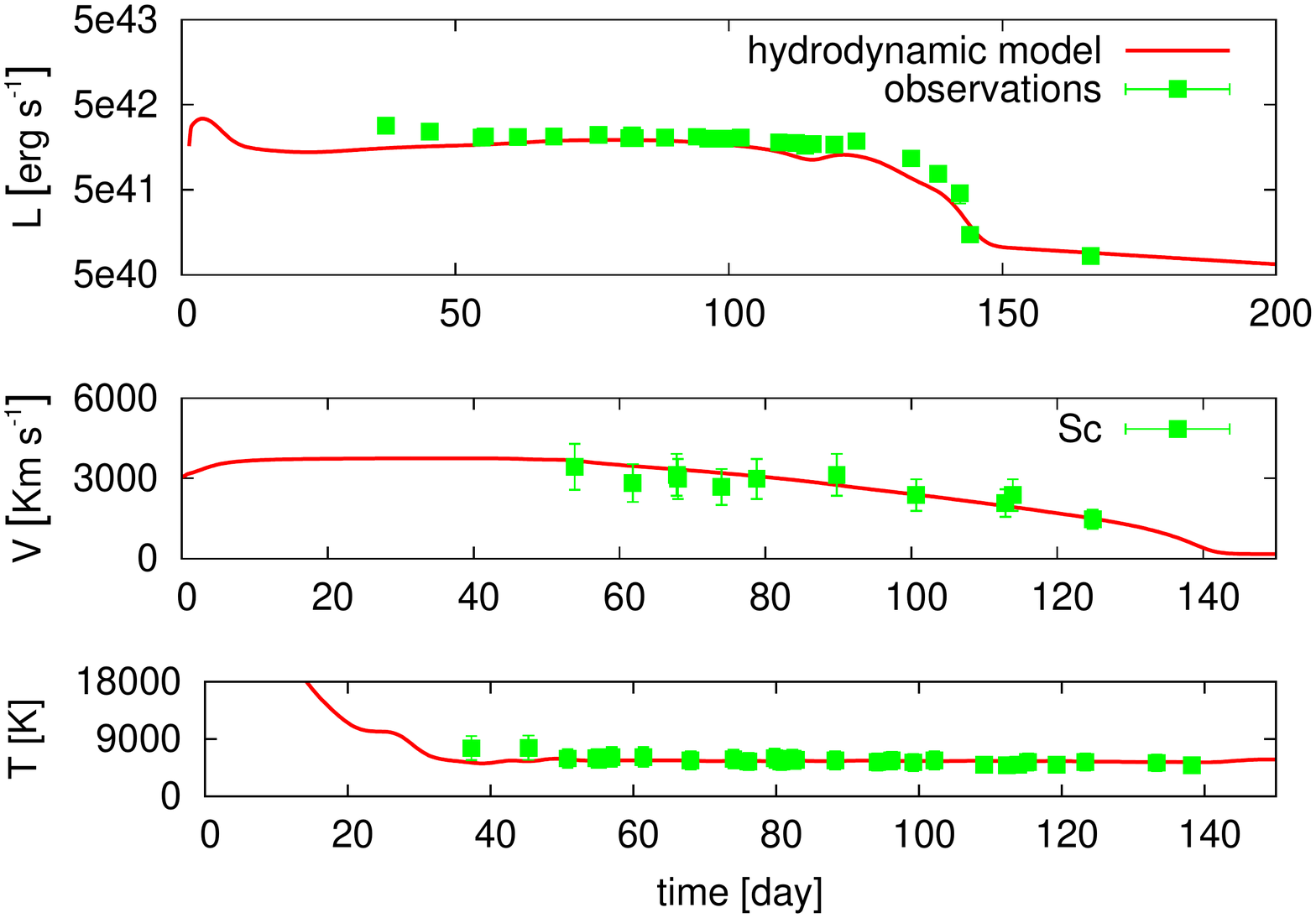}
\caption{Comparison of the evolution of the main observables of SN 2015ba with the best-fit model computed with the general-relativistic, radiation-hydrodynamics code. The best-fit model parameters are $E$= 1.6 foe, $R$= 4.8 $\times$ 10$^{13}$ cm, and $M_{ej}$= 24 M$_\odot$. Top, middle, and bottom panels show the bolometric light curve, the photospheric velocity, and the photospheric temperature as a function of time. To estimate the photosphere velocity from observations, we use the minima of the profile of the Sc lines.} 
\label{fig:model_15ba_hyd}
\end{figure}

Based on the adopted explosion epochs (Section~\ref{exp_ep}), bolometric luminosities (Section~\ref{sec8.1}) and nickel masses (Section~\ref{ni_mass}), we find the best-fitting model shown in Figure \ref{fig:model_15ba_hyd}. It has $E$= 1.6 foe, $R$= 4.8$\times$ 10$^{13}$ cm ($\sim$ 690 R$_\odot$), and $M_{ej}$= 24 M$_\odot$.  
Considering the 2$\sigma$ confidence intervals for one parameter based on the $\chi^2$ distributions produced by the semi-analytical models, we estimate that the error due to the $\chi^2$ fitting procedure is about 15-20 per cent for $M_{ej}$ and $R$ and 30-40 per cent for $E$ \citep[see][for details on the procedure used for the error estimate]{2017MNRAS.464.3013P}.

The values of the best-fit model parameters are consistent with the explosion of a relatively massive red supergiant (RSG) star. In particular, adding the mass of the compact remnant ($\sim$ 1.3-2 M$_\odot$) to that of the ejected material, we obtain a total stellar mass at the time of the explosion of $\sim$ 25.3-26 M$_\odot$. This value seems to indicate a progenitor mass clearly higher than the observational limit of 16.5 $\pm$ 1.5M$_\odot$ of the Type IIP events that raised the so-called \textquotedblleft RSG problem\textquotedblright{} \citep{2009MNRAS.395.1409S}. High progenitor mass has also been estimated for SN 2012aw from the hydrodynamical modelling \citep[$\sim$ 22 M$_\odot$;][]{2014ApJ...787..139D}. Moreover the inferred progenitor\textquoteright s mass is fully consistent (within the errors) with the maximum mass for a SN IIP progenitor found more recently by \cite{2012MNRAS.419.2054W} although the dependability of such higher limit values is strongly debated \citep[e.g.][]{2012ApJ...759...20K}. We also note that the moderately small amount of oxygen and $^{56}$Ni found in the ejecta of SN 2015ba should be not in contrast with a relatively massive progenitor scenario given 1) the uncertainties in modelling the mixing processes of the stellar inner layers that may lead to a not clear connection between the progenitor mass and the nucleosynthesis efficiency \citep[e.g.][]{2014MNRAS.439.3694J}, and 2) the non-monotonical link among the pre-supernova structure, the explosion properties and the progenitor mass \citep{2002RvMP...74.1015W,2014ApJ...783...10S,2016ApJ...818..124E}. 
Finally we point out that, although the progenitor masses estimated from the hydrodynamical modelling are generally higher to be consistent with those determined from direct observations of SN progenitors, the code used here gives lower progenitor masses (compared to other hydrodynamical codes) that are often consistent with mass estimates from the direct progenitor detection method and, more in general, from other independent methods such as the modelling of the observed nebular spectra \citep[e.g.][]{2013MNRAS.434.1636T,2017MNRAS.464.3013P}. 
Here the inferred progenitor\textquoteright s mass is in good agreement with that found using the independent analytical modelling reported in Section \ref{analy}.

\section{Summary}
\label{sec9}

The photometric and spectroscopic data and analysis of SN 2015ba in IC 1029 is presented in this paper. This is a relatively bright SN IIP with a strikingly long plateau ($\sim$ 123~d). The absolute {\it V}-band magnitude at mid-plateau of SN 2015ba is similar to SN 2004et \citep[M$_V^p$=$-$17.1~mag,][]{2006MNRAS.372.1315S}, while significantly lower ($\sim$ 0.7~mag) than the brightest SNe 2007od and DLT16am \citep[$-$17.70 and $-$17.73 mag, respectively,][]{2011MNRAS.417..261I,2018ApJ...853...62T}. The mean distance to the SN is estimated to be 34.8 $\pm$ 0.7 Mpc. The $^{56}$Ni mass obtained from the tail of the light curve is 0.032 $\pm$ 0.006 M$_\odot$. Compared to its brightness, the $^{56}$Ni production in SN 2015ba is much lower than SN 2004et (0.06 M$_\odot$). However, brighter events with comparatively lower $^{56}$Ni masses have been reported for the cases of SNe 2007od and 2009bw \citep{2011MNRAS.417..261I,2012MNRAS.422.1122I}. The drop in magnitude from the plateau to the radioactive tail is also relatively larger and steeper ($\sim$ 3 mag in 20 days in {\it V}-band) than normal SNe IIP, however, smaller than the sub-luminous SNe IIP \citep[e.g. SN 2005cs, $\sim$ 3.8 mag in 20 days in {\it V}-band;][]{2009MNRAS.394.2266P}. We mapped the temperature evolution by fitting a blackbody to the SED obtained from the photometric fluxes of SN 2015ba, resulting in temperature as high as $\sim$20000 K at early epochs, falling to $\sim$6300 K at 50~d and finally settling to $\sim$4800 K at late epochs. 

The early spectra of SN 2015ba exhibit a blue continuum with some broad Balmer P-Cygni profiles. The initial velocities of  H$_\alpha$ and H$_\beta$ are $\sim$9200 km s$^{-1}$ and $\sim$8100  km s$^{-1}$ respectively. The emergence of weak CSM interaction in SN 2015ba has been observed in the late photospheric stages evident from the appearance of HV features blueward of H I lines. These non-evolving, narrow features at similar epochs have been reported in SNe 1999em, 2004dj and 2009bw \citep{2007ApJ...662.1136C,2012MNRAS.422.1122I}.

The analytical and hydrodynamical modelling yield parameters of the progenitor and the explosion. While both analytical and hydrodynamical modelling gives consistent values of ejecta mass M$_{ej}$, $^{56}$Ni mass and pre-explosion radius, the inferred values of explosion energy from the analytical modelling (2.3 foe) is higher than the hydrodynamical modelling (1.6 foe). The mass of the progenitor of SN 2015ba estimated from the modelling ($\sim$ 26 M$_\odot$) significantly overshoots the observational limit of 16.5 $\pm$ 1.5 M$_\odot$ of SNe IIP progenitors, that gave rise to the \textquotedblleft RSG problem\textquotedblright{} \citep{2009MNRAS.395.1409S}. Also, we note the complete absence of [O I] 6300,6364 doublet in the nebular spectra of SN 2015ba, which is commonly found in massive progenitors. SN 2009bw is another SN IIP, coming from a relatively massive progenitor, which exhibited oxygen deprived nebular spectra. However, the non-monotonic variation of O-core masses with the ZAMS mass in massive pre-supernova stars \citep[$\sim$ 9 M$_\odot$,][]{2002RvMP...74.1015W,2014ApJ...783...10S,2016ApJ...818..124E} and uncertainties in the mixing scenario, may give rise to such an explosion.
While hydrodynamical models have been known to produce higher estimates of progenitor mass, the code used in the present work gives lower progenitor masses which are often found to be consistent with masses derived from other prevalent methods, like direct progenitor detection in pre-explosion images or modelling of nucleosynthesis yields in the nebular spectra \citep[e.g.][]{2013MNRAS.434.1636T,2017MNRAS.464.3013P}.

\section*{Acknowledgments}

Based on observations made with: Copernico 1.82 m Telescope operated by INAF Osservatorio Astronomico di Padova at Asiago, Italy. The Gran Telescopio Canarias (GTC) operated on the island of La Palma at the Spanish Observatorio del Roque de los Muchachos of the Instituto de Astrofisica de Canarias. The Nordic Optical Telescope (NOT), operated by the NOT Scientific Association at the Spanish Observatorio del Roque de los Muchachos of the Instituto de Astrofisica de Canarias.
We thank the observing staff and observing assistants of all the telescopes used in this work for their support during observations of SN 2015ba.  We also thank S. Taubenberger and A. Floers for their support. We acknowledge Wiezmann Interactive Supernova data REPository \citep[http://wiserep.weizmann.ac.il (WISeREP),][]{2012PASP..124..668Y}. This research has made use of the NASA/IPAC Extragalactic Database (NED) which is operated by the Jet Propulsion Laboratory, California Institute of Technology, under contract with the National
Aeronautics and Space Administration. We acknowledge the usage of the HyperLeda data base (http://leda.univ-lyon1.fr). This work makes use of observations with Las Cumbres Observatory from the Supernova Key Project. DAH, GH, and CM are supported by NSF grant AST-1313484. Support for IA was provided by NASA through the Einstein Fellowship Program, grant PF6-170148. AP, LT, SB PO and MT are partially supported by the PRIN-INAF 2017 \textquotedblleft Towards the SKA and CTA era: discovery, localisation and physics of transient sources\textquotedblright (PI M. Giroletti). AMG acknowledges financial support by the University of C$\acute{a}$diz grant PR2017-64. SBP and KM acknowledges BRICS grant DST/IMRCD/BRICS/Pilotcall/ProFCheap/2017(G) for the present work.



\begin{figure}
	\begin{center} 
			\includegraphics[scale=0.7]{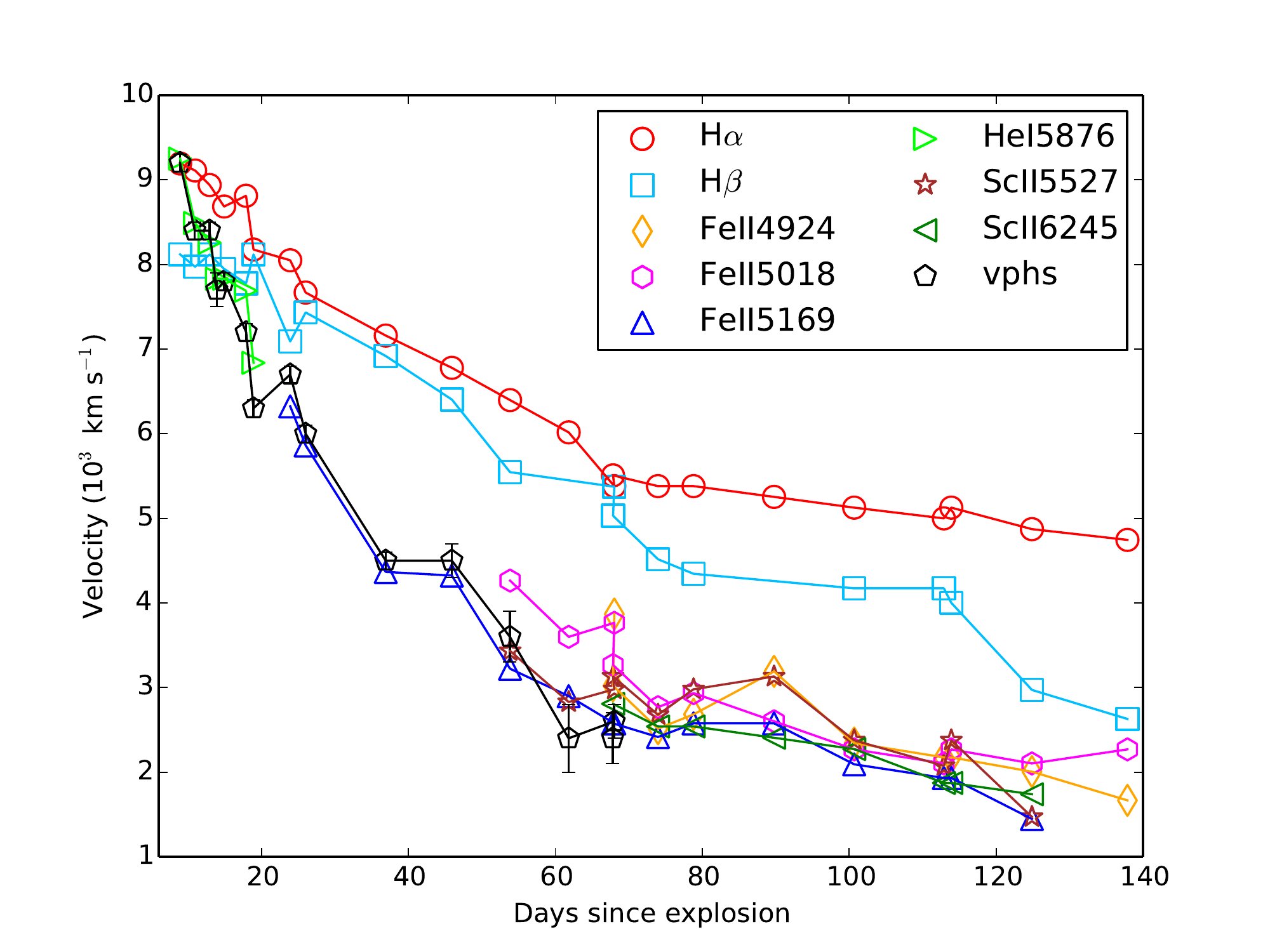} 
	\end{center}
	\caption{Line velocity evolution of H$_\alpha$, H$_\beta$, He I, Fe II and Sc II lines. The velocities are estimated from the absorption minima. The SYN++ modelled photospheric velocities (v$_{phs}$) follow roughly the He I velocities at early phases and Fe II/ Sc II velocities at late phases.}
\end{figure} 

\afterpage{
\begin{table}
\caption{Basic information on SN 2015ba and the host galaxy IC 1029. The host galaxy parameters are taken from NED.}
\centering
\smallskip
\begin{tabular}{c c }
\hline \hline
Host galaxy & IC 1029    \\
Galaxy type & SABb \\  
Redshift & 0.007949 \\ 
Major Diameter & 3.34 arcmin \\
Minor Diameter & 0.61 arcmin \\
Helio. Radial Velocity &  2383$\pm$3 km s$^{-1}$ \\
\hline
R.A.(J2000.0) & 14$^h$32$^m$29.19$^s$ \\
Dec.(J2000.0) & +49$^d$53$^m$34.5$^s$ \\
Distance modulus  & 32.73 $\pm$ 0.04 mag \\
Total Extinction E(B-V) & 0.46 $\pm$ 0.05 mag \\
SN type & II\\
Offset from nucleus & 19$^{''}$.2 E,41$^{''}$.7 S \\
Date of Discovery & 2457355.3 (JD) \\
Estimated date of explosion & 2457349.7 (JD)\\
\hline 
\end{tabular}
\label{tab:sn15ba_ic1029_detail}      
\end{table}

\begin{table}
 \centering
 \begin{minipage}{84mm}
  \caption{The distances to the host galaxy of SN~2015ba determined in this paper. The mean of the values is also shown.}
  \label{ave_dist}
  \begin{tabular}{@{}lccccc@{}}
  \hline
  \hline
Method & $D$ & $\mu$ & References  \\
& (Mpc) & (mag) & \\
  \hline
Virgo-infall  & 36.4 (0.9) & 32.80 (0.05) &  \citet{2014AA...570A..13M}\\
EPM & 36.6 (1.9) & 32.82 (0.11) & this paper\\
SCM & 30.1 (1.4) & 32.39 (0.10) & this paper\\
\hline
Weighted Mean & 34.8 (0.7) & 32.73 (0.04) & \\
\hline				  
\end{tabular}		  
\end{minipage}
\end{table}
\begin{table}
 \centering
 \begin{minipage}{84mm}
  \caption{The best three matches to the spectrum obtained on 2015 December 4 along with the rlap parameter to estimate the age of SN 2015ba.}
  
  \begin{tabular}{@{}lccccc@{}}
  \hline
  \hline
SN & rlap$^\dagger$ & Age since V$_{max}$ (days) & Age since explosion (days)  \\
 \hline
SN 2004et  & 10.2 & -1.9 &  14.2 $\pm$3 \\
SN 2004et & 9.2 & -2.9 & 13.2 $\pm$3 \\
SN 1999em & 9.1 & 0.2 & 10.1 $\pm$ 5\\
\hline
Mean & & &12.5 $\pm$ 7.3 \\
\hline				  
\end{tabular}			  
$^\dagger$Quality of fit
     \label{expl_epoch}
\end{minipage}
\end{table}
\begin{table}
 \centering
 \begin{minipage}{84mm}
  \caption{The equivalent width (EW) of the narrow Na I D absorption feature due to host galaxy from the spectra of SN 2015ba.}
  \label{15ba_EW}
  \begin{tabular}{@{}cccc@{}}
  \hline
  \hline 
 Date & Phase$^a$ & SNR$^b$ & EW  \\  
 (yyyy-mm-dd) &  (days)            &  & (\AA) \\ 
  \hline
 2015-12-01 & 8.9 & 126.3 & 1.8 (0.2)\\
 2015-12-05 & 12.9 & 80.6 & 2.1 (0.4)\\
 2015-12-08 & 14.9 &  86.0 & 1.9 (0.1)\\
 2015-12-30 & 36.9 &  55.8 & 1.9 (0.7)\\
 2016-02-04 & 74.0 &  71.9 & 1.6 (0.7)\\
 \hline 
 Weighted Mean & & & 1.88 (0.06) \\
\hline
\end{tabular}			  
 \begin{tablenotes}
      \item[a]{$^a$since explosion t$_0$ = 2457349.7 JD}
       \item[a]{$^b$ at 0.6 $\mu$m}
     \end{tablenotes}
\end{minipage}
\end{table}

\begin{table}
 \centering
 \begin{minipage}{84mm}
  \caption{The host galaxy reddening derived for SN 2015ba.}
  \label{ebv}
  \begin{tabular}{@{}cccccc@{}}
  \hline
  \hline 
  Method  & Formula & E(B-V)$_{host}$ & Reference\\  
                    &                  &  mag                      &\\ 
  \hline
 Na I D & E(B-V) = 0.25EW             & 0.48$\pm$0.20 & \citet{1990AA...237...79B} \\
 Na I D & E(B-V) = 0.16EW - 0.01 & 0.30$\pm$0.20 & \citet{2003fthp.conf..200T} \\
 Na I D & E(B-V) = 0.43EW - 0.08 & 0.74$\pm$0.30 & \citet{2011MNRAS.415L..81P} \\
 \hline
 Colour curve & $A_V(V-I) = 2.518[(V-I)-0.656]$  & 0.44$\pm$0.06 & \cite{2010ApJ...715..833O} \\
 \hline 
 Weighted Mean & & 0.44 (0.05) & \\
 \hline
\end{tabular}			  
\end{minipage}
\end{table}

\begin{table*}
\caption{Parameters of the SNe IIP sample.}
\scalebox{0.8}{

\begin{tabular}{@{}llllllllllll}\hline \hline
&Parent & Distance$^\dagger$ & $A_V^{tot}$ & M$^{V}$ & $t_{p}$ & E (10$^{51})$ & R & M$_{ej}$ & M$_{ms}$ & $^{56}$Ni & Ref.\\

Supernova& Galaxy & (Mpc) & (mag) & (mag) & (days) & (ergs) & (R$_{\odot}$) & (M$_{\odot}$) & (M$_{\odot}$) &(M$_{\odot}$)\\
\hline
&&&&&&&&&&\\
1987A & LMC & 0.05 & 0.60 & - & 40 & 1.3 & 40 & 15 & 20 & 0.075 & 1\\
&&&&&&&&&&\\
1999em & NGC 1637 & 11.7 (0.1) & 0.31 & $-$16.71 & 95 & 1.2$^{+0.6}_{-0.3}$ & 249$^{+243}_{-150}$ & 27$^{+14}_{-8}$ & 12$\pm$1 & 0.042$^{+0.027}_{-0.019}$ & 2,3,4\\
&&&&&&&&&&\\
1999gi & NGC 3184 & 11.1 & 0.65 & $-$15.68 & 95 & 1.5$^{+0.7}_{-0.5}$ & 81$^{+110}_{-51}$ & 43$^{+24}_{-14}$& 15$^{+5}_{-3}$ & 0.018$^{+0.013}_{-0.009}$ & 5\\
&&&&&&&&&&\\
2004dj & NGC 2403 & 3.5 (0.3) & 0.22 & $-$15.86 & 100$\pm$20 & 0.86$^{+0.89}_{-0.49}$ & 155$^{+150}_{-75}$ & 19$^{+20}_{-10}$ & $>$20 & 0.02$\pm$0.01 & 6\\
&&&&&&&&&&\\
2004et &  NGC 6946 &  5.4 (1.0) &  1.27 &  $-$17.04 &  110$\pm$10 &  0.98$\pm$0.25 & 530$\pm$280 &  16$\pm$5 &  $\sim$ 20 & 0.06$\pm$0.03 & 7,8,9\\
&&&&&&&&&&\\
2005cs & M51 & 7.1 (1.2) & 0.34 & $-$14.83 & 90-120 & 0.3 & 100 & 8-13 & 10-15 & 0.003-0.004 & 10,11\\
&&&&&&&&&&\\
2007od & UGC~12846 & 25.3 (0.8) & 0.12 & $-$17.64 & 25 & 0.5 & 670 & 5-7 & 7-9 & 0.02 & 12\\
&&&&&&&&&&\\
2008in & NGC 4303/M61 & 12.6 (1.0) & 0.3 & $-$15.25 & 98 & 0.54 & 126 & 16.7 & \textgreater 20 & 0.015 & 13\\
&&&&&&&&&&\\
2009bw & UGC~2890 & 20.2 (0.6) & 0.96 & $-$17.24 & 100 & 0.3 & 510-1000 & 8.3-12 & 10-14 & 0.022 & 14\\
&&&&&&&&&&\\
2009ib & NGC 1559 & 19.5 (2.8) & 0.496 & $-$15.77 & 118 & 0.55 & 400 & 15 & 16.5-17 & 0.046 $\pm$ 0.015 & 15\\
&&&&&&&&&&\\
2009N & NGC 4487 & 19.8 (1.1) & 0.403 & $-$15.52 & 110 & 0.48 & 287 & 11.5 & 13-13.5 & 0.020$\pm$0.004 & 16\\ 
&&&&&&&&&&\\
2012aw & NGC 3351 & 9.9 (0.1) & 0.23 & $-$16.67 & $\sim$90 & 1-2 & 337 & 9-14 & 11-16 & 0.06$\pm$0.01& 17\\
&&&&&&&&&&\\
2013ab & NGC 5669 & 23.9 (0.9) & 0.14 & $-$16.69 & 80 & 0.35 & $\sim$600 & $\sim$7 & $\sim$9 & 0.064 & 18\\
&&&&&&&&&&\\
2014cx & NGC 337 & 18.5 (3.7) & 0.3 & $-$16.56 & $\sim$100 & 0.4 & 640 & 8 & 10 & 0.056$\pm$0.008 & 19\\
&&&&&&&&&&\\
ASASSN-14gm & NGC 337 & 19.97 (1.8) & 0.31 & $-$16.86 & $\sim$90 & - & - & - & - & - & 20\\
&&&&&&&&&&\\
DLT16am & NGC 1532 & 20.3 (4.1) & 6.04 (0.46) & $-$17.76 & $\sim$80 & - & - & & - & 0.208$\pm$0.044 & 21\\
&&&&&&&&&&\\
2016X & UGC 08041 & 15.6 (1.3) & 0.05 (0.21) & $-$16.36 & $\sim$90 & - & 930$\pm$70 & - & 18.5-19.7 & 0.034$\pm$0.006 & 22\\
&&&&&&&&&&\\
{\bf 2015ba} & {\bf IC 1029} & {\bf 34.8 (0.7)} & {\bf 1.42 (0.16)} & {\bf $-$17.00} & {\bf $\sim$123} & {\bf 1.6} & {\bf 690} & {\bf 24} & - & {\bf 0.032$\pm$0.006} & {\bf This paper}\\
\hline
\end{tabular}}
\label{parameter_SNIIP_sample}
\flushleft
$^\dagger$ In the H$_0$ = 73.24 km s$^{-1}$ Mpc$^{-1}$ scale.\\
References: (1) \cite{1990AJ.....99.1146H},
(2) \cite{2001ApJ...558..615H},
(3) \cite{2002PASP..114...35L},
(4) \cite{2003ApJ...594..247L},
(5)~\cite{2002AJ....124.2490L},
(6) \cite{2006MNRAS.369.1780V},
(7) \cite{2006MNRAS.372.1315S},
(8) \cite{2007MNRAS.381..280M},
(9) \cite{2010MNRAS.404..981M},
(10)~\cite{2006MNRAS.370.1752P},
(11) \cite{2009MNRAS.394.2266P},
(12) \cite{2011MNRAS.417..261I},
(13) \cite{2011ApJ...736...76R},
(14) \cite{2012MNRAS.422.1122I},
(15) \cite{2015MNRAS.450.3137T},
(16) \cite{2014MNRAS.438..368T},
(17) \cite{2013MNRAS.433.1871B,2014ApJ...787..139D},
(18) \cite{2015MNRAS.450.2373B},
(19) \cite{2016ApJ...832..139H},
(20) \cite{2016MNRAS.459.3939V},
(21) \cite{2018ApJ...853...62T},
(22) \cite{2018MNRAS.475.3959H}.
\end{table*}
\begin{table}
  \caption{Best fit values of t$_{PT}$, a$_0$ and w$_0$ using \citep{2016MNRAS.459.3939V} expression.}
  \label{lc_par_comp}
  \begin{tabular}{@{}lccccc@{}}
  \hline
  \hline
SN &  t$_0$ & t$_0$-t$_{PT}$ & a$_0$ & w$_0$ & References \\
      &  JD (2400000+)&  (days)     & (mag) &          &    \\
  \hline
SN 1999em &  51477.0 (1.0) & 116.8 $\pm$ 1.8 & 1.84 $\pm$ 0.03 & 4.1 $\pm$ 0.6 & 1,2\\
SN 1999gi   & 51518.3 (3.0) & 126.5 $\pm$ 0.6 & 2.2 $\pm$ 0.1 & 5.1 $\pm$ 0.6 & 3  \\
SN 2004et & 53270.5 (0.2) & 122.9 $\pm$ 0.6 & 1.55 $\pm$ 0.03 & 3.6 $\pm$ 0.8  & 4,5,6\\
SN 2005cs & 53549.0 (0.5) & 125.4 $\pm$ 0.2 & 4.07 $\pm$ 0.05 & 2.2 $\pm$ 0.2 & 7\\
SN 2008in & 54825.6 (0.8) & 107.8 $\pm$ 0.3 & 1.93 $\pm$ 0.03 & 2.2 $\pm$ 0.3 & 10\\
SN 2009N &  54848.1 (1.2) & 108.1 $\pm$ 0.2 & 1.87 $\pm$ 0.03 & 2.6 $\pm$ 0.19 & 11\\
SN 2009ib & 55041.8 (3.1) & 139.7 $\pm$ 0.3 & 1.61 $\pm$ 0.04 & 5.4 $\pm$ 0.3 & 12\\
SN 2013ab & 56340.0 (1.0) & 102.1 $\pm$ 0.2 & 1.52 $\pm$ 0.03 & 3.6 $\pm$ 0.2 & 14\\
ASASSN-14gm & 56901.0 (1.5) & 110.8 $\pm$ 0.6 & 1.61 $\pm$ 0.04 & 5.4 $\pm$ 0.5 & 15\\	
SN 2014cx & 56902.4 (0.5) & 109.5 $\pm$ 0.4 & 1.67 $\pm$ 0.04 & 6.0$\pm$0.4 &  16 \\
SN 2016X & 57406.4 (0.6) & 95.0 $\pm$ 0.3 & 1.33 $\pm$ 0.04 & 2.28 $\pm$ 0.33 & 17\\
SN 2015ba & 57349.7 (1.0) & 140.7 $\pm$ 0.2 & 3.01 $\pm$ 0.09 & 2.3 $\pm$ 0.2 & This paper\\ 
\hline 
\end{tabular}	

References: (1) \cite{2001ApJ...558..615H}, (2) \cite{2002PASP..114...35L},
(3) \cite{2002AJ....124.2490L},
(4) \cite{2006MNRAS.372.1315S}
(5) \cite{2007MNRAS.381..280M},
(6) \cite{2010MNRAS.404..981M},
(7) \cite{2006MNRAS.370.1752P},
(8) \cite{2009MNRAS.394.2266P},
(9) \cite{2011MNRAS.417..261I},
(10) \cite{2011ApJ...736...76R},
(11) \cite{2014MNRAS.438..368T},
(12) \cite{2015MNRAS.450.3137T},
(13) \cite{2013ApJ...767..166G},
(14) \cite{2015MNRAS.450.2373B},
(15) \cite{2016MNRAS.459.3939V},
(16) \cite{2016ApJ...832..139H},
(17) \cite{2018MNRAS.475.3959H}.  
\end{table}
\begin{table}
 \centering
 \begin{minipage}{84mm}
  \caption{The derived parameters of SN~2015ba needed to apply EPM: the angular size ($\theta$), photospheric temperature ($T$), dilution factor ($\zeta_\lambda(T)$), and the interpolated photospheric velocity ($v_{ph}$).}
 
  \begin{tabular}{@{}lcllll@{}}
  \hline
  \hline 
  $t^\dagger$ & $\theta$ & $T$ & $\zeta_\lambda(T)$ & $v_{ph}$ \\  
   (days) & ($10^8$ km Mpc$^{-1}$) & (K) & & (km s$^{-1}$) \\ 
  \hline
 4.71 & 1.870 (0.247) & 23019 (2794) & 0.519 (0.007) & 8757 (56) \\
 5.72 & 2.330 (0.192) & 18766 (1350) & 0.509 (0.003) & 8512 (47) \\
 8.72 & 2.149 (0.237) & 19641 (1876) & 0.511 (0.005) & 7841 (38) \\
 11.70 & 2.787 (0.232) & 15115 (1906) & 0.503 (0.001) & 7250 (43) \\
 12.33 & 3.069 (0.037) & 13404 (169) & 0.504 (0.0004) & 7138 (44) \\
 12.72 & 3.208 (0.239) & 12840 (1232) & 0.506 (0.004) & 7067 (45) \\
 20.42 & 3.750 (0.306) & 10367 (703) & 0.527 (0.01) & 5909 (52) \\
 21.71 & 3.204 (0.256) & 11671 (786) & 0.512 (0.006) & 5746 (53) \\
 31.72 & 4.181 (0.368) &  8550 (434) & 0.572 (0.02) & 4750 (82) \\
 39.72 & 4.248 (0.258) &  7648 (252) & 0.616 (0.02) & 4127 (108)\\
 45.20 & 4.346 (0.063)& 7238 (508) & 0.645 (0.004) & 3701 (132)\\
 \hline
\end{tabular}	
 \label{epm}		  
 \begin{tablenotes}
       \item[a]{$^\dagger$ since discovery, $t_0$=2457355.3 JD}
     \end{tablenotes}
\end{minipage}
\end{table}

\begin{table}
 \centering
 \begin{minipage}{84mm}
  \caption{The line velocities of Fe II $\lambda$5169 and He I $\lambda$5876 estimated with SYN++ modelling of the observed spectra of SN 2015ba.}
  \label{syn_vel}
  \begin{tabular}{@{}cccccc@{}}
  \hline
  \hline 
 Date & Phase$^\dagger$ & v( He I ) & v( Fe II ) \\  
  (yyyy-mm-dd) & (days) & 10$^3$ km s$^{-1}$ & 10$^3$ km s$^{-1}$ \\ 
  \hline
 2015-12-02 & 8.9 & 9.2 $\pm$ 0.1 & -- \\
 2015-12-04 & 10.9 & 8.4 $\pm$ 0.1 & --\\
 2015-12-06 & 12.9 & 8.4 $\pm$ 0.1 & --\\
 2015-12-07 & 13.9 & 7.7 $\pm$ 0.2 & --\\
 2015-12-08 & 14.9 & 7.8 $\pm$ 0.1 & --\\
 2015-12-11 & 17.9 & 7.2 $\pm$ 0.1 & --\\
 2015-12-12 & 18.9 & 6.3 $\pm $ 0.1 & --\\
 2015-12-17 & 23.9 & -- & 6.7 $\pm$ 0.1 \\
 2015-12-19 & 26.0 & -- & 6.0 $\pm$ 0.1 \\
 2015-12-30 & 36.9 & -- & 4.5 $\pm$ 0.1\\
 2016-01-08 & 45.9 & -- & 4.5 $\pm$ 0.2\\
 2016-01-16 & 53.8 & -- & 3.6 $\pm$ 0.3\\
 2016-01-24 & 61.8 & -- & 2.4 $\pm$ 0.4\\
 2016-01-29 & 68.0 & -- & 2.6 $\pm$ 0.2 \\
\hline
\end{tabular}			  
 \begin{tablenotes}
       \item[a]{$^\dagger$ relative to the date of explosion, $t_0$=2457349.7 JD}
     \end{tablenotes}
\end{minipage}
\end{table}

\begin{table}
 \centering
 \begin{minipage}{84mm}
  \caption{The best fit core and shell parameters for true bolometric light curve of SN 2015ba using \citet{2016A&A...589A..53N}.}
  
  \begin{tabular}{@{}lccc@{}}
  \hline
  \hline
Parameter & Core & Shell & Remarks  \\
 \hline
R$_0$ (cm)  & 4.8 $\times$ 10$^{13}$ & 2.6 $\times$ 10$^{13}$ & Initial radius of ejecta\\
T$_{rec} (K)$ & 9500 & -  & Recombination temperature\\
M$_{ej}$ (M$_\odot$) & 22 & 0.4 & Ejecta mass\\
E$_{th} (foe)$ & 0.8 & 0.09 & Initial thermal energy \\
E$_{kin} (foe)$ & 1.4 & 0.05 & Initial kinetic energy \\
M$_{Ni}$ (M$_\odot$) & 0.032 & - & Initial $^{56}$Ni mass\\
$\kappa$ (cm$^2$ g$^{-1}$) & 0.14 & 0.4 &Opacity\\
A$_g$ (day$^2$) & 3e4 & 1e10 & Gamma-ray leakage \\
\hline
     \label{Nagy}
     \end{tabular}
\end{minipage}
\end{table}
}
\bibliographystyle{mnras}
\bibliography{refag}

\begin{thebibliography}{}
\makeatletter
\relax
\def\mn@urlcharsother{\let\do\@makeother \do\$\do\&\do\#\do\^\do\_\do\%\do\~}
\def\mn@doi{\begingroup\mn@urlcharsother \@ifnextchar [ {\mn@doi@}
  {\mn@doi@[]}}
\def\mn@doi@[#1]#2{\def\@tempa{#1}\ifx\@tempa\@empty \href
  {http://dx.doi.org/#2} {doi:#2}\else \href {http://dx.doi.org/#2} {#1}\fi
  \endgroup}
\def\mn@eprint#1#2{\mn@eprint@#1:#2::\@nil}
\def\mn@eprint@arXiv#1{\href {http://arxiv.org/abs/#1} {{\tt arXiv:#1}}}
\def\mn@eprint@dblp#1{\href {http://dblp.uni-trier.de/rec/bibtex/#1.xml}
  {dblp:#1}}
\def\mn@eprint@#1:#2:#3:#4\@nil{\def\@tempa {#1}\def\@tempb {#2}\def\@tempc
  {#3}\ifx \@tempc \@empty \let \@tempc \@tempb \let \@tempb \@tempa \fi \ifx
  \@tempb \@empty \def\@tempb {arXiv}\fi \@ifundefined
  {mn@eprint@\@tempb}{\@tempb:\@tempc}{\expandafter \expandafter \csname
  mn@eprint@\@tempb\endcsname \expandafter{\@tempc}}}

\bibitem[\protect\citeauthoryear{{Anderson} et~al.,}{{Anderson}
  et~al.}{2014a}]{2014MNRAS.441..671A}
{Anderson} J.~P.,  et~al., 2014a, \mn@doi [\mnras] {10.1093/mnras/stu610},
  \href {http://adsabs.harvard.edu/abs/2014MNRAS.441..671A} {441, 671}

\bibitem[\protect\citeauthoryear{{Anderson} et~al.,}{{Anderson}
  et~al.}{2014b}]{2014ApJ...786...67A}
{Anderson} J.~P.,  et~al., 2014b, \mn@doi [\apj] {10.1088/0004-637X/786/1/67},
  \href {http://adsabs.harvard.edu/abs/2014ApJ...786...67A} {786, 67}

\bibitem[\protect\citeauthoryear{{Anderson} et~al.,}{{Anderson}
  et~al.}{2018}]{2018NatAs.tmp...53A}
{Anderson} J.~P.,  et~al., 2018, \mn@doi [Nature Astronomy]
  {10.1038/s41550-018-0458-4}, \href
  {http://adsabs.harvard.edu/abs/2018NatAs.tmp...53A} {}

\bibitem[\protect\citeauthoryear{{Arcavi} et~al.,}{{Arcavi}
  et~al.}{2012}]{2012ApJ...756L..30A}
{Arcavi} I.,  et~al., 2012, \mn@doi [\apjl] {10.1088/2041-8205/756/2/L30},
  \href {http://adsabs.harvard.edu/abs/2012ApJ...756L..30A} {756, L30}

\bibitem[\protect\citeauthoryear{{Arnett}}{{Arnett}}{1980}]{1980ApJ...237..541A}
{Arnett} W.~D.,  1980, \mn@doi [\apj] {10.1086/157898}, \href
  {http://adsabs.harvard.edu/abs/1980ApJ...237..541A} {237, 541}

\bibitem[\protect\citeauthoryear{{Arnett} \& {Fu}}{{Arnett} \&
  {Fu}}{1989}]{1989ApJ...340..396A}
{Arnett} W.~D.,  {Fu} A.,  1989, \mn@doi [\apj] {10.1086/167402}, \href
  {http://adsabs.harvard.edu/abs/1989ApJ...340..396A} {340, 396}

\bibitem[\protect\citeauthoryear{{Barbarino} et~al.,}{{Barbarino}
  et~al.}{2015}]{2015MNRAS.448.2312B}
{Barbarino} C.,  et~al., 2015, \mn@doi [\mnras] {10.1093/mnras/stv106}, \href
  {http://adsabs.harvard.edu/abs/2015MNRAS.448.2312B} {448, 2312}

\bibitem[\protect\citeauthoryear{{Barbon}, {Benetti}, {Rosino}, {Cappellaro}
  \& {Turatto}}{{Barbon} et~al.}{1990}]{1990AA...237...79B}
{Barbon} R.,  {Benetti} S.,  {Rosino} L.,  {Cappellaro} E.,   {Turatto} M.,
  1990, \aap, \href {http://adsabs.harvard.edu/abs/1990A%26A...237...79B} {237,
  79}

\bibitem[\protect\citeauthoryear{{Baron} et~al.,}{{Baron}
  et~al.}{2000}]{2000ApJ...545..444B}
{Baron} E.,  et~al., 2000, \mn@doi [\apj] {10.1086/317795}, \href
  {http://adsabs.harvard.edu/abs/2000ApJ...545..444B} {545, 444}

\bibitem[\protect\citeauthoryear{{Benetti} et~al.,}{{Benetti}
  et~al.}{2015}]{2015ATel.8353....1B}
{Benetti} S.,  et~al., 2015, The Astronomer's Telegram, \href
  {http://adsabs.harvard.edu/abs/2015ATel.8353....1B} {8353}

\bibitem[\protect\citeauthoryear{{Bersten} \& {Hamuy}}{{Bersten} \&
  {Hamuy}}{2009}]{2009ApJ...701..200B}
{Bersten} M.~C.,  {Hamuy} M.,  2009, \mn@doi [\apj]
  {10.1088/0004-637X/701/1/200}, \href
  {http://adsabs.harvard.edu/abs/2009ApJ...701..200B} {701, 200}

\bibitem[\protect\citeauthoryear{{Bersten}, {Benvenuto}  \& {Hamuy}}{{Bersten}
  et~al.}{2011}]{2011ApJ...729...61B}
{Bersten} M.~C.,  {Benvenuto} O.,   {Hamuy} M.,  2011, \mn@doi [\apj]
  {10.1088/0004-637X/729/1/61}, \href
  {http://adsabs.harvard.edu/abs/2011ApJ...729...61B} {729, 61}

\bibitem[\protect\citeauthoryear{{Bessell}, {Castelli}  \& {Plez}}{{Bessell}
  et~al.}{1998}]{1998A&A...333..231B}
{Bessell} M.~S.,  {Castelli} F.,   {Plez} B.,  1998, \aap, \href
  {http://adsabs.harvard.edu/abs/1998A%26A...333..231B} {333, 231}

\bibitem[\protect\citeauthoryear{{Blinnikov} \& {Bartunov}}{{Blinnikov} \&
  {Bartunov}}{1993}]{1993A&A...273..106B}
{Blinnikov} S.~I.,  {Bartunov} O.~S.,  1993, \aap, \href
  {http://adsabs.harvard.edu/abs/1993A%26A...273..106B} {273, 106}

\bibitem[\protect\citeauthoryear{{Blondin} \& {Tonry}}{{Blondin} \&
  {Tonry}}{2007}]{2007ApJ...666.1024B}
{Blondin} S.,  {Tonry} J.~L.,  2007, \mn@doi [\apj] {10.1086/520494}, \href
  {http://adsabs.harvard.edu/abs/2007ApJ...666.1024B} {666, 1024}

\bibitem[\protect\citeauthoryear{{Bose} et~al.,}{{Bose}
  et~al.}{2013}]{2013MNRAS.433.1871B}
{Bose} S.,  et~al., 2013, \mn@doi [\mnras] {10.1093/mnras/stt864}, \href
  {http://adsabs.harvard.edu/abs/2013MNRAS.433.1871B} {433, 1871}

\bibitem[\protect\citeauthoryear{{Bose} et~al.,}{{Bose}
  et~al.}{2015a}]{2015MNRAS.450.2373B}
{Bose} S.,  et~al., 2015a, \mn@doi [\mnras] {10.1093/mnras/stv759}, \href
  {http://adsabs.harvard.edu/abs/2015MNRAS.450.2373B} {450, 2373}

\bibitem[\protect\citeauthoryear{{Bose} et~al.,}{{Bose}
  et~al.}{2015b}]{2015ApJ...806..160B}
{Bose} S.,  et~al., 2015b, \mn@doi [\apj] {10.1088/0004-637X/806/2/160}, \href
  {http://adsabs.harvard.edu/abs/2015ApJ...806..160B} {806, 160}

\bibitem[\protect\citeauthoryear{{Cardelli}, {Clayton}  \& {Mathis}}{{Cardelli}
  et~al.}{1989}]{1989ApJ...345..245C}
{Cardelli} J.~A.,  {Clayton} G.~C.,   {Mathis} J.~S.,  1989, \mn@doi [\apj]
  {10.1086/167900}, \href {http://adsabs.harvard.edu/abs/1989ApJ...345..245C}
  {345, 245}

\bibitem[\protect\citeauthoryear{{Chevalier}}{{Chevalier}}{1976}]{1976ApJ...207..872C}
{Chevalier} R.~A.,  1976, \mn@doi [\apj] {10.1086/154557}, \href
  {http://adsabs.harvard.edu/abs/1976ApJ...207..872C} {207, 872}

\bibitem[\protect\citeauthoryear{{Chugai}}{{Chugai}}{1988}]{1988SvAL...14..334C}
{Chugai} N.~N.,  1988, Soviet Astronomy Letters, \href
  {http://adsabs.harvard.edu/abs/1988SvAL...14..334C} {14, 334}

\bibitem[\protect\citeauthoryear{{Chugai}, {Chevalier}  \& {Utrobin}}{{Chugai}
  et~al.}{2007}]{2007ApJ...662.1136C}
{Chugai} N.~N.,  {Chevalier} R.~A.,   {Utrobin} V.~P.,  2007, \mn@doi [\apj]
  {10.1086/518160}, \href {http://adsabs.harvard.edu/abs/2007ApJ...662.1136C}
  {662, 1136}

\bibitem[\protect\citeauthoryear{{Dall'Ora} et~al.,}{{Dall'Ora}
  et~al.}{2014}]{2014ApJ...787..139D}
{Dall'Ora} M.,  et~al., 2014, \mn@doi [\apj] {10.1088/0004-637X/787/2/139},
  \href {http://adsabs.harvard.edu/abs/2014ApJ...787..139D} {787, 139}

\bibitem[\protect\citeauthoryear{{Dessart} \& {Hillier}}{{Dessart} \&
  {Hillier}}{2005a}]{2005A&A...437..667D}
{Dessart} L.,  {Hillier} D.~J.,  2005a, \mn@doi [\aap]
  {10.1051/0004-6361:20042525}, \href
  {http://adsabs.harvard.edu/abs/2005A%26A...437..667D} {437, 667}

\bibitem[\protect\citeauthoryear{{Dessart} \& {Hillier}}{{Dessart} \&
  {Hillier}}{2005b}]{2005A&A...439..671D}
{Dessart} L.,  {Hillier} D.~J.,  2005b, \mn@doi [\aap]
  {10.1051/0004-6361:20053217}, \href
  {http://adsabs.harvard.edu/abs/2005A%26A...439..671D} {439, 671}

\bibitem[\protect\citeauthoryear{{Dessart} \& {Hillier}}{{Dessart} \&
  {Hillier}}{2006}]{2006A&A...447..691D}
{Dessart} L.,  {Hillier} D.~J.,  2006, \mn@doi [\aap]
  {10.1051/0004-6361:20054044}, \href
  {http://adsabs.harvard.edu/abs/2006A%26A...447..691D} {447, 691}

\bibitem[\protect\citeauthoryear{{Dessart}, {Livne}  \& {Waldman}}{{Dessart}
  et~al.}{2010}]{2010MNRAS.405.2113D}
{Dessart} L.,  {Livne} E.,   {Waldman} R.,  2010, \mn@doi [\mnras]
  {10.1111/j.1365-2966.2010.16626.x}, \href
  {http://adsabs.harvard.edu/abs/2010MNRAS.405.2113D} {405, 2113}

\bibitem[\protect\citeauthoryear{{Eastman}, {Schmidt}  \& {Kirshner}}{{Eastman}
  et~al.}{1996}]{1996ApJ...466..911E}
{Eastman} R.~G.,  {Schmidt} B.~P.,   {Kirshner} R.,  1996, \mn@doi [\apj]
  {10.1086/177563}, \href {http://adsabs.harvard.edu/abs/1996ApJ...466..911E}
  {466, 911}

\bibitem[\protect\citeauthoryear{{Eldridge} \& {Tout}}{{Eldridge} \&
  {Tout}}{2004}]{2004MNRAS.353...87E}
{Eldridge} J.~J.,  {Tout} C.~A.,  2004, \mn@doi [\mnras]
  {10.1111/j.1365-2966.2004.08041.x}, \href
  {http://adsabs.harvard.edu/abs/2004MNRAS.353...87E} {353, 87}

\bibitem[\protect\citeauthoryear{{Ertl}, {Janka}, {Woosley}, {Sukhbold}  \&
  {Ugliano}}{{Ertl} et~al.}{2016}]{2016ApJ...818..124E}
{Ertl} T.,  {Janka} H.-T.,  {Woosley} S.~E.,  {Sukhbold} T.,   {Ugliano} M.,
  2016, \mn@doi [\apj] {10.3847/0004-637X/818/2/124}, \href
  {http://adsabs.harvard.edu/abs/2016ApJ...818..124E} {818, 124}

\bibitem[\protect\citeauthoryear{{Falk} \& {Arnett}}{{Falk} \&
  {Arnett}}{1977}]{1977ApJS...33..515F}
{Falk} S.~W.,  {Arnett} W.~D.,  1977, \mn@doi [\apjs] {10.1086/190440}, \href
  {http://adsabs.harvard.edu/abs/1977ApJS...33..515F} {33, 515}

\bibitem[\protect\citeauthoryear{{Faran} et~al.,}{{Faran}
  et~al.}{2014a}]{2014MNRAS.442..844F}
{Faran} T.,  et~al., 2014a, \mn@doi [\mnras] {10.1093/mnras/stu955}, \href
  {http://adsabs.harvard.edu/abs/2014MNRAS.442..844F} {442, 844}

\bibitem[\protect\citeauthoryear{{Faran} et~al.,}{{Faran}
  et~al.}{2014b}]{2014MNRAS.445..554F}
{Faran} T.,  et~al., 2014b, \mn@doi [\mnras] {10.1093/mnras/stu1760}, \href
  {http://adsabs.harvard.edu/abs/2014MNRAS.445..554F} {445, 554}

\bibitem[\protect\citeauthoryear{{Fraser}}{{Fraser}}{2016}]{2016MNRAS.456L..16F}
{Fraser} M.,  2016, \mn@doi [\mnras] {10.1093/mnrasl/slv168}, \href
  {http://adsabs.harvard.edu/abs/2016MNRAS.456L..16F} {456, L16}

\bibitem[\protect\citeauthoryear{{Fukugita}, {Ichikawa}, {Gunn}, {Doi},
  {Shimasaku}  \& {Schneider}}{{Fukugita} et~al.}{1996}]{1996AJ....111.1748F}
{Fukugita} M.,  {Ichikawa} T.,  {Gunn} J.~E.,  {Doi} M.,  {Shimasaku} K.,
  {Schneider} D.~P.,  1996, \mn@doi [\aj] {10.1086/117915}, \href
  {http://adsabs.harvard.edu/abs/1996AJ....111.1748F} {111, 1748}

\bibitem[\protect\citeauthoryear{{Gandhi} et~al.,}{{Gandhi}
  et~al.}{2013}]{2013ApJ...767..166G}
{Gandhi} P.,  et~al., 2013, \mn@doi [\apj] {10.1088/0004-637X/767/2/166}, \href
  {http://adsabs.harvard.edu/abs/2013ApJ...767..166G} {767, 166}

\bibitem[\protect\citeauthoryear{{Grassberg}, {Imshennik}  \&
  {Nadyozhin}}{{Grassberg} et~al.}{1971}]{1971Ap&SS..10...28G}
{Grassberg} E.~K.,  {Imshennik} V.~S.,   {Nadyozhin} D.~K.,  1971, \mn@doi
  [\apss] {10.1007/BF00654604}, \href
  {http://adsabs.harvard.edu/abs/1971Ap%26SS..10...28G} {10, 28}

\bibitem[\protect\citeauthoryear{{Graur}, {Bianco}, {Modjaz}, {Shivvers},
  {Filippenko}, {Li}  \& {Smith}}{{Graur} et~al.}{2017}]{2017ApJ...837..121G}
{Graur} O.,  {Bianco} F.~B.,  {Modjaz} M.,  {Shivvers} I.,  {Filippenko} A.~V.,
   {Li} W.,   {Smith} N.,  2017, \mn@doi [\apj] {10.3847/1538-4357/aa5eb7},
  \href {http://adsabs.harvard.edu/abs/2017ApJ...837..121G} {837, 121}

\bibitem[\protect\citeauthoryear{{Guti{\'e}rrez} et~al.,}{{Guti{\'e}rrez}
  et~al.}{2017}]{2017ApJ...850...89G}
{Guti{\'e}rrez} C.~P.,  et~al., 2017, \mn@doi [\apj]
  {10.3847/1538-4357/aa8f52}, \href
  {http://adsabs.harvard.edu/abs/2017ApJ...850...89G} {850, 89}

\bibitem[\protect\citeauthoryear{{Hamuy}}{{Hamuy}}{2003}]{2003ApJ...582..905H}
{Hamuy} M.,  2003, \mn@doi [\apj] {10.1086/344689}, \href
  {http://adsabs.harvard.edu/abs/2003ApJ...582..905H} {582, 905}

\bibitem[\protect\citeauthoryear{{Hamuy} \& {Pinto}}{{Hamuy} \&
  {Pinto}}{2002}]{2002ApJ...566L..63H}
{Hamuy} M.,  {Pinto} P.~A.,  2002, \mn@doi [\apjl] {10.1086/339676}, \href
  {http://adsabs.harvard.edu/abs/2002ApJ...566L..63H} {566, L63}

\bibitem[\protect\citeauthoryear{{Hamuy} \& {Suntzeff}}{{Hamuy} \&
  {Suntzeff}}{1990}]{1990AJ.....99.1146H}
{Hamuy} M.,  {Suntzeff} N.~B.,  1990, \mn@doi [\aj] {10.1086/115403}, \href
  {http://adsabs.harvard.edu/abs/1990AJ.....99.1146H} {99, 1146}

\bibitem[\protect\citeauthoryear{{Hamuy} et~al.,}{{Hamuy}
  et~al.}{2001}]{2001ApJ...558..615H}
{Hamuy} M.,  et~al., 2001, \mn@doi [\apj] {10.1086/322450}, \href
  {http://adsabs.harvard.edu/abs/2001ApJ...558..615H} {558, 615}

\bibitem[\protect\citeauthoryear{{Heger}, {Fryer}, {Woosley}, {Langer}  \&
  {Hartmann}}{{Heger} et~al.}{2003}]{2003ApJ...591..288H}
{Heger} A.,  {Fryer} C.~L.,  {Woosley} S.~E.,  {Langer} N.,   {Hartmann} D.~H.,
   2003, \mn@doi [\apj] {10.1086/375341}, \href
  {http://adsabs.harvard.edu/abs/2003ApJ...591..288H} {591, 288}

\bibitem[\protect\citeauthoryear{{Hendry} et~al.,}{{Hendry}
  et~al.}{2005}]{2005MNRAS.359..906H}
{Hendry} M.~A.,  et~al., 2005, \mn@doi [\mnras]
  {10.1111/j.1365-2966.2005.08928.x}, \href
  {http://adsabs.harvard.edu/abs/2005MNRAS.359..906H} {359, 906}

\bibitem[\protect\citeauthoryear{{Huang} et~al.,}{{Huang}
  et~al.}{2016}]{2016ApJ...832..139H}
{Huang} F.,  et~al., 2016, \mn@doi [\apj] {10.3847/0004-637X/832/2/139}, \href
  {http://adsabs.harvard.edu/abs/2016ApJ...832..139H} {832, 139}

\bibitem[\protect\citeauthoryear{{Huang} et~al.,}{{Huang}
  et~al.}{2018}]{2018MNRAS.475.3959H}
{Huang} F.,  et~al., 2018, \mn@doi [\mnras] {10.1093/mnras/sty066}, \href
  {http://adsabs.harvard.edu/abs/2018MNRAS.475.3959H} {475, 3959}

\bibitem[\protect\citeauthoryear{{Inserra} et~al.,}{{Inserra}
  et~al.}{2011}]{2011MNRAS.417..261I}
{Inserra} C.,  et~al., 2011, \mn@doi [\mnras]
  {10.1111/j.1365-2966.2011.19128.x}, \href
  {http://adsabs.harvard.edu/abs/2011MNRAS.417..261I} {417, 261}

\bibitem[\protect\citeauthoryear{{Inserra} et~al.,}{{Inserra}
  et~al.}{2012}]{2012MNRAS.422.1122I}
{Inserra} C.,  et~al., 2012, \mn@doi [\mnras]
  {10.1111/j.1365-2966.2012.20685.x}, \href
  {http://adsabs.harvard.edu/abs/2012MNRAS.422.1122I} {422, 1122}

\bibitem[\protect\citeauthoryear{{Jeffery} \& {Branch}}{{Jeffery} \&
  {Branch}}{1990}]{1990sjws.conf..149J}
{Jeffery} D.~J.,  {Branch} D.,  1990, in {Wheeler} J.~C.,  {Piran} T.,
  {Weinberg} S.,  eds, Supernovae, Jerusalem Winter School for Theoretical
  Physics. p.~149

\bibitem[\protect\citeauthoryear{{Jerkstrand}, {Fransson}, {Maguire}, {Smartt},
  {Ergon}  \& {Spyromilio}}{{Jerkstrand} et~al.}{2012}]{2012A&A...546A..28J}
{Jerkstrand} A.,  {Fransson} C.,  {Maguire} K.,  {Smartt} S.,  {Ergon} M.,
  {Spyromilio} J.,  2012, \mn@doi [\aap] {10.1051/0004-6361/201219528}, \href
  {http://adsabs.harvard.edu/abs/2012A%26A...546A..28J} {546, A28}

\bibitem[\protect\citeauthoryear{{Jerkstrand}, {Smartt}, {Fraser}, {Fransson},
  {Sollerman}, {Taddia}  \& {Kotak}}{{Jerkstrand}
  et~al.}{2014}]{2014MNRAS.439.3694J}
{Jerkstrand} A.,  {Smartt} S.~J.,  {Fraser} M.,  {Fransson} C.,  {Sollerman}
  J.,  {Taddia} F.,   {Kotak} R.,  2014, \mn@doi [\mnras]
  {10.1093/mnras/stu221}, \href
  {http://adsabs.harvard.edu/abs/2014MNRAS.439.3694J} {439, 3694}

\bibitem[\protect\citeauthoryear{{Jerkstrand} et~al.,}{{Jerkstrand}
  et~al.}{2015}]{2015MNRAS.448.2482J}
{Jerkstrand} A.,  et~al., 2015, \mn@doi [\mnras] {10.1093/mnras/stv087}, \href
  {http://adsabs.harvard.edu/abs/2015MNRAS.448.2482J} {448, 2482}

\bibitem[\protect\citeauthoryear{{Jordi}, {Grebel}  \& {Ammon}}{{Jordi}
  et~al.}{2006}]{2006A&A...460..339J}
{Jordi} K.,  {Grebel} E.~K.,   {Ammon} K.,  2006, \mn@doi [\aap]
  {10.1051/0004-6361:20066082}, \href
  {http://adsabs.harvard.edu/abs/2006A%26A...460..339J} {460, 339}

\bibitem[\protect\citeauthoryear{{Kirshner} \& {Kwan}}{{Kirshner} \&
  {Kwan}}{1974}]{1974ApJ...193...27K}
{Kirshner} R.~P.,  {Kwan} J.,  1974, \mn@doi [\apj] {10.1086/153123}, \href
  {http://adsabs.harvard.edu/abs/1974ApJ...193...27K} {193, 27}

\bibitem[\protect\citeauthoryear{{Kochanek}, {Khan}  \& {Dai}}{{Kochanek}
  et~al.}{2012}]{2012ApJ...759...20K}
{Kochanek} C.~S.,  {Khan} R.,   {Dai} X.,  2012, \mn@doi [\apj]
  {10.1088/0004-637X/759/1/20}, \href
  {http://adsabs.harvard.edu/abs/2012ApJ...759...20K} {759, 20}

\bibitem[\protect\citeauthoryear{{Landolt}}{{Landolt}}{2009}]{2009AJ....137.4186L}
{Landolt} A.~U.,  2009, \mn@doi [\aj] {10.1088/0004-6256/137/5/4186}, \href
  {http://adsabs.harvard.edu/abs/2009AJ....137.4186L} {137, 4186}

\bibitem[\protect\citeauthoryear{{Leonard} et~al.,}{{Leonard}
  et~al.}{2002a}]{2002PASP..114...35L}
{Leonard} D.~C.,  et~al., 2002a, \mn@doi [\pasp] {10.1086/324785}, \href
  {http://adsabs.harvard.edu/abs/2002PASP..114...35L} {114, 35}

\bibitem[\protect\citeauthoryear{{Leonard} et~al.,}{{Leonard}
  et~al.}{2002b}]{2002AJ....124.2490L}
{Leonard} D.~C.,  et~al., 2002b, \mn@doi [\aj] {10.1086/343771}, \href
  {http://adsabs.harvard.edu/abs/2002AJ....124.2490L} {124, 2490}

\bibitem[\protect\citeauthoryear{{Leonard}, {Kanbur}, {Ngeow}  \&
  {Tanvir}}{{Leonard} et~al.}{2003}]{2003ApJ...594..247L}
{Leonard} D.~C.,  {Kanbur} S.~M.,  {Ngeow} C.~C.,   {Tanvir} N.~R.,  2003,
  \mn@doi [\apj] {10.1086/376831}, \href
  {http://adsabs.harvard.edu/abs/2003ApJ...594..247L} {594, 247}

\bibitem[\protect\citeauthoryear{{Li} et~al.,}{{Li}
  et~al.}{2011}]{2011MNRAS.412.1441L}
{Li} W.,  et~al., 2011, \mn@doi [\mnras] {10.1111/j.1365-2966.2011.18160.x},
  \href {http://adsabs.harvard.edu/abs/2011MNRAS.412.1441L} {412, 1441}

\bibitem[\protect\citeauthoryear{{Lusk} \& {Baron}}{{Lusk} \&
  {Baron}}{2017}]{2017PASP..129d4202L}
{Lusk} J.~A.,  {Baron} E.,  2017, \mn@doi [\pasp] {10.1088/1538-3873/aa5e49},
  \href {http://adsabs.harvard.edu/abs/2017PASP..129d4202L} {129, 044202}

\bibitem[\protect\citeauthoryear{{Lyman}, {Bersier}  \& {James}}{{Lyman}
  et~al.}{2014}]{2014MNRAS.437.3848L}
{Lyman} J.~D.,  {Bersier} D.,   {James} P.~A.,  2014, \mn@doi [\mnras]
  {10.1093/mnras/stt2187}, \href
  {http://adsabs.harvard.edu/abs/2014MNRAS.437.3848L} {437, 3848}

\bibitem[\protect\citeauthoryear{{Maguire} et~al.,}{{Maguire}
  et~al.}{2010}]{2010MNRAS.404..981M}
{Maguire} K.,  et~al., 2010, \mn@doi [\mnras]
  {10.1111/j.1365-2966.2010.16332.x}, \href
  {http://adsabs.harvard.edu/abs/2010MNRAS.404..981M} {404, 981}

\bibitem[\protect\citeauthoryear{{Maguire} et~al.,}{{Maguire}
  et~al.}{2012}]{2012MNRAS.420.3451M}
{Maguire} K.,  et~al., 2012, \mn@doi [\mnras]
  {10.1111/j.1365-2966.2011.20276.x}, \href
  {http://adsabs.harvard.edu/abs/2012MNRAS.420.3451M} {420, 3451}

\bibitem[\protect\citeauthoryear{{Makarov}, {Prugniel}, {Terekhova}, {Courtois}
   \& {Vauglin}}{{Makarov} et~al.}{2014}]{2014AA...570A..13M}
{Makarov} D.,  {Prugniel} P.,  {Terekhova} N.,  {Courtois} H.,   {Vauglin} I.,
  2014, \mn@doi [\aap] {10.1051/0004-6361/201423496}, \href
  {http://adsabs.harvard.edu/abs/2014A%26A...570A..13M} {570, A13}

\bibitem[\protect\citeauthoryear{{Misra}, {Pooley}, {Chandra}, {Bhattacharya},
  {Ray}, {Sagar}  \& {Lewin}}{{Misra} et~al.}{2007}]{2007MNRAS.381..280M}
{Misra} K.,  {Pooley} D.,  {Chandra} P.,  {Bhattacharya} D.,  {Ray} A.~K.,
  {Sagar} R.,   {Lewin} W.~H.~G.,  2007, \mn@doi [\mnras]
  {10.1111/j.1365-2966.2007.12258.x}, \href
  {http://adsabs.harvard.edu/abs/2007MNRAS.381..280M} {381, 280}

\bibitem[\protect\citeauthoryear{{Munari} \& {Zwitter}}{{Munari} \&
  {Zwitter}}{1997}]{1997A&A...318..269M}
{Munari} U.,  {Zwitter} T.,  1997, \aap, \href
  {http://adsabs.harvard.edu/abs/1997A%26A...318..269M} {318, 269}

\bibitem[\protect\citeauthoryear{{Nadyozhin}}{{Nadyozhin}}{2003}]{2003MNRAS.346...97N}
{Nadyozhin} D.~K.,  2003, \mn@doi [\mnras] {10.1046/j.1365-2966.2003.07070.x},
  \href {http://adsabs.harvard.edu/abs/2003MNRAS.346...97N} {346, 97}

\bibitem[\protect\citeauthoryear{{Nagy} \& {Vink{\'o}}}{{Nagy} \&
  {Vink{\'o}}}{2016}]{2016A&A...589A..53N}
{Nagy} A.~P.,  {Vink{\'o}} J.,  2016, \mn@doi [\aap]
  {10.1051/0004-6361/201527931}, \href
  {http://adsabs.harvard.edu/abs/2016A%26A...589A..53N} {589, A53}

\bibitem[\protect\citeauthoryear{{Nagy}, {Ordasi}, {Vink{\'o}}  \&
  {Wheeler}}{{Nagy} et~al.}{2014}]{2014A&A...571A..77N}
{Nagy} A.~P.,  {Ordasi} A.,  {Vink{\'o}} J.,   {Wheeler} J.~C.,  2014, \mn@doi
  [\aap] {10.1051/0004-6361/201424237}, \href
  {http://adsabs.harvard.edu/abs/2014A%26A...571A..77N} {571, A77}

\bibitem[\protect\citeauthoryear{{Nugent} et~al.,}{{Nugent}
  et~al.}{2006}]{2006ApJ...645..841N}
{Nugent} P.,  et~al., 2006, \mn@doi [\apj] {10.1086/504413}, \href
  {http://adsabs.harvard.edu/abs/2006ApJ...645..841N} {645, 841}

\bibitem[\protect\citeauthoryear{{Olivares} et~al.,}{{Olivares}
  et~al.}{2010}]{2010ApJ...715..833O}
{Olivares} E.~F.,  et~al., 2010, \mn@doi [\apj] {10.1088/0004-637X/715/2/833},
  \href {http://adsabs.harvard.edu/abs/2010ApJ...715..833O} {715, 833}

\bibitem[\protect\citeauthoryear{{Pastorello} et~al.,}{{Pastorello}
  et~al.}{2004}]{2004MNRAS.347...74P}
{Pastorello} A.,  et~al., 2004, \mn@doi [\mnras]
  {10.1111/j.1365-2966.2004.07173.x}, \href
  {http://adsabs.harvard.edu/abs/2004MNRAS.347...74P} {347, 74}

\bibitem[\protect\citeauthoryear{{Pastorello} et~al.,}{{Pastorello}
  et~al.}{2006}]{2006MNRAS.370.1752P}
{Pastorello} A.,  et~al., 2006, \mn@doi [\mnras]
  {10.1111/j.1365-2966.2006.10587.x}, \href
  {http://adsabs.harvard.edu/abs/2006MNRAS.370.1752P} {370, 1752}

\bibitem[\protect\citeauthoryear{{Pastorello} et~al.,}{{Pastorello}
  et~al.}{2009}]{2009MNRAS.394.2266P}
{Pastorello} A.,  et~al., 2009, \mn@doi [\mnras]
  {10.1111/j.1365-2966.2009.14505.x}, \href
  {http://adsabs.harvard.edu/abs/2009MNRAS.394.2266P} {394, 2266}

\bibitem[\protect\citeauthoryear{{Pastorello} et~al.,}{{Pastorello}
  et~al.}{2012}]{2012A&A...537A.141P}
{Pastorello} A.,  et~al., 2012, \mn@doi [\aap] {10.1051/0004-6361/201118112},
  \href {http://adsabs.harvard.edu/abs/2012A%26A...537A.141P} {537, A141}

\bibitem[\protect\citeauthoryear{{Patat}, {Barbon}, {Cappellaro}  \&
  {Turatto}}{{Patat} et~al.}{1994}]{1994A&A...282..731P}
{Patat} F.,  {Barbon} R.,  {Cappellaro} E.,   {Turatto} M.,  1994, \aap, \href
  {http://adsabs.harvard.edu/abs/1994A%26A...282..731P} {282, 731}

\bibitem[\protect\citeauthoryear{{Poznanski} et~al.,}{{Poznanski}
  et~al.}{2009}]{2009ApJ...694.1067P}
{Poznanski} D.,  et~al., 2009, \mn@doi [\apj] {10.1088/0004-637X/694/2/1067},
  \href {http://adsabs.harvard.edu/abs/2009ApJ...694.1067P} {694, 1067}

\bibitem[\protect\citeauthoryear{{Poznanski}, {Ganeshalingam}, {Silverman}  \&
  {Filippenko}}{{Poznanski} et~al.}{2011}]{2011MNRAS.415L..81P}
{Poznanski} D.,  {Ganeshalingam} M.,  {Silverman} J.~M.,   {Filippenko} A.~V.,
  2011, \mn@doi [\mnras] {10.1111/j.1745-3933.2011.01084.x}, \href
  {http://adsabs.harvard.edu/abs/2011MNRAS.415L..81P} {415, L81}

\bibitem[\protect\citeauthoryear{{Pumo} \& {Zampieri}}{{Pumo} \&
  {Zampieri}}{2011}]{2011ApJ...741...41P}
{Pumo} M.~L.,  {Zampieri} L.,  2011, \mn@doi [\apj]
  {10.1088/0004-637X/741/1/41}, \href
  {http://adsabs.harvard.edu/abs/2011ApJ...741...41P} {741, 41}

\bibitem[\protect\citeauthoryear{{Pumo} \& {Zampieri}}{{Pumo} \&
  {Zampieri}}{2013}]{2013MNRAS.434.3445P}
{Pumo} M.~L.,  {Zampieri} L.,  2013, \mn@doi [\mnras] {10.1093/mnras/stt1256},
  \href {http://adsabs.harvard.edu/abs/2013MNRAS.434.3445P} {434, 3445}

\bibitem[\protect\citeauthoryear{{Pumo}, {Zampieri}  \& {Turatto}}{{Pumo}
  et~al.}{2010}]{2010MSAIS..14..123P}
{Pumo} M.~L.,  {Zampieri} L.,   {Turatto} M.,  2010, Memorie della Societa
  Astronomica Italiana Supplementi, \href
  {http://adsabs.harvard.edu/abs/2010MSAIS..14..123P} {14, 123}

\bibitem[\protect\citeauthoryear{{Pumo}, {Zampieri}, {Spiro}, {Pastorello},
  {Benetti}, {Cappellaro}, {Manic{\`o}}  \& {Turatto}}{{Pumo}
  et~al.}{2017}]{2017MNRAS.464.3013P}
{Pumo} M.~L.,  {Zampieri} L.,  {Spiro} S.,  {Pastorello} A.,  {Benetti} S.,
  {Cappellaro} E.,  {Manic{\`o}} G.,   {Turatto} M.,  2017, \mn@doi [\mnras]
  {10.1093/mnras/stw2625}, \href
  {http://adsabs.harvard.edu/abs/2017MNRAS.464.3013P} {464, 3013}

\bibitem[\protect\citeauthoryear{{Riess} et~al.,}{{Riess}
  et~al.}{2016}]{2016ApJ...826...56R}
{Riess} A.~G.,  et~al., 2016, \mn@doi [\apj] {10.3847/0004-637X/826/1/56},
  \href {http://adsabs.harvard.edu/abs/2016ApJ...826...56R} {826, 56}

\bibitem[\protect\citeauthoryear{{Roy} et~al.,}{{Roy}
  et~al.}{2011}]{2011ApJ...736...76R}
{Roy} R.,  et~al., 2011, \mn@doi [\apj] {10.1088/0004-637X/736/2/76}, \href
  {http://adsabs.harvard.edu/abs/2011ApJ...736...76R} {736, 76}

\bibitem[\protect\citeauthoryear{{Sahu}, {Anupama}, {Srividya}  \&
  {Muneer}}{{Sahu} et~al.}{2006}]{2006MNRAS.372.1315S}
{Sahu} D.~K.,  {Anupama} G.~C.,  {Srividya} S.,   {Muneer} S.,  2006, \mn@doi
  [\mnras] {10.1111/j.1365-2966.2006.10937.x}, \href
  {http://adsabs.harvard.edu/abs/2006MNRAS.372.1315S} {372, 1315}

\bibitem[\protect\citeauthoryear{{Schlafly} \& {Finkbeiner}}{{Schlafly} \&
  {Finkbeiner}}{2011}]{2011ApJ...737..103S}
{Schlafly} E.~F.,  {Finkbeiner} D.~P.,  2011, \mn@doi [\apj]
  {10.1088/0004-637X/737/2/103}, \href
  {http://adsabs.harvard.edu/abs/2011ApJ...737..103S} {737, 103}

\bibitem[\protect\citeauthoryear{{Silverman} et~al.,}{{Silverman}
  et~al.}{2017}]{2017MNRAS.467..369S}
{Silverman} J.~M.,  et~al., 2017, \mn@doi [\mnras] {10.1093/mnras/stx058},
  \href {http://adsabs.harvard.edu/abs/2017MNRAS.467..369S} {467, 369}

\bibitem[\protect\citeauthoryear{{Smartt}}{{Smartt}}{2009}]{2009ARA&A..47...63S}
{Smartt} S.~J.,  2009, \mn@doi [\araa] {10.1146/annurev-astro-082708-101737},
  \href {http://adsabs.harvard.edu/abs/2009ARA%26A..47...63S} {47, 63}

\bibitem[\protect\citeauthoryear{{Smartt}}{{Smartt}}{2015}]{2015PASA...32...16S}
{Smartt} S.~J.,  2015, \mn@doi [\pasa] {10.1017/pasa.2015.17}, \href
  {http://adsabs.harvard.edu/abs/2015PASA...32...16S} {32, e016}

\bibitem[\protect\citeauthoryear{{Smartt}, {Maund}, {Hendry}, {Tout},
  {Gilmore}, {Mattila}  \& {Benn}}{{Smartt} et~al.}{2004}]{2004Sci...303..499S}
{Smartt} S.~J.,  {Maund} J.~R.,  {Hendry} M.~A.,  {Tout} C.~A.,  {Gilmore}
  G.~F.,  {Mattila} S.,   {Benn} C.~R.,  2004, \mn@doi [Science]
  {10.1126/science.1092967}, \href
  {http://adsabs.harvard.edu/abs/2004Sci...303..499S} {303, 499}

\bibitem[\protect\citeauthoryear{{Smartt}, {Eldridge}, {Crockett}  \&
  {Maund}}{{Smartt} et~al.}{2009}]{2009MNRAS.395.1409S}
{Smartt} S.~J.,  {Eldridge} J.~J.,  {Crockett} R.~M.,   {Maund} J.~R.,  2009,
  \mn@doi [\mnras] {10.1111/j.1365-2966.2009.14506.x}, \href
  {http://adsabs.harvard.edu/abs/2009MNRAS.395.1409S} {395, 1409}

\bibitem[\protect\citeauthoryear{{Spiro} et~al.,}{{Spiro}
  et~al.}{2014}]{2014MNRAS.439.2873S}
{Spiro} S.,  et~al., 2014, \mn@doi [\mnras] {10.1093/mnras/stu156}, \href
  {http://adsabs.harvard.edu/abs/2014MNRAS.439.2873S} {439, 2873}

\bibitem[\protect\citeauthoryear{{Stalin}, {Hegde}, {Sahu}, {Parihar},
  {Anupama}, {Bhatt}  \& {Prabhu}}{{Stalin} et~al.}{2008}]{2008BASI...36..111S}
{Stalin} C.~S.,  {Hegde} M.,  {Sahu} D.~K.,  {Parihar} P.~S.,  {Anupama} G.~C.,
   {Bhatt} B.~C.,   {Prabhu} T.~P.,  2008, Bulletin of the Astronomical Society
  of India, \href {http://adsabs.harvard.edu/abs/2008BASI...36..111S} {36, 111}

\bibitem[\protect\citeauthoryear{{Stetson}}{{Stetson}}{1987}]{1987PASP...99..191S}
{Stetson} P.~B.,  1987, \mn@doi [\pasp] {10.1086/131977}, \href
  {http://adsabs.harvard.edu/abs/1987PASP...99..191S} {99, 191}

\bibitem[\protect\citeauthoryear{{Stetson}}{{Stetson}}{1992}]{1992JRASC..86...71S}
{Stetson} P.~B.,  1992, \jrasc, \href
  {http://adsabs.harvard.edu/abs/1992JRASC..86...71S} {86, 71}

\bibitem[\protect\citeauthoryear{{Sukhbold} \& {Woosley}}{{Sukhbold} \&
  {Woosley}}{2014}]{2014ApJ...783...10S}
{Sukhbold} T.,  {Woosley} S.~E.,  2014, \mn@doi [\apj]
  {10.1088/0004-637X/783/1/10}, \href
  {http://adsabs.harvard.edu/abs/2014ApJ...783...10S} {783, 10}

\bibitem[\protect\citeauthoryear{{Swartz}, {Wheeler}  \& {Harkness}}{{Swartz}
  et~al.}{1991}]{1991ApJ...374..266S}
{Swartz} D.~A.,  {Wheeler} J.~C.,   {Harkness} R.~P.,  1991, \mn@doi [\apj]
  {10.1086/170115}, \href {http://adsabs.harvard.edu/abs/1991ApJ...374..266S}
  {374, 266}

\bibitem[\protect\citeauthoryear{{Tak{\'a}ts} \& {Vink{\'o}}}{{Tak{\'a}ts} \&
  {Vink{\'o}}}{2012}]{2012MNRAS.419.2783T}
{Tak{\'a}ts} K.,  {Vink{\'o}} J.,  2012, \mn@doi [\mnras]
  {10.1111/j.1365-2966.2011.19921.x}, \href
  {http://adsabs.harvard.edu/abs/2012MNRAS.419.2783T} {419, 2783}

\bibitem[\protect\citeauthoryear{{Tak{\'a}ts} et~al.,}{{Tak{\'a}ts}
  et~al.}{2014}]{2014MNRAS.438..368T}
{Tak{\'a}ts} K.,  et~al., 2014, \mn@doi [\mnras] {10.1093/mnras/stt2203}, \href
  {http://adsabs.harvard.edu/abs/2014MNRAS.438..368T} {438, 368}

\bibitem[\protect\citeauthoryear{{Tak{\'a}ts} et~al.,}{{Tak{\'a}ts}
  et~al.}{2015}]{2015MNRAS.450.3137T}
{Tak{\'a}ts} K.,  et~al., 2015, \mn@doi [\mnras] {10.1093/mnras/stv857}, \href
  {http://adsabs.harvard.edu/abs/2015MNRAS.450.3137T} {450, 3137}

\bibitem[\protect\citeauthoryear{{Tartaglia} et~al.,}{{Tartaglia}
  et~al.}{2018}]{2018ApJ...853...62T}
{Tartaglia} L.,  et~al., 2018, \mn@doi [\apj] {10.3847/1538-4357/aaa014}, \href
  {http://adsabs.harvard.edu/abs/2018ApJ...853...62T} {853, 62}

\bibitem[\protect\citeauthoryear{{Terreran} et~al.,}{{Terreran}
  et~al.}{2017}]{2017NatAs...1..713T}
{Terreran} G.,  et~al., 2017, \mn@doi [Nature Astronomy]
  {10.1038/s41550-017-0228-8}, \href
  {http://adsabs.harvard.edu/abs/2017NatAs...1..713T} {1, 713}

\bibitem[\protect\citeauthoryear{{Thomas}, {Nugent}  \& {Meza}}{{Thomas}
  et~al.}{2011}]{2011PASP..123..237T}
{Thomas} R.~C.,  {Nugent} P.~E.,   {Meza} J.~C.,  2011, \mn@doi [\pasp]
  {10.1086/658673}, \href {http://adsabs.harvard.edu/abs/2011PASP..123..237T}
  {123, 237}

\bibitem[\protect\citeauthoryear{{Tomasella} et~al.,}{{Tomasella}
  et~al.}{2013}]{2013MNRAS.434.1636T}
{Tomasella} L.,  et~al., 2013, \mn@doi [\mnras] {10.1093/mnras/stt1130}, \href
  {http://adsabs.harvard.edu/abs/2013MNRAS.434.1636T} {434, 1636}

\bibitem[\protect\citeauthoryear{{Tomasella} et~al.,}{{Tomasella}
  et~al.}{2018}]{2018MNRAS.475.1937T}
{Tomasella} L.,  et~al., 2018, \mn@doi [\mnras] {10.1093/mnras/stx3220}, \href
  {http://adsabs.harvard.edu/abs/2018MNRAS.475.1937T} {475, 1937}

\bibitem[\protect\citeauthoryear{{Turatto} et~al.,}{{Turatto}
  et~al.}{1998}]{1998ApJ...498L.129T}
{Turatto} M.,  et~al., 1998, \mn@doi [\apjl] {10.1086/311324}, \href
  {http://adsabs.harvard.edu/abs/1998ApJ...498L.129T} {498, L129}

\bibitem[\protect\citeauthoryear{{Turatto}, {Benetti}  \&
  {Cappellaro}}{{Turatto} et~al.}{2003}]{2003fthp.conf..200T}
{Turatto} M.,  {Benetti} S.,   {Cappellaro} E.,  2003, in {Hillebrandt} W.,
  {Leibundgut} B.,  eds, From Twilight to Highlight: The Physics of Supernovae.
  p.~200 (\mn@eprint {} {astro-ph/0211219}), \mn@doi{10.1007/10828549_26}

\bibitem[\protect\citeauthoryear{{Utrobin} \& {Chugai}}{{Utrobin} \&
  {Chugai}}{2008}]{2008A&A...491..507U}
{Utrobin} V.~P.,  {Chugai} N.~N.,  2008, \mn@doi [\aap]
  {10.1051/0004-6361:200810272}, \href
  {http://adsabs.harvard.edu/abs/2008A%26A...491..507U} {491, 507}

\bibitem[\protect\citeauthoryear{{Utrobin} \& {Chugai}}{{Utrobin} \&
  {Chugai}}{2009}]{2009A&A...506..829U}
{Utrobin} V.~P.,  {Chugai} N.~N.,  2009, \mn@doi [\aap]
  {10.1051/0004-6361/200912273}, \href
  {http://adsabs.harvard.edu/abs/2009A%26A...506..829U} {506, 829}

\bibitem[\protect\citeauthoryear{{Valenti} et~al.,}{{Valenti}
  et~al.}{2014}]{2014MNRAS.438L.101V}
{Valenti} S.,  et~al., 2014, \mn@doi [\mnras] {10.1093/mnrasl/slt171}, \href
  {http://adsabs.harvard.edu/abs/2014MNRAS.438L.101V} {438, L101}

\bibitem[\protect\citeauthoryear{{Valenti} et~al.,}{{Valenti}
  et~al.}{2016}]{2016MNRAS.459.3939V}
{Valenti} S.,  et~al., 2016, \mn@doi [\mnras] {10.1093/mnras/stw870}, \href
  {http://adsabs.harvard.edu/abs/2016MNRAS.459.3939V} {459, 3939}

\bibitem[\protect\citeauthoryear{{Van Dyk}, {Li}  \& {Filippenko}}{{Van Dyk}
  et~al.}{2003}]{2003PASP..115.1289V}
{Van Dyk} S.~D.,  {Li} W.,   {Filippenko} A.~V.,  2003, \mn@doi [\pasp]
  {10.1086/378308}, \href {http://adsabs.harvard.edu/abs/2003PASP..115.1289V}
  {115, 1289}

\bibitem[\protect\citeauthoryear{{Vink{\'o}} et~al.,}{{Vink{\'o}}
  et~al.}{2006}]{2006MNRAS.369.1780V}
{Vink{\'o}} J.,  et~al., 2006, \mn@doi [\mnras]
  {10.1111/j.1365-2966.2006.10416.x}, \href
  {http://adsabs.harvard.edu/abs/2006MNRAS.369.1780V} {369, 1780}

\bibitem[\protect\citeauthoryear{{Walmswell} \& {Eldridge}}{{Walmswell} \&
  {Eldridge}}{2012}]{2012MNRAS.419.2054W}
{Walmswell} J.~J.,  {Eldridge} J.~J.,  2012, \mn@doi [\mnras]
  {10.1111/j.1365-2966.2011.19860.x}, \href
  {http://adsabs.harvard.edu/abs/2012MNRAS.419.2054W} {419, 2054}

\bibitem[\protect\citeauthoryear{{Woosley}, {Heger}  \& {Weaver}}{{Woosley}
  et~al.}{2002}]{2002RvMP...74.1015W}
{Woosley} S.~E.,  {Heger} A.,   {Weaver} T.~A.,  2002, \mn@doi [Reviews of
  Modern Physics] {10.1103/RevModPhys.74.1015}, \href
  {http://adsabs.harvard.edu/abs/2002RvMP...74.1015W} {74, 1015}

\bibitem[\protect\citeauthoryear{{Yaron} \& {Gal-Yam}}{{Yaron} \&
  {Gal-Yam}}{2012}]{2012PASP..124..668Y}
{Yaron} O.,  {Gal-Yam} A.,  2012, \mn@doi [\pasp] {10.1086/666656}, \href
  {http://adsabs.harvard.edu/abs/2012PASP..124..668Y} {124, 668}

\bibitem[\protect\citeauthoryear{{Zampieri}, {Pastorello}, {Turatto},
  {Cappellaro}, {Benetti}, {Altavilla}, {Mazzali}  \& {Hamuy}}{{Zampieri}
  et~al.}{2003}]{2003MNRAS.338..711Z}
{Zampieri} L.,  {Pastorello} A.,  {Turatto} M.,  {Cappellaro} E.,  {Benetti}
  S.,  {Altavilla} G.,  {Mazzali} P.,   {Hamuy} M.,  2003, \mn@doi [\mnras]
  {10.1046/j.1365-8711.2003.06082.x}, \href
  {http://adsabs.harvard.edu/abs/2003MNRAS.338..711Z} {338, 711}

\bibitem[\protect\citeauthoryear{{Zhang} et~al.,}{{Zhang}
  et~al.}{2014}]{2014ApJ...797....5Z}
{Zhang} J.,  et~al., 2014, \mn@doi [\apj] {10.1088/0004-637X/797/1/5}, \href
  {http://adsabs.harvard.edu/abs/2014ApJ...797....5Z} {797, 5}

\bibitem[\protect\citeauthoryear{{van Dokkum}}{{van
  Dokkum}}{2001}]{2001PASP..113.1420V}
{van Dokkum} P.~G.,  2001, \mn@doi [\pasp] {10.1086/323894}, \href
  {http://adsabs.harvard.edu/abs/2001PASP..113.1420V} {113, 1420}

\makeatother
\end{thebibliography}


\appendix
\afterpage{

\begin{figure}
  \centering
  \includegraphics[width=0.32\textwidth,clip, trim= 5cm 7.5cm 3.4cm 9cm]{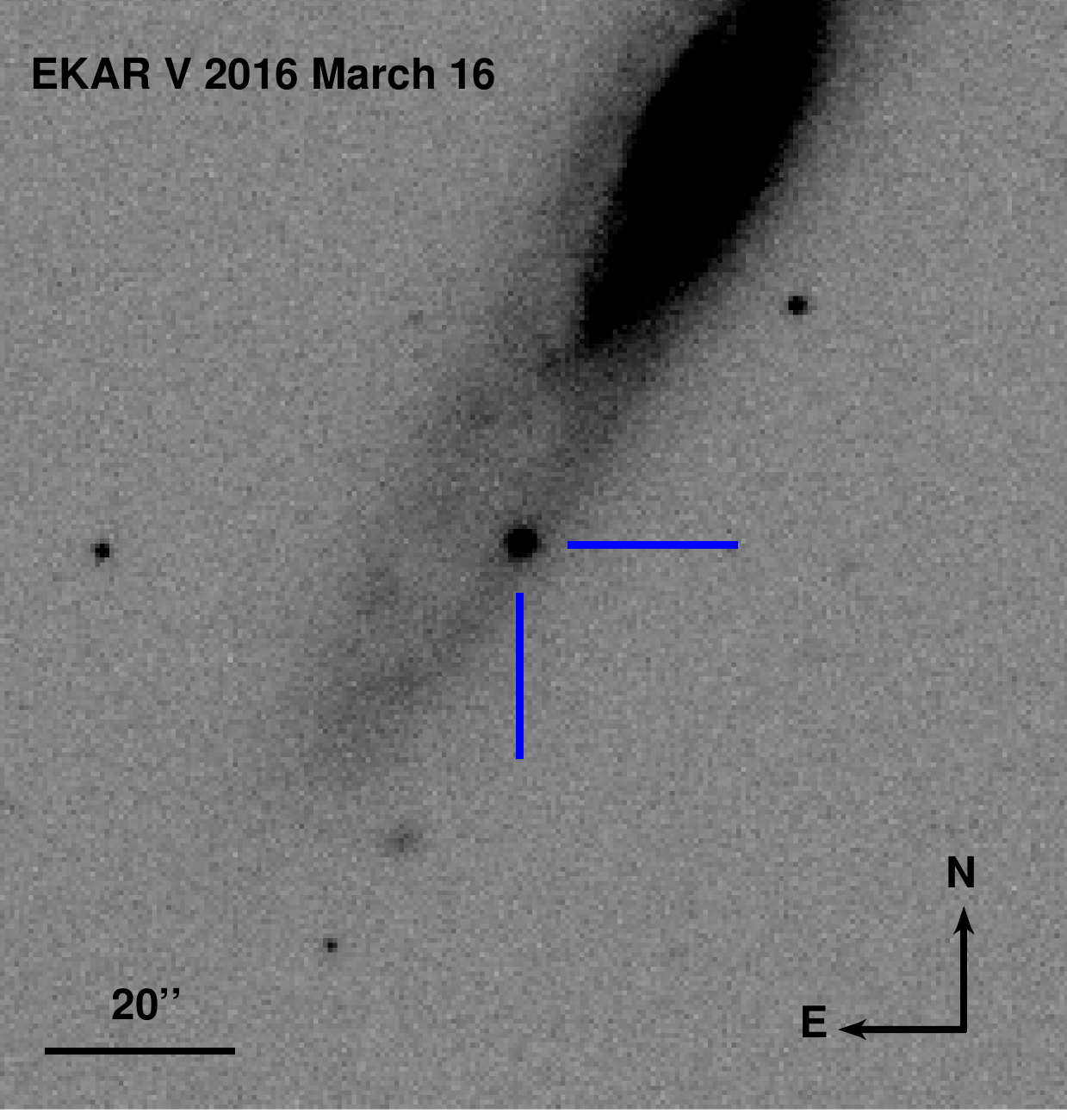}\label{fig:f1}
  \includegraphics[width=0.32\textwidth,clip, trim= 5cm 7.2cm 3.4cm 9cm]{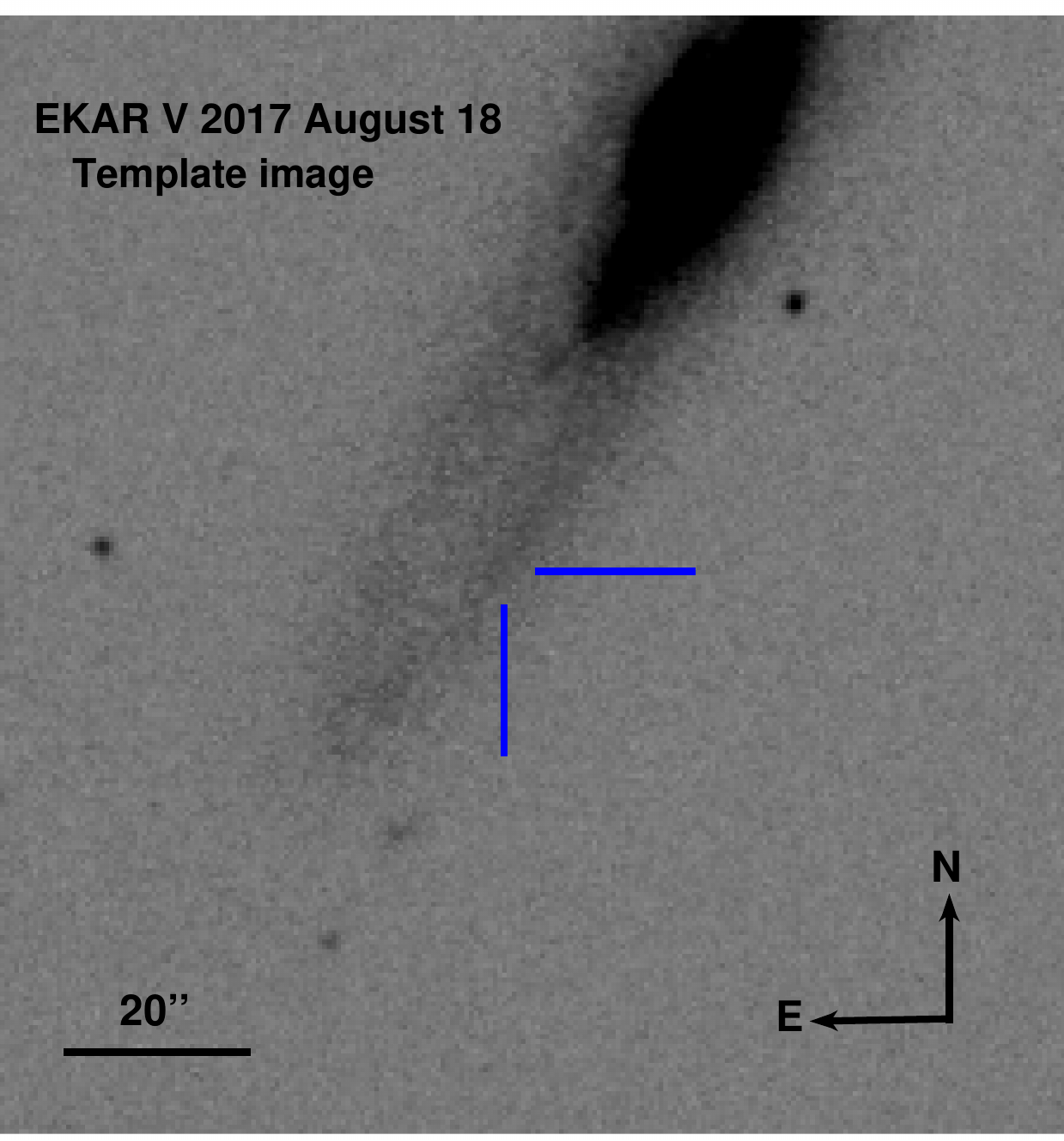}\label{fig:f2}
    \includegraphics[width=0.32\textwidth,clip, trim= 5cm 7.5cm 3.4cm 9cm]{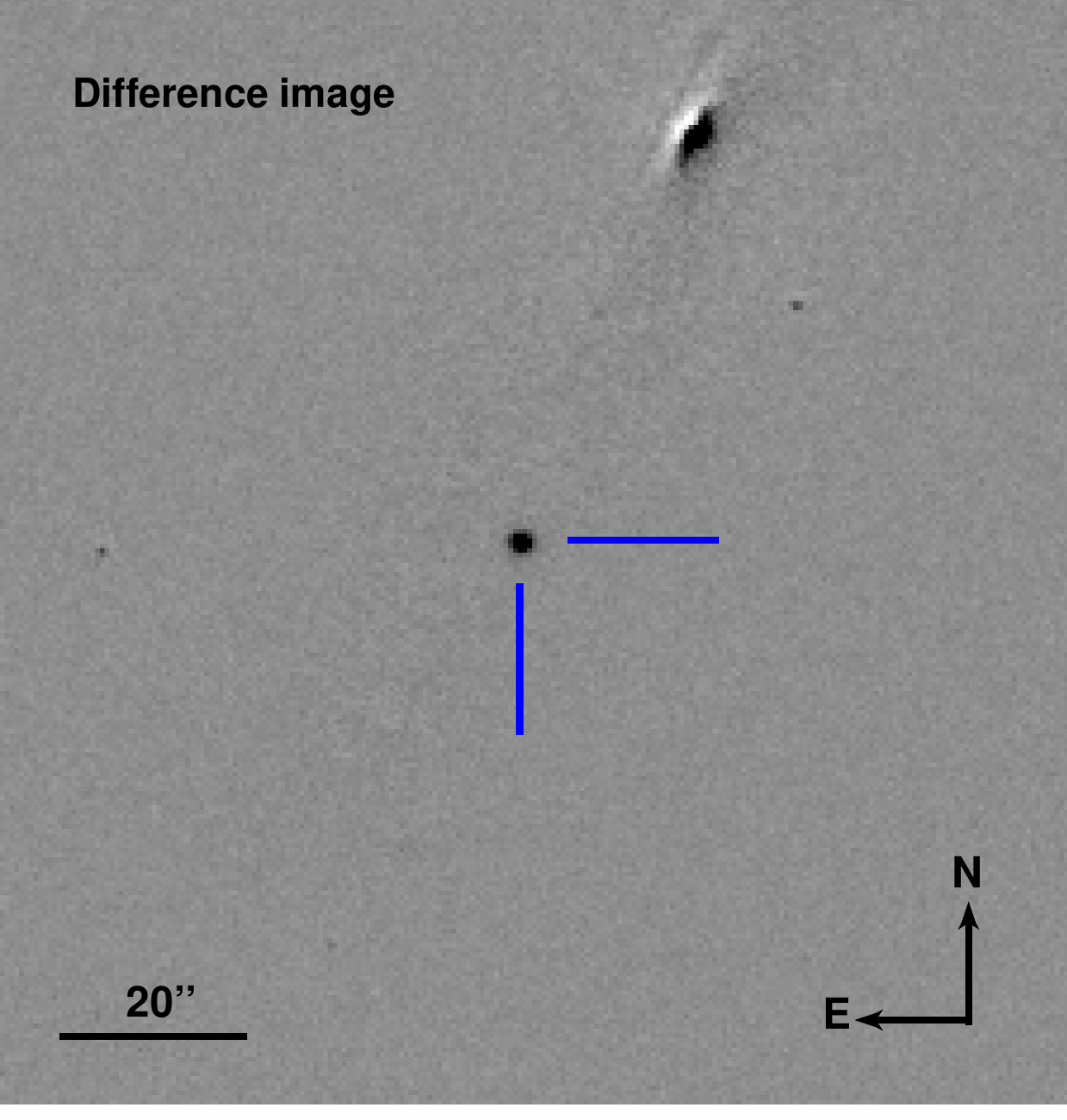}\label{fig:f3}
  \caption{The left panel taken with the 1.82m EKAR telescope in V band on 2016 March 16 shows the presence of SN 2015ba. The middle panel is  acquired with the EKAR telescope in V band on 2017 August 18 ($\sim$ 634 days since explosion) and shows the disappearance of the SN. The right panel shows the subtracted image. }
  \label{temp_sub}
\end{figure}
    
\begin{table}
\caption{Summary of the instruments used for the follow-up campaign of SN 2015ba.}
\centering
\smallskip
\flushleft
\begin{tabular}{l l l l l l l}
\hline \hline

Telescope & Location & Instrument  & Pixel Scale    & Imaging & Dispersers/ Grisms   &   \\
                        &  &   & ($''$/pixel) & bands\\
\hline
1m LCO & McDonald Observatory, USA & SBIG/Sinistro & 0.464/0.389 & {\it BVgri} & -& 1 \\
1.04m Sampurnanand & ARIES Observatory, India & Tek 1k$\times$1k   & 0.53  & {\it VRI}  & - & 2\\
Telescope (ST)         &  &  & & \\
1.3m Devasthal Fast & ARIES Observatory, India  & Andor 512$\times$512 & 0.64  & {\it BVRI} & - & 3\\
Optical Telescope (DFOT)         &  &  & & \\
1.82m Copernico Telescope & Asiago, Mount Ekar (Italy) & AFOSC & 0.48 & {\it uBVgriz} & Gr04, VPH7 & 4 \\
2m LCO Faulkes Telescope    & Haleakala Observatory, USA  & Spectral,  & 0.34 & {\it BVgri} & Cross disperser & 5\\
North (FTN)        & & FLOYDS & & \\
2.01m Himalayan  & Indian Astronomical   & HFOSC  & 0.17    & {\it BVRI} & - & 6\\
Chandra Telescope (HCT)         & Observatory, Hanle, India  &  & & \\
2.56m Nordic Optical & Roque de los Muchachos,  & ALFOSC  & 0.19 &  {\it uBVgriz}   & Gr4 & 7\\
 Telescope (NOT)        & La Palma, Canaris (Spain) &  & & \\
10.4m Gran Telescopio    &  Roque de los Muchachos, & OSIRIS &    0.25    & {\it r} & R300B, R300R & 8 \\
CANARIS (GTC) & La Palma, Canaris (Spain) &  & & \\
\hline                                   
\end{tabular}

\label{tab:details_instrument_detectors}      
\end{table}

\begin{table*}
\setlength{\tabcolsep}{1.5pt}.
\caption{Coordinates and photometry of the local sequence reference stars in the $BVRI$ and in the $ugriz$ systems.}
\scalebox{0.9}{
\begin{tabular}{llcccccccccc}
\hline
 ID & $ \alpha_{J2000.0} $ & $ \delta_{J2000.0} $ & $B$ & $V$  & $R$  & $I$ &$u$ &$g$ &$r$ &$i$ &$z$ \\
 & (hh:mm:ss) &(dd:mm:ss) & mag & mag & mag & mag & mag & mag & mag & mag & mag \\
\hline 
A        & 14:32:37.5  & +49:54:12.4     &  17.690 (.060)  & 17.060 (.040) & 	16.560 (.005) & 16.245 (.007) & 18.390 (.010) & 17.215 (.004) & 16.781 (.004) & 16.650 (.005) & 16.600 (.009) \\
B        & 14:32:34.0  & +49:53:32.0     &   20.300 (.400) & 19.410 (.060) & 	18.460 (.010) & 17.850 (.010) & 22.900 (.400) & 20.030 (.020) & 18.780 (.010) & 18.291 (.009) & 17.990 (.020) \\
C        & 14:32:36.8  & +49:52:23.6     &  19.800 (.100)  & 18.840 (.080) & 	18.137 (.011) & 17.691 (.007) & 21.220 (.090) & 19.210 (.010) & 18.368 (.008) & 18.100 (.008) & 17.910 (.020) \\
D        & 14:32:33.7  & +49:51:31.3     &  20.000 (.100) & 18.680 (.060) & 	17.683 (.005) & 16.967 (.006) & 21.800 (.200) & 19.370 (.010) & 18.017 (.007) & 17.413 (.006) & 17.090 (.010) \\
E        & 14:32:22.8  & +49:51:15.2     &  18.860 (.060)  & 18.200 (.060) & 	17.644 (.006) & 17.277 (.007) & 19.650 (.030) & 18.396 (.007)  & 17.893 (.007) & 17.723 (.007) & 17.700 (.020) \\
F        & 14:32:17.1  & +49:51:52.5     &  17.918 (.100)  & 17.400 (.040) & 	16.949 (.003) & 16.632 (.005) & 18.510 (.010) & 17.488 (.005)  & 17.171 (.005) & 17.068 (.006) & 17.060 (.010) \\
G        & 14:32:14.3  & +49:51:47.5     &  19.900 (.200)  & 19.100 (.100) & 	18.508 (.008) & 18.110 (.010) & 21.200 (.100) & 19.390 (.010)  & 18.740 (.010) & 18.510 (.010) & 18.400 (.030) \\
H        & 14:32:09.9  & +49:53:34.2     &  17.630 (.060)  & 16.960 (.030) & 	16.435 (.004) & 16.094 (.003) & 18.460 (.020) & 17.133 (.004)  & 16.673 (.004) & 16.490 (.005) & 16.466 (.008) \\
\hline

\end{tabular}
}
\label{tab:local}   
\end{table*}


\begin{ThreePartTable}
\small \setlength{\tabcolsep}{3.5pt}
  \begin{TableNotes}\footnotesize
\item[a] since the explosion epoch t$_0$ = 2457349.7 JD

  \end{TableNotes}
 
{\footnotesize
\begin{longtable}{ l l l c c c c c l }
\caption{Optical photometry of SN 2015ba.}
 \label{photometry}\\
\hline \multicolumn{1}{c}{UT Date} & \multicolumn{1}{c}{JD} & \multicolumn{1}{l}{Phase$^a$} & \multicolumn{1}{c}{B} & \multicolumn{1}{c}{V} & \multicolumn{1}{c}{R} & \multicolumn{1}{c}{I} & \multicolumn{1}{c}{} & \multicolumn{1}{c}{Tel}\\
\multicolumn{1}{c}{(yyyy-mm-dd)} & \multicolumn{1}{l}{2457000+} & \multicolumn{1}{c}{(days)} & \multicolumn{1}{c}{(mag)} & \multicolumn{1}{c}{(mag)} & \multicolumn{1}{c}{(mag)} & \multicolumn{1}{c}{(mag)} & \multicolumn{1}{c}{} & \multicolumn{1}{c}{}\\ 
\midrule
\endfirsthead
\multicolumn{9}{l}%
{{\bfseries \tablename\ \thetable{} -- continued}} \\
\midrule
\endhead
\midrule
\endfoot
\insertTableNotes
\endlastfoot

2015-12-02.18 & 358.68 & 8.98     & 16.74  $\pm$ 0.06  &         -          &       -            &      -             & & 4 \\
2015-12-03.21 & 359.71 & 10.01    & 16.76  $\pm$ 0.02  & 16.51  $\pm$ 0.06  &       -            &      -             & & 4 \\
2015-12-03.52 & 360.02 & 10.32    & 16.69  $\pm$ 0.02  & 16.47  $\pm$ 0.02  &       -            &      -             & & 1 \\
2015-12-04.16 & 360.66 & 10.96    & 16.74  $\pm$ 0.02  & 16.48  $\pm$ 0.02  &       -            &      -             & & 4 \\
2015-12-04.52 & 361.02 & 11.32    & 16.69  $\pm$ 0.02  & 16.44  $\pm$ 0.02  &       -            &      -             & & 1 \\
2015-12-06.24 & 362.74 & 13.04    & 16.82 $\pm$ 0.01 & 16.45  $\pm$ 0.02  &       -            &      -             & & 4 \\
2015-12-07.52 & 364.02 & 14.32    & 16.77  $\pm$ 0.02  & 16.50  $\pm$ 0.02  &       -            &      -             & & 1 \\
2015-12-10.50 & 367.00 & 17.3     & 16.80  $\pm$ 0.07  & 16.68  $\pm$ 0.07  &       -            &      -             & & 1 \\
2015-12-11.24 & 367.63 & 17.93    & 16.96  $\pm$ 0.01  & 16.61 $\pm$ 0.01 &       -            &      -             & & 4 \\
2015-12-11.53 & 368.03 & 18.33    & 16.95  $\pm$ 0.07  & 16.81  $\pm$ 0.08  &       -            &      -             & & 1 \\
2015-12-16.53 & 373.03 & 23.33    & 17.09  $\pm$ 0.02  & 16.68  $\pm$ 0.02  &       -            &      -             & & 1 \\
2015-12-19.13 & 375.71 & 26.01    & 17.34  $\pm$ 0.09  & 16.74 $\pm$ 0.10   &       -            &      -             & & 1 \\
2015-12-20.52 & 377.02 & 27.32    & 17.36  $\pm$ 0.04  & 16.79  $\pm$ 0.02  &       -            &      -             & & 1 \\
2015-12-30.52 & 387.02 & 37.32    & 17.73  $\pm$ 0.05  & 16.90  $\pm$ 0.04  &       -            &      -             & & 1 \\
2016-01-01.96 & 389.47 & 39.77    &          -         & 16.91 $\pm$ 0.02  & 16.32 $\pm$ 0.01 & 15.98 $\pm$ 0.01 & & 2 \\
2016-01-02.97 & 390.48 & 40.78    &          -         & 17.06  $\pm$ 0.02  & 16.36 $\pm$ 0.01 & 16.01 $\pm$ 0.01 & & 2 \\
2016-01-07.52 & 395.02 & 45.32    & 18.11  $\pm$ 0.03  & 17.05  $\pm$ 0.02  &      -             &      -             & & 1 \\
2016-01-13.02 & 400.50 & 50.8     & 18.20  $\pm$ 0.01  & 17.07 $\pm$ 0.01 & 16.38 $\pm$ 0.01 & 16.12  $\pm$ 0.01  & & 3 \\
2016-01-14.02 & 401.53 & 51.83    &          -         & 17.12 $\pm$ 0.01 & 16.39 $\pm$ 0.01 & 16.05 $\pm$ 0.01 & & 3 \\
2016-01-16.03 & 403.53 & 53.83    &          -         & 17.29 $\pm$ 0.03  & 16.40 $\pm$ 0.01 & 16.28  $\pm$ 0.01  & & 3 \\
2016-01-17.03 & 404.53 & 54.83    & 17.58 $\pm$ 0.01 & 17.24 $\pm$ 0.01 & 16.41 $\pm$ 0.01 & 16.50  $\pm$ 0.03  & & 3 \\
2016-01-17.64 & 405.14 & 55.44    & 18.27  $\pm$ 0.02  & 17.15  $\pm$ 0.02  &      -             &      -             & & 5 \\
2016-01-19.21 & 406.68 & 56.98    & 18.48  $\pm$ 0.01  & 17.16 $\pm$ 0.01 &      -             &      -             & & 4 \\
2016-01-20.94 & 408.45 & 58.75    &          -         & 17.34 $\pm$ 0.04  & 16.38 $\pm$ 0.01 & 16.10 $\pm$ 0.01 & & 2 \\
2016-01-23.62 & 411.12 & 61.42    & 18.23  $\pm$ 0.04  &  17.26 $\pm$ 0.03  &       -            &       -            & & 5 \\
2016-01-30.22 & 417.72 & 68.02    & 18.82  $\pm$ 0.05  & 17.17  $\pm$ 0.02  &       -            &       -            & & 7 \\
2016-01-31.89 & 419.40 & 69.7     &          -         & 17.24  $\pm$ 0.01  & 16.41 $\pm$ 0.01 & 16.02 $\pm$ 0.01 & & 2 \\
2016-02-02.87 & 421.38 & 71.68    &          -         & 17.28 $\pm$ 0.03  & 16.41 $\pm$ 0.01 & 16.00 $\pm$ 0.01 & & 2 \\
2016-02-05.21 & 423.72 & 74.02    & 18.71  $\pm$ 0.04  & 17.19  $\pm$ 0.01  &      -             &      -             & & 4 \\
2016-02-07.33 & 425.83 & 76.13    & 18.42  $\pm$ 0.03  & 17.08  $\pm$ 0.02  &      -             &      -             & & 1 \\
2016-02-10.98 & 429.48 & 79.78    & 18.52 $\pm$ 0.01 & 17.23 $\pm$ 0.01 & 16.41 $\pm$ 0.01 & 16.10 $\pm$ 0.01 & & 3 \\
2016-02-11.91 & 430.41 & 80.71    & 18.58  $\pm$ 0.01  & 17.23 $\pm$ 0.01 & 16.41 $\pm$ 0.01 & 15.96 $\pm$ 0.01 & & 3 \\
2016-02-12.93 & 431.43 & 81.73    & 18.62  $\pm$ 0.01  & 17.28 $\pm$ 0.01 & 16.41 $\pm$ 0.01 & 15.99 $\pm$ 0.01 & & 3 \\
2016-02-13.97 & 432.47 & 82.77    & 18.64  $\pm$ 0.01  & 17.25 $\pm$ 0.01 & 16.41 $\pm$ 0.01 & 15.99 $\pm$ 0.01 & & 3 \\
2016-02-13.47 & 431.97 & 82.27    & 18.66  $\pm$ 0.02  & 17.22  $\pm$ 0.02  &       -            &       -            & & 1 \\
2016-02-19.48 & 437.98 & 88.28    & 18.67  $\pm$ 0.03  & 17.24  $\pm$ 0.02  &       -            &       -            & & 1 \\
2016-02-25.35 & 443.85 & 94.15    & 18.65  $\pm$ 0.03  & 17.21  $\pm$ 0.02  &       -            &       -            & & 1 \\
2016-02-27.42 & 445.92 & 96.22    & 18.73  $\pm$ 0.03  & 17.25  $\pm$ 0.01  &       -            &       -            & & 1 \\
2016-03-01.38 & 448.88 & 99.18    & 18.74  $\pm$ 0.02  & 17.26  $\pm$ 0.01  &       -            &       -            & & 1 \\
2016-03-04.35 & 451.85 & 102.15   & 18.68  $\pm$ 0.05  & 17.24  $\pm$ 0.02  &       -            &       -            & & 1 \\
2016-03-07.96 & 455.46 & 105.66 & 18.82  $\pm$ 0.01  & 17.33  $\pm$ 0.01  & 16.42 $\pm$ 0.01 & 16.08 $\pm$ 0.01 & & 3 \\
2016-03-11.32 & 458.82 & 109.12   & 18.79  $\pm$ 0.02  & 17.31  $\pm$ 0.01  &       -            &       -            & & 1 \\
2016-03-14.48 & 461.98 & 112.28   & 18.83  $\pm$ 0.02  & 17.34  $\pm$ 0.02  &       -            &       -            & & 1 \\
2016-03-15.88 & 463.37 & 113.67   &          -         &         -          & 16.45 $\pm$ 0.02   & -  & & 2 \\
2016-03-16.21 & 463.60 & 113.9    & 18.90  $\pm$ 0.02  & 17.40  $\pm$ 0.01  &       -            &       -            & & 7 \\
2016-03-17.48 & 464.98 & 115.28   & 18.93  $\pm$ 0.02  & 17.40  $\pm$ 0.01  &       -            &       -            & & 1 \\
2016-03-18.89 & 466.42 & 116.72   &          -         & 17.37  $\pm$ 0.02  &       -            & 16.06  $\pm$ 0.02  & & 2 \\
2016-03-19.96 & 467.45 & 117.75   &          -         &         -          & 16.51 $\pm$ 0.01 & 16.10  $\pm$ 0.02  & & 2 \\
2016-03-21.47 & 468.97 & 119.27   & 18.89  $\pm$ 0.04  & 17.40  $\pm$ 0.03  &      -             &       -            & & 1 \\
2016-03-25.47 & 472.97 & 123.27   & 18.70   $\pm$ 0.10   & 17.41  $\pm$ 0.08  &      -             &       -            & & 1 \\
2016-04-04.46 & 482.96 & 133.26   & 19.36  $\pm$ 0.03  & 17.86  $\pm$ 0.02  &      -             &       -            & & 1 \\
2016-04-09.42 & 487.92 & 138.22   & 19.92  $\pm$ 0.07  & 18.26  $\pm$ 0.05  &      -             &       -            & & 1 \\
2016-04-13.43 & 491.93 & 142.23   & 20.29  $\pm$ 0.10   & 19.40   $\pm$ 0.10   &      -             &       -            & & 1 \\
2016-04-14.91 & 493.41 & 143.61   & -   & 20.17   $\pm$ 0.02   & 19.14 $\pm$ 0.01  & 18.62 $\pm$ 0.01 & & 3 \\
2016-04-15.23 & 493.73 & 144.03   &          -         & 20.16  $\pm$ 0.07  &      -             &       -            & & 1 \\
2016-05-02.81 & 511.31 & 161.61   &          -         &          -         & 19.56  $\pm$ 0.08  & 19.03  $\pm$ 0.04  & & 2 \\
\hline
 &  &             &       u        &              g       &      r      &     i         & z          &   Tel\\
 &  &             &   (mag)    &         (mag)   & (mag)  & (mag)  & (mag) & \\
\hline
2015-12-02.18 & 358.68 & 8.98  & 16.82 $\pm$ 0.06 & 16.53 $\pm$ 0.05 & 16.29 $\pm$ 0.04 & 16.23 $\pm$ 0.01 & 16.20 $\pm$ 0.04 &   4\\
2015-12-03.21 & 359.70 & 10    & 16.74 $\pm$ 0.02 & 16.48 $\pm$ 0.05 & 16.27 $\pm$ 0.05 & 16.24 $\pm$ 0.04 & 16.20 $\pm$ 0.06  &   4\\
2015-12-04.16 & 360.66 & 10.96 & 16.87 $\pm$ 0.01 & 16.63 $\pm$ 0.02 & 16.27 $\pm$ 0.02  & 16.16 $\pm$ 0.03 & 16.17 $\pm$ 0.03  &   4\\
2015-12-06.24 & 362.74 & 13.04 & 16.90 $\pm$ 0.01 & - & - & - & - & 4\\
2015-12-07.51 & 364.01 & 14.31 & - & 16.90 $\pm$ 0.01    & 16.27 $\pm$ 0.03 & 16.23 $\pm$ 0.04 & - & 1 \\
2015-12-11.24 & 367.62 & 17.92 & 17.13 $\pm$ 0.06 & 16.663 $\pm$ 0.005 & 16.29 $\pm$ 0.01 & 16.23 $\pm$ 0.01 & 16.20 $\pm$ 0.01 &   4\\
2015-12-14.50 & 371.00 & 21.3  & - & 16.55 $\pm$ 0.07 & 16.31 $\pm$ 0.03 &  16.32 $\pm$ 0.04 & - & 1 \\
2015-12-19.13 & 375.73 & 26.03 & 19.50 $\pm$ 0.10 & 16.82 $\pm$ 0.01 & 16.49 $\pm$ 0.02 & 16.30 $\pm$ 0.01 & 18.60 $\pm$ 0.10 &   4\\
2015-12-20.53 & 377.03 & 27.33 & - & 16.87 $\pm$ 0.08 & 16.44 $\pm$ 0.03 &  16.44 $\pm$ 0.04 & - & 1 \\
2015-12-30.53 & 387.03 & 37.33 & - & 17.22 $\pm$ 0.08 & 16.52 $\pm$ 0.03 &   16.38 $\pm$ 0.04 & - & 1\\
2016-01-07.53 & 395.03 & 45.33 & - & 17.45 $\pm$0.07 & 16.63 $\pm$ 0.02 &    16.56 $\pm$ 0.03 & - & 1 \\
2016-01-17.64 & 405.14 & 55.44 & - & 17.58 $\pm$ 0.07 & 16.68 $\pm$ 0.02 &    16.51 $\pm$ 0.03 & - & 5 \\
2016-01-19.21 & 406.69 & 56.99 & 19.86 $\pm$ 0.08 & 17.61 $\pm$ 0.01 & 16.67 $\pm$ 0.01 & 16.49 $\pm$ 0.01 & 16.36 $\pm$ 0.01 & 4\\
2016-01-23.62 & 411.12 & 61.42 & - & 17.48 $\pm$ 0.08 & 16.71 $\pm$ 0.02 & 16.44 $\pm$ 0.02 & - & 5 \\
2016-01-25.47 & 412.97 & 63.27 & - & - & 16.63 $\pm$ 0.01 & - & - &  1 \\
2016-01-30.22 & 417.72 & 68.02 & - & 17.78 $\pm$ 0.01 & 16.67 $\pm$ 0.01 & 16.52 $\pm$ 0.01 & 16.46 $\pm$ 0.04  &  4\\
2016-02-05.21 & 423.72 & 74.02 & 19.7 $\pm$ 0.1 & 17.77 $\pm$ 0.02       & 16.68 $\pm$ 0.01 & 16.48 $\pm$ 0.01 & 16.38 $\pm$ 0.01  &   4\\
2016-02-07.34 & 425.84 & 76.14 & - & 17.60 $\pm$ 0.08 & 16.66 $\pm$ 0.02 & 16.43 $\pm$ 0.03 & - & 1\\
2016-02-13.48 & 431.98 & 82.28 & - & 17.84 $\pm$ 0.01 & 16.62 $\pm$ 0.01 & 16.45 $\pm$ 0.01 & -& 1 \\
2016-02-19.48 & 437.98 & 88.28 & - & 17.86 $\pm$ 0.01 & 16.69 $\pm$ 0.01 & 16.46 $\pm$ 0.01 &- & 1 \\
2016-02-25.36 & 443.86 & 94.16 & - & 17.87 $\pm$ 0.01 & 16.70 $\pm$ 0.01 & 16.46 $\pm$ 0.01 & -& 1 \\
2016-02-27.44 & 445.94 & 96.24 &  - & 17.89 $\pm$ 0.01 & 16.71 $\pm$ 0.01 & 16.46 $\pm$ 0.01 & -& 1 \\ 
2016-03-01.39 & 448.89 & 99.19 & - & 17.89 $\pm$ 0.01 &  16.72 $\pm$ 0.01 & 16.49 $\pm$ 0.01 & -& 1 \\ 
2016-03-04.36 & 451.86 & 102.16 & - & 17.90 $\pm$ 0.01 & 16.73 $\pm$ 0.01 & 16.50 $\pm$ 0.01 & -& 1 \\
2016-03-11.33 & 458.83 & 109.13 & - & 17.89 $\pm$ 0.07 & 16.85 $\pm$ 0.02 & 16.55 $\pm$ 0.02 & -& 1 \\
2016-03-14.50 & 462.00 & 112.30 & - & 17.92 $\pm$ 0.07 & 16.82 $\pm$ 0.02 & 16.57 $\pm$ 0.02 & -& 1 \\
2016-03-16.21 & 463.60 & 113.90 & 19.50 $\pm$ 0.10 & 18.03 $\pm$ 0.01 & 16.86 $\pm$ 0.01 & 16.74 $\pm$ 0.02 & 16.49 $\pm$ 0.03  &   4\\
2016-03-17.49 & 464.99 & 115.29 & - & 18.05 $\pm$ 0.01 & 16.85 $\pm$ 0.01 & 16.62 $\pm$ 0.01 & -& 1 \\
2016-03-21.49 & 468.99 & 119.29 & - & 17.99 $\pm$ 0.08 & 16.88 $\pm$ 0.03 & 16.64 $\pm$ 0.03 & -& 1 \\
2016-03-25.48 & 472.98 & 123.28 & - & 18.17 $\pm$ 0.03 & 16.90 $\pm$ 0.01 & 16.66 $\pm$ 0.01 & -& 1 \\
2016-04-04.48 & 482.98 & 133.28 & - & 18.56 $\pm$ 0.01 & 17.24 $\pm$ 0.01 & 17.01 $\pm$ 0.01 & -& 1 \\
2016-04-09.44 & 487.94 & 138.24 & - & 19.04 $\pm$ 0.08 & 17.74 $\pm$ 0.02 & 17.50 $\pm$ 0.03 & -& 1 \\
2016-04-13.44 & 491.94 & 142.24 & - & 20.25 $\pm$ 0.04 & 18.95 $\pm$ 0.02 & 19.06 $\pm$ 0.04 & -& 1 \\
2016-04-15.25 & 493.75 & 144.05 & - & 20.67 $\pm$ 0.06 & 19.53 $\pm$ 0.03 & 19.46 $\pm$ 0.05 & -& 1 \\
2016-05-05.40 & 513.90 & 164.20 & - & - & 20.22 $\pm$ 0.09 & - & - & 1 \\
2016-05-07.33 & 515.83 & 166.13 & - & 21.20 $\pm$ 0.10 & 20.27 $\pm$ 0.08 & 20.19 $\pm$ 0.09 & -& 1 \\
2016-05-12.42 & 520.92 & 171.22 & - & - & 20.20 $\pm$ 0.10 &  - & - & 1 \\
2016-06-17.96 & 557.46 & 207.76 & - & - & 21.20 $\pm$ 0.20 & 20.50 $\pm$ 0.10 & 20.10 $\pm$ 0.10 & 4 \\
2016-08-02.90 & 602.90 & 253.20 & - & - & 20.30 $\pm$ 0.10 & - & - & 8 \\
2016-08-08.92 & 609.42 & 259.72 & - & - & 21.40 $\pm$ 0.10 & 21.40 $\pm$ 0.10 &    20.29 $\pm$ 0.09 & 4 \\
2016-08-21.87 & 621.87 & 272.17 & - & - & 20.80 $\pm$ 0.10 & - & - & 8 \\
\hline

\end{longtable}}

\end{ThreePartTable}

    
\begin{figure}
	\begin{center}
		\includegraphics[scale=0.6]{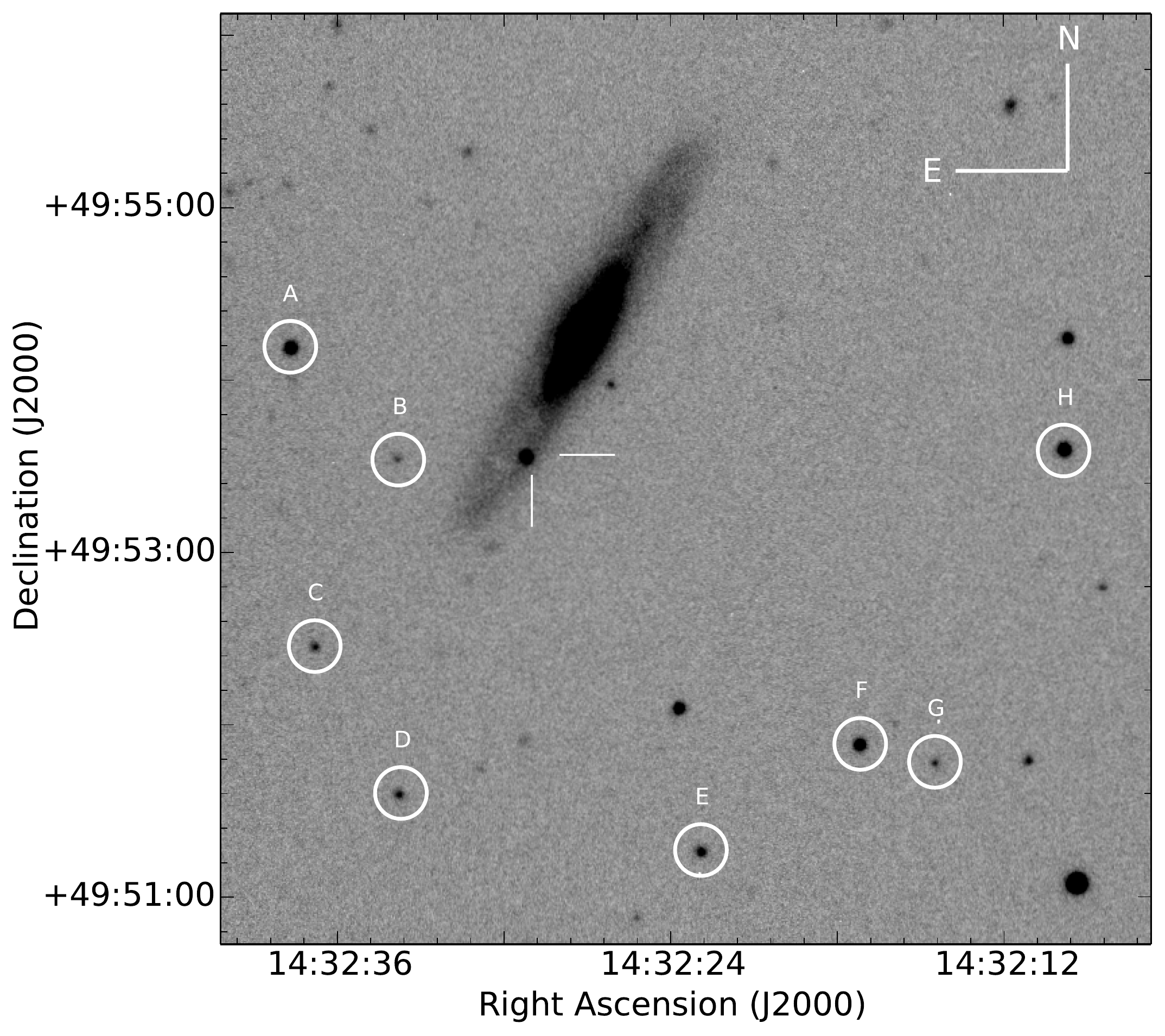}
	\end{center}
	\caption{SN 2015ba in IC 1029 along with 8 local standard stars. This {\it V}-band image was taken on 2015 December 15 with an 1m LCO telescope.}
	\label{fig:id_chart}
\end{figure}

}
\section{Photometry}
\label{phot}
The images were pre-processed to eliminate detector signatures  including overscan correction, bias subtraction, flat-field correction and trimming, following standard procedures in IRAF\footnote{IRAF stands for Image Reduction and Analysis Facility distributed by the National Optical Astronomy Observatories which is operated by the Association of Universities for research in Astronomy, Inc., under cooperative agreement with the National Science Foundation.}. Cosmic hits were detected and removed using the laplacian detection algorithm L.A.Cosmic \citep{2001PASP..113.1420V}. When multiple frames were taken on the same night, the frames were median-combined in order to improve the signal-to-noise ratio. Point spread function fitting photometry was performed on the images using DAOPHOT II \citep{1987PASP...99..191S} to derive the SN magnitudes. The mean FWHM of the frame was used as the first aperture and a 8 pixel wide sky background annulus was used, with an inner radius of four times the FWHM. 

Considering the location of the SN, we performed template subtraction to eliminate any possible contamination of the host galaxy flux in the SN magnitudes. The template images used for subtraction were obtained on 2017 February 26, May 05 and August 18 for the different instrumental setups listed in Table \ref{tab:details_instrument_detectors}. For template subtraction, the images were first registered astrometrically to the template image and subsequently subtracted using High Order Transform of PSF ANd Template Subtraction (HOTPANTS)\footnote{https://github.com/acbecker/hotpants}. We used the point spread function constructed using the unsubtracted SN frames to derive the SN magnitudes from the subtracted frames. Figure \ref{temp_sub} displays 2016 March 16 image before and after performing template subtraction.

We observed 3 Landolt standard fields PG 918, PG 942 and PG 1525 \citep{2009AJ....137.4186L} on 2017 February 26 under photometric night conditions (FWHM seeing in {\it V} $\sim$ 2$^{\prime\prime}$) with the 2.01 m Himalayan Chandra Telescope (HCT) at Indian Astronomical Observatory (IAO), Hanle, to convert the instrumental magnitudes in {\it BVRI} to standard magnitudes. The {\it V}-band magnitudes of the stars in the standard field lies in the range 12.27 mag to 16.4 mag and $B-V$ colour varies between $-$0.271 to 1.109 mag. These fields were observed at three altitudes covering airmass from 1.15 to 1.58.  With the aid of the instrumental and standard magnitudes of the Landolt field stars, the zero points and colour coefficients of the transformation equations were fitted using the least square linear regression technique as described in \cite{1992JRASC..86...71S} and the values thus obtained are given below:
$$ b = B + (1.632 \pm 0.001) + (0.022 \pm 0.003)(B-V) + k_b.X$$
$$ v = V + (1.001 \pm 0.001) + (-0.057 \pm 0.001)(B-V) + k_v.X$$
$$ r = R + (1.018 \pm 0.004) + (0.03 \pm 0.03)(V-R) +k_r.X$$
$$ i  = I + (1.264 \pm 0.003) + (0.065 \pm 0.004)(V-I) +k_i.X$$
where X is the airmass and the site extinction coefficients $k_{\lambda}$ are 0.21, 0.12, 0.08 and 0.05 mag per unit airmass for the {\it B}, {\it V}, {\it R} and {\it I} bands respectively \citep{2008BASI...36..111S}. The rms scatter between the observed and standard magnitudes of the Landolt field stars are $\sim$ 0.08 mag in {\it B}, $\sim$ 0.04 mag in {\it V} and {\it I} and $\sim$ 0.03 mag in {\it R}. 
We used these coefficients to generate a photometric sequence of 8 non-variable stars in the SN field. Observations of standard fields and subsequent calibrations were also performed with the Las Cumbres Observatory (LCO) telescopes on multiple photometric nights. We picked the 8 local standards chosen before and calibrated their magnitudes in {\it ugriz} and {\it BV} filters. Since, the {\it B} and {\it V} bands are in common to ARIES, LCO and ASIAGO data, to maintain uniformity we preferred to use LCO secondary standard magnitudes for all {\it BV} bands data. Similarly, to calibrate the {\it ugriz} data obtained from LCO and ASIAGO, we used the LCO secondary standard {\it ugriz} magnitudes. The 8 local standards are marked in Figure \ref{fig:id_chart}, and their magnitudes are tabulated in Table \ref{tab:local}. Night to night zero points were estimated and applied for obtaining the final SN magnitudes. The combined PSF fitting and photometric calibration errors propagated in quadrature are listed in Table \ref{photometry}.

\section{Spectroscopy}

We performed the initial pre-processing steps on each frame, like bias subtraction, flat-fielding, trimming and eliminating the cosmic hits from the frames using the L.A. Cosmic routine \citep{2001PASP..113.1420V}. One dimensional spectra were extracted using the APALL task, and then wavelength and flux calibrated using arc lamps and spectrophotometric standard-star spectra respectively observed at a similar airmass during the same night. Night sky emission lines were used to check the wavelength calibration, and shifts were applied when necessary. On the nights when standard star observations were unavailable, we used the sensitivity function, generated using STANDARD and SENSFUNC tasks, of the standard star observed in a close by night for flux calibration. The calibrated spectra were compared with the photometry and, when necessary, multiplied by a constant factor to correct for atmospheric transparency, bad-seeing conditions and slit losses to match the spectroscopic and photometric continuum flux. Finally, each spectra were de-redshifted to the heliocentric frame using the DOPCOR task.

\begin{table}
\caption{Log of the spectroscopic observations.}
\centering
\smallskip
\begin{tabular}{c c c c c c c}
\hline \hline
Date          & Phase$^\dagger$          & Grism      & Spectral Range        & Resolution & Instrument & Telescope       \\
              &(Days)                    &            & (\AA)                 &               &   & \\
\hline
2015-12-02      &   8.91 & Gr04       & 3360-7740       & 311 & AFOSC & EKAR  \\
2015-12-04      &  10.95 & Gr04      & 3360-7740       & 311 & AFOSC & EKAR  \\
2015-12-06      &  12.93 & Gr04      &  3360-7740      & 311 & AFOSC & EKAR  \\
2015-12-07      &  13.92 & --     & 3200-10000       & 400-700 & FLOYDS & FTN  \\
2015-12-08      &  14.90 & --	  & 3200-10000        & 400-700 & FLOYDS & FTN  \\
2015-12-11      &  17.88 & Gr04                  & 3360-7740       & 311 & AFOSC & EKAR  \\
2015-12-12      &  18.90 & --                & 3200-10000 & 400-700 & FLOYDS & FTN  \\
2015-12-17      &  23.90 & -- & 3200-10000     & 400-700 & FLOYDS & FTN  \\
2015-12-19      &  26.00 & VPH7                 & 3200-7000    & 311& AFOSC & EKAR   \\
2015-12-30      &  36.91 & --  & 3200-10000  & 400-700 & FLOYDS & FTN    \\
2016-01-08      &  45.91 & --        & 3200-10000     &  400-700 &  FLOYDS & FTN \\
2016-01-16      &  53.83 & --   & 3200-10000   & 400-700 &   FLOYDS & FTN \\
2016-01-24      &  61.81 & --    & 5400-10000     & 400-700 & FLOYDS & FTN     \\
2016-01-30      &  68.02 & Gr04              & 3200-9600  & 311 & ALFOSC & NOT     \\
2016-01-30      &  67.85 & --   & 3200-10000  & 400-700 & FLOYDS & FTN     \\
2016-02-05      &  73.97 & Gr04               & 3360-7740  & 311     & AFOSC & EKAR     \\
2016-02-10      &  78.81 & --   & 3200-10000  & 400-700 & FLOYDS & FTN     \\
2016-02-21      &  89.77 & --   & 5400-10000  & 400-700 & FLOYDS & FTN     \\
2016-03-03      & 100.68 & --   & 3200-10000 & 400-700 & FLOYDS & FTN    \\
2016-03-15      & 112.89 & --  & 3200-10000 & 400-700 & FLOYDS & FTN     \\
2016-03-16      & 113.90 & Gr04                 & 3200-9600 & 311 & ALFOSC & NOT     \\
2016-03-27      & 124.88 & -- & 3200-10000 & 400-700 & FLOYDS & FTN\\
2016-04-09      & 137.88 & -- & 3200-10000 & 400-700 & FLOYDS & FTN \\
2016-04-11      & 139.83 & -- & 3200-10000 & 400-700 & FLOYDS & FTN\\
2016-04-12      & 140.61 & -- & 5400-10000 & 400-700 & FLOYDS & FTN\\
2016-08-02      & 253.23 & R300B & 3600-7200 & 360 & OSIRIS & GTC \\
2016-08-21      & 272.16 & R300R & 4800-10000 & 348 & OSIRIS & GTC \\
\hline                                   
\end{tabular}
\newline
\begin{tablenotes}
\item[a]$^\dagger$since explosion epoch t$_0$ = 2457349.7 JD (2015 November 23)
     \end{tablenotes}

\label{tab:spectra_log}      
\end{table}



\bsp	
\label{lastpage}
\end{document}